\documentstyle[prc,preprint,aps,epsf]{revtex}
\def\pd{\partial}
\def\oln{\overline}
\def\olft{\overleftarrow}
\def\ds{\displaystyle}

\def\pslash#1{\slash \hspace{-2.5mm} #1}
\newcommand{\ignorethis}[1]{}

\def\beq{\begin{equation}}
\def\eeq{\end{equation} }
\def\bea{\begin{eqnarray}}
\def\eea{\end{eqnarray}}
\def\eqref#1{Eq.~(\ref{eq:#1})}
\def\eqlab#1{\label{eq:#1}}

\def\figlab#1{\label{fig:#1}}

\newcommand{\vslash}[1]{#1 \hspace{-0.5 em} /}
\begin{document}
\tighten
\title{Non-perturbative model for the half-off-shell $\gamma N N$ vertex}
\author{S. Kondratyuk, O. Scholten}
\address{Kernfysisch Versneller Instituut, 9747 AA Groningen, The Netherlands.}
\date{\today}
\maketitle
\begin{abstract}

Form factors in the nucleon-photon vertex with one off-shell nucleon are
calculated by dressing the vertex with pion loops up to infinite order. 
Cutting rules and dispersion relations are implemented in the model.
Using the prescription of minimal substitution we construct a 
$\gamma \pi N N$ vertex and show that it has to be included
in the model in order that the Ward-Takahashi identity for the $\gamma N N$ 
vertex be fulfilled. 
The vertex is to be applied in a 
coupled-channel K-matrix formalism for Compton scattering, pion photoproduction
and pion scattering. The form factors show a pronounced cusp structure at the
pion threshold. As an illustration of a consistent application of the 
model, we calculate the cross section of Compton scattering. 
To provide gauge invariance in Compton scattering, a four-point 
$\gamma \gamma N N$ 
contact term is constructed using minimal substitution.

\end{abstract}

\vspace{2cm}

\noindent {\bf 1999 PACS}:
13.40.Gp, 13.60.-r, 14.20.Dh, 13.60.Fz

\bigskip
\noindent
Corresponding author: \\
S. Kondratyuk, Kernfysisch Versneller Instituut \\
Zernikelaan 25, 9747 AA Groningen, The Netherlands  \\
e-mail:  KONDRAT@KVI.NL, \\
phone:  +31-(0)50-3636192, fax:    +31-(0)50-3634003 \\

- - - - - - - - - - - - - - - - -  \today\ - - - - - - - - - - - - - - - - -


\section{Introduction}


The nucleon-photon vertex function is parametrized by form factors that
depend on relevant Lorentz-invariants. For instance, the
$\gamma N N$ vertex with both nucleons on the mass shell has two form factors 
that are functions
of only one argument, the four-momentum squared of the photon. A more
complicated form of the vertex is encountered if one of the nucleons is allowed
to be off-shell. In general, such a vertex contains 6 form factors which depend 
on the four-momentum squared of the off-shell nucleon \cite{bincer}
(so-called half-off-shell form factors).

Electromagnetic vertices of the nucleon with one or both off-shell nucleons have
been studied in the past. The method of dispersion relations was applied in 
Refs.\cite{bincer,drellpag,nyman,hare}. Dynamical models based on a perturbative
dressing of the vertex with meson loops, within effective Lagrangian approaches,
were developed in articles 
\cite{nauskoch,tiemtjon,bosscherkoch,boskoch,dongmos}.  
The role of off-shell nucleon-photon form factors has been investigated, for
example, in models for proton-proton bremsstrahlung \cite{nyman_br,kondsch,li}
and virtual Compton scattering \cite{korschjong}.

In this paper the half-off-shell electromagnetic form factors of the nucleon are
calculated in a model based on an integral equation for the $\gamma N N$ vertex.
This non-perturbative approach comprises nucleons and pions as hadronic
degrees of freedom, and the vertex is dressed with pion loops up to infinite 
order. The dressed half-off-shell nucleon-pion vertex and the nucleon
propagator, necessary for the equation, were calculated
in \cite{us_pion} within a framework consistent with the
present model. The equation is solved using an iterative
procedure, without introducing bare form factors in the zeroth 
iteration vertex. At each iteration step, the
imaginary parts of the loop integrals are found by applying 
Cutkosky rules \cite{cut}. In doing so, only the discontinuities related to 
the intermediate states with one nucleon and one pion are taken into account 
in the loop integrals. The real parts are constructed using dispersion relations
\cite{bincer}. The dispersion integrals converge due to the sufficiently fast
falloff of the $\pi N N$ form factors in the loop diagrams.
The resulting $\gamma N N$ vertex is normalized in such 
a way that, at the point where both nucleons are on-shell, it reproduces the
physical anomalous magnetic moment of the nucleon.  

The $\gamma N N$ vertex must satisfy the Ward-Takahashi identity \cite{takah}.
This is achieved in our model by including in the equation a loop diagram with
a four-point $\gamma \pi N N$ vertex (the contact term). The latter is
constructed based on the dressed $\pi N N$ vertex using the minimal 
substitution prescription (various constructions of contact 
terms can be found in, e.g.\ \cite{ohta,haber,nag}). 
Here we use the technique of variational 
derivatives applied to
an action functional with higher derivatives corresponding to the
half-off-shell $\pi N N$ vertex. Such a procedure leads to a unique 
result only for the
longitudinal (with respect to the four-momentum of the photon) part of the 
four-point vertex. To investigate the role of the transverse terms, we
calculated the electromagnetic form factors utilizing two different 
$\gamma \pi N N$ vertices. We found that the transverse terms affect chiefly
the form factors related to negative-energy states of the off-shell nucleon.

The principle behind our approach can be outlined as follows. Given 
an interaction Lagrangian in terms of hadronic fields, one can 
use it to 
calculate the scattering amplitude for certain scattering processes (to be
specific, we consider Compton scattering, pion photoproduction
and pion scattering). 
In general, both real and imaginary parts of the loop integrals will be included
in this amplitude. Instead of the above approach, we follow an alternative
scheme. Namely, first we construct dressed half-off-shell $\gamma N N$ and 
$\pi N N$ vertices as well as the dressed nucleon propagator. This
corresponds to constructing effective nucleon-photon and nucleon-pion 
interactions accounting for the real parts of loop integrals in the
amplitude. To calculate the full amplitude, these dressed vertices and 
nucleon propagator have to be used in a K-matrix formalism
\cite{gouds,korsch,korschtim} since there 
the imaginary parts of the loop integrals are generated as part of the
method. It is important to note that in such a scheme one avoids double counting
of the loop diagrams. This aspect is discussed 
in Section V.       

To provide current conservation in the description
of Compton scattering, one needs a contact $\gamma \gamma N N$ term. 
We build such a term by minimal substitution, applying a technique
somewhat different (though equivalent) to that used for the construction of
the $\gamma \pi N N$ vertex.
As an application of the model, we calculate the cross section
for Compton scattering, where the K-matrix is constructed using the
dressed $\gamma N N$ and $\pi N N$ vertices and the nucleon propagator.
The result is compared with the cross section computed using the K-matrix in
which bare vertices and the free nucleon propagator are substituted. These
calculations both satisfy the low-energy theorem \cite{gelgold,low}. 
It follows from
this theorem that in an expansion of the cross section in the photon
energy, only the leading term -- the Thomson cross section -- does not 
depend on a particular model used for the description of the process. For the
simplified case where no magnetic form factors are included in the vertex, 
we calculated the next-to-leading
order term (quadratic in the photon energy) in the expansion of the forward 
scattering cross section. It is expressed in terms of two scalar 
functions parametrizing the renormalized 
dressed nucleon propagator.

In connection with applications of off-shell form factors for calculation of
physical observables, the following caveats are in order. In principle, 
one should utilize off-shell vertices to obtain various reaction amplitudes 
and calculate the observables from the corresponding S-matrix. However, the
off-shell form factors by themselves cannot be directly measured. In 
particular,
they can be changed by a redefinition of the nucleon field. 
At the same time, the field redefinition will in
general also change the four- and higher-point vertices. It is known that the 
S-matrix is independent of the representation of the fields \cite{kamef}.
Therefore, in a consistent calculation of the observables     
three- and higher-point Green's functions should be treated 
using the same model assumptions and representation of the fields. 
The link between off-shell effects 
and contact interactions was emphasized in \cite{scherfear}, where 
Compton scattering by a pion was considered in the framework of 
chiral perturbation theory. 
Also, pion-nucleon scattering in a particular model was used recently 
\cite{davpoul} 
to demonstrate how a redefinition of the nucleon field changes the off-shell 
dependence in three-point $\pi N N$ vertices and leads to occurrence of
a four-point $\pi \pi N N$ vertex, such that the total scattering amplitude
calculated using the two field representations remains the same. Symmetries
of the theory can relate Green's functions with different numbers of
external lines, the electromagnetic Ward-Takahashi identities
\cite{takah,kazes} being a relevant example. The four-point 
$\gamma \pi N N$ and $\gamma \gamma N N$ vertices constructed in this paper are
derived 
observing the 
requirement of current conservation for the pion photoproduction and Compton
scattering amplitudes.   

The paper is organized as follows. The general structure of the nucleon-photon
vertex and its reduction to the half-off-shell cases are given in 
Section II. The equation for the vertex is presented in Section III.A. The 
solution procedure is described in detail in Section III.B. In particular,
Section III.B.1 explains a mathematical method used to project the imaginary
parts of the form factors from the general expressions for the discontinuities
of the loop integrals on the right-hand side of the equation. The projection
method is applied in Sections III.B.2, III.B.3 to two of the three loop 
integrals. The minimal substitution procedure is employed in Section III.B.4 to
construct a contact $\gamma \pi N N$ vertex based on the general expression for 
the half-off-shell nucleon-pion vertex, followed in Section III.B.5 by the 
calculation of the discontinuity of the remaining third loop integral.    
The dispersion relations utilized 
to find the real parts of the form factors are given in Section III.B.6. 
Results for the form factors are discussed in Section IV. In Section V we
argue
that the half-off-shell $\gamma N N$ and $\pi N N$ vertices and the dressed 
nucleon propagator calculated in the framework of our model can be consistently
applied in a coupled-channel K-matrix formalism in such a way that double 
counting is avoided. To apply the model to (real)
Compton scattering, we need first to construct a contact $\gamma \gamma N N$
vertex. Such a vertex (or rather its matrix element between on-shell nucleon
states) is given in Section VI.A. Some results of the application of 
the model to 
Compton scattering are presented in Section VI.B. Section VI.B.1 contains 
a calculation of 
the next-to-leading order term in the low-energy expansion of the 
cross section for an illustrative case where the magnetic 
form factors are excluded form the
vertex. Section VII concludes the paper. Appendices A and B contain technical
details of the minimal substitution procedure utilized to construct the
contact $\gamma \pi N N$ and $\gamma \gamma N N$ vertices. In Appendix C we
prove that a gauge invariant amplitude for
pion photoproduction can be constructed using the contact $\gamma \pi N N$ term
and use this result to show that the equation for the nucleon-photon vertex is
consistent with the Ward-Takahashi identity.

\section{Structure of the $\gamma N N$ vertex\label{sec:Struc}}

We consider the (irreducible) $\gamma N N$ vertex operator 
which, in principle, is the sum of all connected Feynman diagrams
with one incoming nucleon (carrying the four-momentum $p$), one outgoing nucleon
($p^\prime$) and one photon ($q=p-p^\prime$), with the propagators for
the external legs stripped away. The most general Lorentz covariant form
can be written as \cite{bincer} \footnote{We use the Dirac representation for
the $\gamma$-matrices (see the appendix in the textbook \cite{itzub})}
\beq
-i e\, \Gamma_{\mu}(p^{\prime},p) = -i e \sum_{k,l=\pm} \Lambda_k(p^{\prime})\,
\left\{ \gamma_{\mu} F_1^{kl} -
i \frac{\sigma_{\mu \nu}q^{\nu}}{2m} F_2^{kl} -
\frac{q_{\mu}}{m} F_3^{kl} \right\}\, \Lambda_l(p).
\eqlab{genstruc}
\eeq
In this formula, $e$ and $m$ are the elementary electric charge 
and the mass of the nucleon, the operators
\beq
\Lambda_{\pm}(p) \equiv \frac{{\pm}\vslash{p}+m}{2m},
\eqlab{proj}
\eeq
and the 12 functions (form factors) $F_i^{kl}$ depend on the momenta squared of
the three particles in the vertex, $F_i^{kl} = F_i^{kl}(p^{\prime 2},p^2,q^2)$.
We will consider only real photons throughout the paper, i.e.\ $q^2=0$, 
and hence omit the argument $q^2$ of the form factors.
In the description of physical processes, the vertex \eqref{genstruc} 
will always 
enter in the scalar product $\Gamma(p^\prime,p)\cdot \epsilon$ with the
polarization vector $\epsilon^\mu$ of the photon. Since $q\cdot \epsilon=0$
for real photons, the form factors $F_3^{kl}$ will be dropped from 
further consideration in this paper.
The isospin structure of the vertex is
$\Gamma_{\mu} = \Gamma_{\mu}^s + {\tau}_3 \Gamma_{\mu}^v$,
where $\Gamma_{\mu}^s$ and $\Gamma_{\mu}^v$ are the isoscalar and isovector
parts, and ${\tau}_3=diag(1,-1)$ is the third Pauli matrix. 
Correspondingly, the form
factors have a similar decomposition in the isospin space,
$F_i^{kl} = (F_i^{kl})^s + {\tau}_3 (F_i^{kl})^v$. Thus, the cases of the
proton-photon and neutron-photon vertices correspond to taking the 
form factors $(F_i^{kl})^s + (F_i^{kl})^v$  and $(F_i^{kl})^s - (F_i^{kl})^v$,
respectively.

Invariance of the theory with respect to charge conjugation and space-time
inversion allows one to write the following relations for the vertex
\cite{gelgold}:
\beq
C^{-1} \Gamma_{\mu}(p^{\prime},p) C = -\Gamma_{\mu}^T(-p,-p^{\prime}),
\eqlab{chargeconj}
\eeq
\beq
\gamma_5 \Gamma_{\mu}(p^{\prime},p) \gamma_5 = -\Gamma_{\mu}(-p^{\prime},-p),
\eqlab{spacetimerev}
\eeq
where $C = i \gamma_2 \gamma_0$ is the charge conjugation matrix. Applying
these two symmetry transformations successively gives
\beq
{\gamma}_1 {\gamma}_3 \Gamma_{\mu}(p^{\prime},p) {\gamma}_3 {\gamma}_1 =
\Gamma_{\mu}^T(p,p^{\prime}).
\eqlab{cpt}
\eeq
On substituting \eqref{genstruc} in \eqref{cpt} one obtains the following
relations among the form factors:
\beq
F_{1,2}^{kl}(p^{\prime 2},p^2) = F_{1,2}^{lk}(p^2,p^{\prime 2}).
\eqlab{symm12}
\eeq

For the course of this paper, we will be interested in the so-called
half-off-shell vertices in which only one of the nucleons is off-shell,
i.e. either $p^{\prime 2}=m^2$ and
$\overline{u}(p^{\prime})\,\vslash{p}^{\prime}=\overline{u}(p^{\prime})\,m$
or $p^2=m^2$ and
$\vslash{p}\, u(p)=m\, u(p)$, where $u(p)$ is the positive energy spinor
of the nucleon with the four-momentum $p$.
These
vertices will be denoted $\Gamma_{\mu}(m,p)$ and $\Gamma_{\mu}(p^{\prime},m)$,
respectively. 
Using Eqs.~(\ref{eq:symm12}) with $p^{\prime 2}=m^2$, one can
relate the form factors in the half-off-shell vertices with the
incoming and outgoing on-shell nucleons. We denote $F_i^{++}(m^2,p^2)
\equiv F_i^{+}(p^2)$ and $F_i^{+-}(m^2,p^2) \equiv F_i^{-}(p^2)$ and call
these functions the half-off-shell form factors. If the outgoing nucleon is
on-shell, one obtains from \eqref{genstruc}
\beq
\oln{u}(p^\prime) \Gamma_{\mu}(m,p) =
\oln{u}(p^{\prime})\sum_{l={\pm}}
\left\{ \gamma_{\mu} F_1^{l}(p^2) -
i \frac{\sigma_{\mu \nu}q^{\nu}}{2m} F_2^{l}(p^2) \right\}\,\Lambda_l(p).
\eqlab{finon}
\eeq
Similarly, for the incoming on-shell nucleon,
\beq
\Gamma_{\mu}(p^{\prime},m) u(p)=
\Bigg( \sum_{k=\pm} \Lambda_k(p^{\prime})\,
\left\{ \gamma_{\mu} F_1^{k}(p^{\prime 2}) -
i \frac{\sigma_{\mu \nu}q^{\nu}}{2m} F_2^{k}(p^{\prime 2})\right\} \Bigg)u(p).
\eqlab{initon}
\eeq

\section{Description of the model \label{sec:III}}

\subsection{Basic ingredients}

The model is based on the following integral equation for the vertex:
\beq
\begin{array}{rl}
\Gamma_{\mu}(m,p) = &\Gamma_{\mu}^0(m,p) 
-i{\ds \int\! \frac{d^4 k}{(2\pi)^4}}\,
\Gamma_{5,\alpha}(m,p-k)\, S(p-k)\, \Gamma_{5,\beta}(p-k,p)\, D((k-q)^2) \\
&\times V_{\mu,\alpha \beta}(k-q,k)\,D(k^2) \; 
-i{\ds \int\! \frac{d^4 k}{(2\pi)^4}}\,
\Gamma_{5,\alpha}(m,p^{\prime}-k)\, S(p^{\prime}-k) \\
&\times \Gamma_{\mu}(p^{\prime}-k,p-k)\,S(p-k) 
\Gamma_{5,\alpha}(p-k,p)\,D(k^2) \\
&-i{\ds \int\! \frac{d^4 k}{(2\pi)^4}}\,
M_{\mu, \alpha}(m,p-k,q)\,S(p-k)\,
\Gamma_{5,\alpha}(p-k,p)\,D(k^2) \\
&-i{\ds \int\! \frac{d^4 k}{(2\pi)^4}}\,
\Gamma_{5,\alpha}(m,p^{\prime}-k)\,S(p^{\prime}-k)\,
M_{\mu, \alpha}(p^{\prime}-k,p,q)\,D(k^2) ,
\end{array}
\eqlab{sys} 
\eeq
represented graphically in Fig.~1. This equation expresses the dressing of a
bare vertex $\Gamma_{\mu}^0(m,p)$ with an infinite series of pion loops.

In \eqref{sys}, $\Gamma_{5,\alpha}(p^{\prime},p)$ denotes the 
nucleon-pion vertex function,
where the incoming and the outgoing nucleons have four-momenta $p$ and
$p^{\prime}$, respectively. Although both nucleons can be off-shell in the
$\pi N N$ vertex, we shall need only half-off-shell vertices in our model
as will become clear below. For example, if the outgoing nucleon is on the mass
shell, one has $p^{\prime 2}=m^2$ and
$\overline{u}(p^{\prime})\,\vslash{p}^{\prime}=\overline{u}(p^{\prime})\,m$.
Acting with the 
nucleon-pion vertex on $\overline{u}(p^{\prime})$ to the left gives
\begin{equation}
\oln{u}(p^\prime)\Gamma_{5,\alpha}(m,p) = 
\oln{u}(p^\prime)\tau_\alpha \Gamma_5(m,p) =
\oln{u}(p^\prime)\,\tau_{\alpha}\,\gamma_5 \Big(G_1(m^2,p^2,k^2)+ 
\frac{\vslash{p}-m}{m}
G_2(m^2,p^2,k^2)\Big),
\eqlab{pinnver}
\end{equation}
where $k=p-p^{\prime}$ is the four-momentum of the pion. The functions
$G_{1,2}(m^2,p^2,k^2)$ are
the (half-off-shell) form factors in the nucleon-pion vertex.
If the incoming nucleon is on-shell and the outgoing one is off-shell,
the action of the vertex on $u(p)$ to the right reads
\begin{equation}
\Gamma_{5,\alpha}(p^\prime,m) u(p)= \tau_\alpha \Gamma_5(p^\prime,m)u(p) =
\tau_{\alpha}\, \Big(G_1(p^{\prime 2},m^2,k^2)+ \frac{\vslash{p}^\prime-m}{m}
G_2(p^{\prime 2},m^2,k^2)\Big)\gamma_5 u(p).
\eqlab{pinnver_out}
\end{equation}

$S(p)$ is the renormalized dressed nucleon propagator for which
the following parametrization is used:
\beq
\big( S(p) \big)^{-1} = \alpha(p^2) \Big( \vslash{p}-\xi(p^2) \Big).
\eqlab{prop_inv}
\eeq
The functions $\alpha(p^2)$, $\xi(p^2)$, as well as
$G_{1,2}(m^2,p^2,m_{\pi}^2)$ ($m_{\pi}$ being the mass of the pion), were
calculated in the model of Ref. \cite{us_pion}. We will use these results here.
The dependence of the nucleon-pion form factors on the four-momentum squared
of the pion $k^2$, at least for $k^2 \le m_{\pi}^2$, can be inferred from a
generalization of the aforementioned model.
This will be discussed below in more detail.

$D(k^2)$ is the pion propagator function. For simplicity, we use the free
Feynman propagator, $\big( D(k^2) \big)^{-1} = k^2 - m_{\pi}^2$.
The pion-photon vertex $V_{\mu,\alpha \beta}(k^{\prime},k)$ is chosen such that
the Ward-Takahashi identity is fulfilled with the free propagator $D(k^2)$,
\beq
(k^\prime-k)^{\mu} V_{\mu,\alpha \beta}(k^{\prime},k) =
(\hat{e}_{\pi})_{\alpha \beta} \left[ \big( D(k^{\prime 2}) \big)^{-1} -
\big( D(k^2) \big)^{-1} \right],
\eqlab{wtipion}
\eeq
giving
$V_{\mu,\alpha \beta}(k^{\prime},k)=(\hat{e}_{\pi})_{\alpha \beta}
(k_{\mu}+k^{\prime}_{\mu})$,
where the pion charge operator
$(\hat{e}_{\pi})_{\alpha \beta} = -i \epsilon_{\alpha \beta 3}$.

The function $M_{\mu, \alpha}(p^{\prime},p,q)$ in \eqref{sys} denotes a contact
$\pi \gamma N N$ vertex, where $p^{\prime}$ and $p$ are the momenta of the
outgoing and incoming nucleons, respectively, and $q$ is the four-momentum of
the (outgoing) photon. The construction of the contact vertex will be described
in Section III.B.4. Here we only mention that the two last terms
in \eqref{sys} are required for the $\gamma N N$ vertex to satisfy the
Ward-Takahashi identity.

Finally, the first term on the right-hand side of \eqref{sys} is a ``bare"
nucleon-photon vertex. We use the following form:
\beq
\oln{u}(p^\prime) \Gamma_{\mu}^0(m,p) = 
\oln{u}(p^\prime) ( \gamma_{\mu} \hat{e}_N - i {\kappa}_B
\frac{{\sigma}_{\mu \nu} q^{\nu}}{2 m} ),
\eqlab{ver_bare}
\eeq
where the nucleon charge operator $\hat{e}_N = (1+{\tau}_3)/2$. The constant
$\kappa_B$ is a bare anomalous magnetic moment of nucleon. It has the standard
isospin decomposition $\kappa_B = (\kappa_B)^s + \tau_3 (\kappa_B)^v$ and is to
be adjusted to provide the normalization of the dressed vertex, which will be
discussed in Section III.B.6.

\subsection{Solution procedure}

To construct a 
solution of \eqref{sys}, an iterative procedure is utilized. 
First, we note that all integrals on the right-hand
side of \eqref{sys}, except the second one, are inhomogeneities of the equation
because they do not depend on the $\gamma N N$ vertex. Therefore, they need to 
be calculated only once. We start with the first integral which we denote
$\Gamma_{\mu}(1)$ (the other three integrals will be denoted
$\Gamma_{\mu}(2)$, $\Gamma_{\mu}(3)$ and $\Gamma_{\mu}(4)$). Consider the pole
contribution, $\Gamma_{\mu, I}(1)$, to this integral. It comes
from cutting the nucleon propagator
$S(p-k)$ and the pion propagator $D(k^2)$, i.e. from putting the
corresponding particles on their mass shell (see the upper left diagram 
in Fig. 2).
The imaginary parts of the form factors  $F_i^{\pm}(p^2)$ are obtained from
such pole contributions to all the loop integrals on the right-hand
side of \eqref{sys}.
According to Cutkosky rules \cite{cut}, we replace $S(p-k)$ with
$-2 i \pi (\vslash{p}-\vslash{k}+m) \delta((p-k)^2-m^2) \Theta(p_0-k_0)$
and $D(k^2)$ with $-2 i \pi \delta(k^2-m_{\pi}^2) \Theta(k_0)$.
One of the advantages of the use of cutting rules is that throughout the
solution procedure we need 
vertices with only one off-shell nucleon.
In other words, the knowledge of {\it full}-off-shell form factors will not be
required for the calculation of the pole contributions to the loop integrals,
as can be seen from Fig.~2.

In evaluating $\Gamma_{\mu, I}(1)$, only real parts of the nucleon-pion form
factors $G_i$ are taken into account. As will be shown in Section V,
this is consistent with the usage of these vertices and nucleon self-energy
in a K-matrix approach to pion-nucleon scattering, pion photoproduction and
Compton scattering \cite{korsch,korschtim}. 
The same qualification will apply to the pole
contributions of all loop integrals in \eqref{sys} (see Fig. 2).

Denoting $g_i \equiv Re G_i$, we have for the pole contribution:
\begin{eqnarray}
\Gamma_{\mu, I}(1) &=& {\displaystyle \frac{2 \tau_3}{8 \pi^2} \int }\! \,
d^4 k\,\gamma_5\, g_1(m^2,m^2,(k-q)^2)\, (\vslash{p}-\vslash{k}+m)\, \gamma_5 
\nonumber \\
&&\times \Big[\, g_1(m^2,p^2,m_{\pi}^2)+
{\displaystyle \frac{\vslash{p}-m}{m}} g_2(m^2,p^2,m_{\pi}^2)\, \Big]\, 
{\displaystyle \frac{2 k_{\mu} - q_{\mu}}{(k-q)^2-m_{\pi}^2}} \nonumber \\
&&\times \delta((p-k)^2-m^2)\,
\Theta(p_0-k_0)\, \delta(k^2-m_{\pi}^2)\, \Theta(k_0).\eqlab{pole_1}
\end{eqnarray}
The isospin factor $2\tau_3$ occurs due to multiplying Pauli 
matrices at the
nucleon-pion vertices with the pion charge operator:
$\tau_\alpha(\hat{e}_\pi)_{\alpha \beta}\tau_\beta=2\tau_3$.

\subsubsection{Projection method}

In this section we describe
a method used to project out the imaginary parts of the form
factors. This method will be applied to the pole
contributions from the integrals in \eqref{sys}.

Suppose we consider the half-off-shell vertex \eqref{finon}. It can be
regarded as an element of a six-dimensional vector space $V_6$ with the basis
\beq
\begin{array}{ll}
{\displaystyle (e_1)_{\mu} = \Lambda_{+}(p^{\prime}) \gamma_{\mu}
\Lambda_{+}(p)}, &
{\displaystyle (e_2)_{\mu} = \Lambda_{+}(p^{\prime}) \gamma_{\mu}
\Lambda_{-}(p) }, \\
{\displaystyle (e_3)_{\mu} = \Lambda_{+}(p^{\prime})
(-i)\frac{\sigma_{\mu \nu} q^{\nu}}{2 m} \Lambda_{+}(p) }, &
{\displaystyle (e_4)_{\mu} = \Lambda_{+}(p^{\prime})
(-i)\frac{\sigma_{\mu \nu} q^{\nu}}{2 m} \Lambda_{-}(p) }, \\
{\displaystyle (e_5)_{\mu} = \Lambda_{+}(p^{\prime}) (-)\frac{q_{\mu}}{m}
\Lambda_{+}(p) }, &
{\displaystyle (e_6)_{\mu} = \Lambda_{+}(p^{\prime}) (-)\frac{q_{\mu}}{m}
\Lambda_{-}(p) },
\end{array}
\eqlab{basis}
\eeq
defined over a ring of complex-valued functions (form factors). For example,
the integral of \eqref{pole_1}, multiplied from the left by 
$\Lambda_{+}(p^{\prime})$, belongs to $V_6$,
$v_{\mu}(1)=\Lambda_{+}(p^{\prime}) \Gamma_{\mu, I}(1) \in V_6$.
Thus, to find contributions to the imaginary parts of the form factors
$F_i^{\pm}$ from the integral $\Gamma_{\mu, I}(1)$ amounts to finding the
coefficients in an expansion of $v_{\mu}(1)$ over the basis \eqref{basis}.

The dual space $V_6^{*}$ can be defined as spanned over the basis
$(\theta^i)^{\mu} = g^{\mu \lambda} \overline{(e_i})_{\lambda}$, where the
over-lining denotes the Dirac conjugate of an operator,
$\overline{A} \equiv \gamma_0 A^{\dagger} \gamma_0$. Explicitly, we have
\beq
\begin{array}{ll}
{\displaystyle (\theta^1)^{\mu} = \Lambda_{+}(p) \gamma^{\mu}
\Lambda_{+}(p^{\prime}) }, &
{\displaystyle (\theta^2)^{\mu} = \Lambda_{-}(p) \gamma^{\mu}
\Lambda_{+}(p^{\prime}) }, \\
{\displaystyle (\theta^3)^{\mu} = \Lambda_{+}(p)
i \frac{\sigma^{\mu \nu} q_{\nu}}{2 m} \Lambda_{+}(p^{\prime}) }, &
{\displaystyle (\theta^4)^{\mu} = \Lambda_{-}(p)
i \frac{\sigma^{\mu \nu} q_{\nu}}{2 m} \Lambda_{+}(p^{\prime}) }, \\
{\displaystyle (\theta^5)^{\mu} = \Lambda_{+}(p) (-)\frac{q^{\mu}}{m}
\Lambda_{+}(p^{\prime}) }, &
{\displaystyle (\theta^6)^{\mu} = \Lambda_{-}(p) (-)\frac{q^{\mu}}{m}
\Lambda_{+}(p^{\prime}) }.
\end{array}
\eqlab{dbasis}
\eeq
For any one-form $\omega^{\mu} \in V_6^{*}$ and any vector 
$v_{\mu} \in V_6$, we define the action of
$\omega^{\mu}$ on $v_{\mu}$ by the pairing
\beq
\langle\, \omega^{\mu},\, v_{\mu}\, \rangle = tr (\omega^{\mu} v_{\mu}),
\eqlab{dual_act}
\eeq
with a tacit summation over $\mu$. Now if
\beq
v_{\mu} = \sum_{i=1}^{6} c^i (e_i)_{\mu},
\eqlab{expans}
\eeq
then the coefficients are obtained from the formula
\beq
c^k = \sum_{l=1}^{6}
(E^{-1})^k_l\, \Big\langle\, (\theta^l)^{\mu},\, v_{\mu}\, \Big\rangle,
\eqlab{coeff}
\eeq
where the matrix
$E^i_j = \langle\, (\theta^i)^{\mu},\, (e_j)_{\mu}\, \rangle$.
The coefficients $c^k$ are the form factors (or, more
precisely, contributions to the imaginary parts of the form factors).
Thus, we identify
\beq
\begin{array}{l}
c^1=Im F_1^{+},\;\;\;\;c^2=Im F_1^{-},\;\;\;\; c^3=Im F_2^{+}, \\
c^4=Im F_2^{-},\;\;\;\; c^5=Im F_3^{+},\;\;\;\;c^6=Im F_3^{-}.
\end{array}
\eqlab{identif}
\eeq

\subsubsection{Calculating $\Gamma_{\mu, I}(1)$}

We consider first the case in which only the contributions from
the integral $\Gamma_{\mu, I}(1)$ are implied in \eqref{identif}, and therefore
we denote the coefficients $c^k$ as $c^k(1)$.
Then for
$v_{\mu}=v_{\mu}(1)=\Lambda_{+}(p^{\prime}) \Gamma_{\mu, I}(1)$,
the right-hand side of
\eqref{coeff} is an integral over a scalar function. The latter is found by
taking traces of $\gamma$-matrices and performing matrix multiplications.
These straightforward but tedious operations were done with the help 
of the algebraic programming
system  REDUCE \cite{red}.

More specifically, using \eqref{pole_1} to find $v_{\mu}(1)$ and
substituting the latter in \eqref{coeff} in the place of $v_{\mu}$, we obtain
\begin{eqnarray}
c^i(1)&=&{\displaystyle \frac{2 \tau_3}{8 \pi^2} \int }\! \,d^4 k\, V^i(k)\,
{\displaystyle \frac{g_1(m^2,m^2,(k-q)^2)}{(k-q)^2-m_{\pi}^2} } \nonumber \\
&& \times\delta((p-k)^2-m^2)\,
\Theta(p_0-k_0)\, \delta(k^2-m_{\pi}^2)\, \Theta(k_0),\eqlab{ff_1}
\end{eqnarray}
where
\begin{eqnarray}
V^i(k)&=&{\displaystyle \sum_{j=1}^{6}}(E^{-1})^i_j\,\Bigg{\langle}\,
(\theta^j)^{\mu}\mbox{\Large ,}\; \Lambda_{+}(p^{\prime})\,
\gamma_5\, (\vslash{p}-\vslash{k}+m)\, \gamma_5\,
\big[\, g_1(m^2,p^2,m_{\pi}^2) \nonumber \\
&&+ {\displaystyle \frac{\vslash{p}-m}{m}} g_2(m^2,p^2,m_{\pi}^2)\, \big]  
(2 k_{\mu} - q_{\mu})\, \Bigg{\rangle} \eqlab{vbig}
\end{eqnarray}
is a scalar function of the four-momentum $k_\mu$, 
the variable of integration. The integral in
\eqref{ff_1} is a Lorentz-scalar and therefore can be evaluated in any frame of
reference. We choose the rest frame of the incoming nucleon, i.e.\ we put
$p_{\mu} = w\,\delta_{\mu 0}$, where $w=\sqrt{p^2}$ is the invariant mass of
the off-shell nucleon. Further, we call $x$ the cosine of the
polar angle between the three-vectors $\overrightarrow{q}$ and
$\overrightarrow{k}$. Then, due to the delta- and theta-functions, the
integral in \eqref{ff_1} can be reduced to the one-dimensional integral
\beq
c^i(1)={\displaystyle \tau_3 \frac{r(p^2)}{16 \pi p^2}\,
\Theta(p^2-(m+m_{\pi})^2)
\int_{-1}^1\! d x\,
\widetilde{V}^i(x)\,
\frac{g_1(m^2,m^2,(k-q)^2)}{(k-q)^2-m_{\pi}^2} },
\eqlab{ff_1x}
\eeq
where $r(p^2)=\sqrt{\lambda(p^2,m^2,m_{\pi}^2)}$, with the K\"all\'en function
defined as $\lambda(x,y,z)\equiv (x-y-z)^2-4yz$.
Here, the function $\widetilde{V}^i(x)$ is obtained from $V^i(k)$ if one
expresses scalar products of vectors in the problem in the variable $x$,
\beq
(k \cdot q)=\frac{(p^2+m_{\pi}^2-m^2)(p^2-m^2)}{4 p^2} -
\frac{r(p^2)(p^2-m^2)}{4 p^2} \,x,
\eqlab{kq}
\eeq
\beq
(k \cdot p^{\prime})=\frac{(p^2+m_{\pi}^2-m^2)(p^2+m^2)}{4 p^2} +
\frac{r(p^2)(p^2-m^2)}{4 p^2} \,x,
\eqlab{kppr}
\eeq
and takes into account that $q=p-p^{\prime}$.
Finally, the integral in \eqref{ff_1x} is done numerically.

One point needs to be addressed here.
The form factor $g_1(m^2,m^2,(k-q)^2)$ in \eqref{ff_1x} is pertinent to the
nucleon-pion vertex with both nucleons on-shell and the pion off-shell.
Taking into account that $p^2 \geq (m+m_{\pi})^2$, it can be seen 
from \eqref{kq} that $(k-q)^2 < m_{\pi}^2$, for any $x \in [-1,1]$.
As mentioned above, the model of Ref. \cite{us_pion} can be generalized to
find the form factors in the nucleon-pion vertex with the pion
as well as one of the nucleons off-shell. Moreover, this generalization is 
possible in the framework of Ref. \cite{us_pion} only for the pion 
four-momenta squared less than
or equal to the pion mass squared (otherwise the absolute values of external
momenta in the nucleon-pion vertex would become complex).
In the approach adopted in \cite{us_pion} we find that the dependence 
of $g_1(m^2,m^2,k^2)$ on $k^2$ is very weak 
(in fact, it is almost constant). A possible reason for this
is that t-channel resonances, such as the $\rho$-meson, have not been
included in the model.
Thus, we have approximated $g_1(m^2,m^2,(k-q)^2)$
in \eqref{ff_1x} by the pion-nucleon coupling constant
(we adopt the value from Ref. \cite{korschtim}, $g=13.02$).

\subsubsection{Calculating $\Gamma_{\mu,I}(2)$}

Now we turn to the pinching pole contribution from the second integral on the
right-hand side of \eqref{sys} (the second diagram in Fig.~2). Contrary
to $\Gamma_{\mu}(1)$ considered above, this integral
depends on the unknown half-off-shell $\gamma N N$ vertex and therefore has to
be considered in the context of the iterative procedure applied to \eqref{sys}.
Let $F^{\pm,n}_i(p^2)$ denote the form factors found after the n$^{th}$
iteration step. When calculating $\Gamma_\mu^{n+1}(2)$, the n+1$^{st}$ iteration
for $\Gamma_{\mu}(2)$, we will retain only
the real parts of $F^{\pm,n}_i(p^2)$ as well as the nucleon-pion form factors
and the functions $\alpha(p^2)$ and $\xi(p^2)$ parametrizing the nucleon 
propagator. The reason for doing so will be explained in Section V.
So we introduce the notation
$t^{\pm}_i(p^2) \equiv Re (F^{\pm})^n_i(p^2)$ and
$g_i(p^2) \equiv g_i(m^2,p^2,m_\pi^2)$. Also, only real parts of the functions
$\alpha(p^2)$ and $\xi(p^2)$ are implied in the following formulas. 
The pole
contribution to $\Gamma_\mu^{n+1}(2)$ can be explicitly written,                          
\begin{eqnarray}
\Gamma_{\mu,I}^{n+1}(2)&=&{\ds \frac{1}{8\pi^2}\int}\!d^4k \tau_\alpha
\gamma_5 \Big[ g_1((p^\prime-k)^2)+
{\ds \frac{\vslash{p}^\prime-\vslash{k}-m}{m}}g_2((p^\prime-k)^2) \Big] 
\Big(\vslash{p}^\prime-\vslash{k}+\xi((p^\prime-k)^2)\Big) \nonumber \\ 
&&\times \Bigg\{ \Lambda_{+}(p^\prime-k) \Big[ \gamma_\mu
t^{+}_1((p^\prime-k)^2)-i{\ds \frac{\sigma_{\mu \nu}q^\nu}{2m}}
t^{+}_2((p^\prime-k)^2) \Big] \nonumber \\ 
&&+ \Lambda_{-}(p^\prime-k)\Big[ \gamma_\mu
t^{-}_1((p^\prime-k)^2)-i{\ds \frac{\sigma_{\mu \nu}q^\nu}{2m}}
t^{-}_2((p^\prime-k)^2)  \Big]\Bigg\} \nonumber \\ 
&&\times (\vslash{p}-\vslash{k}+m)\tau_\alpha \gamma_5
\Big[ g_1(p^2)+{\ds \frac{\vslash{p}-m}{m}}g_2(p^2) \Big] \nonumber \\
&&\times{\ds \frac{1}
{\alpha((p^\prime-k)^2)[(p^\prime-k)^2-\xi^2((p^\prime-k)^2)]}} \nonumber \\   
&&\times \delta((p-k)^2-m^2)\,
\Theta(p_0-k_0)\, \delta(k^2-m_{\pi}^2)\, \Theta(k_0). \eqlab{pole_2}
\end{eqnarray}
Note that this expression contains both isoscalar and isovector parts.
Indeed, since $t^{\pm}_i=(t^{\pm}_i)^s+\tau_3(t^{\pm}_i)^v$, the isospin
structure of the integrand is 
$\tau_\alpha [(t^{\pm}_i)^s+\tau_3(t^{\pm}_i)^v] \tau_\alpha =
3(t^{\pm}_i)^s - (t^{\pm}_i)^v$.

In terms of the projection method described above, we have now
$v_\mu=v_\mu(2)=\Lambda_{+}(p^\prime)\Gamma_{\mu,I}(2)$. The contributions to
the imaginary parts of the form factors coming form $\Gamma_{\mu,I}^{n+1}(2)$
are found from \eqref{coeff},
\begin{eqnarray}
c^{i,n+1}(2)&=&{\ds \frac{1}{8\pi^2}\int}\!d^4k\, 
{\ds\frac{U^i(k)}{\alpha((p^\prime-k)^2)[(p^\prime-k)^2-
\xi^2((p^\prime-k)^2)]}}  \nonumber \\
&&\times \delta((p-k)^2-m^2)\,
\Theta(p_0-k_0)\, \delta(k^2-m_{\pi}^2)\, \Theta(k_0), \eqlab{ff_2}
\end{eqnarray}
where
\begin{eqnarray}
U^i(k)&= &\sum_{j=1}^{6} (E^{-1})^i_j \Bigg\langle\;(\theta^j)^\mu
\mbox{\Large ,}\,\Lambda_{+}(p^\prime)\Bigg\{ \tau_\alpha
\gamma_5 \Big[ g_1((p^\prime-k)^2)+
{\ds \frac{\vslash{p}^\prime-\vslash{k}-m}{m}}g_2((p^\prime-k)^2) \Big]
\nonumber \\
&&\times \Big(\vslash{p}^\prime-\vslash{k}+\xi((p^\prime-k)^2)\Big) \Bigg\}\,
\Bigg\{ \Lambda_{+}(p^\prime-k) \Big[ \gamma_\mu
t^{+}_1((p^\prime-k)^2) \nonumber \\
&&-i{\ds \frac{\sigma_{\mu \nu}q^\nu}{2m}}
t^{+}_2((p^\prime-k)^2) \Big]  
+ \Lambda_{-}(p^\prime-k)\Big[ \gamma_\mu
t^{-}_1((p^\prime-k)^2)-i{\ds \frac{\sigma_{\mu \nu}q^\nu}{2m}}
t^{-}_2((p^\prime-k)^2) \Big]\Bigg\}  \nonumber \\
&&\times (\vslash{p}-\vslash{k}+m)\tau_\alpha \gamma_5
\Big[ g_1(p^2)+{\ds \frac{\vslash{p}-m}{m}}g_2(p^2) \Big] \Bigg\}\;
\Bigg\rangle.  \eqlab{ubig}
\end{eqnarray}
We calculate the integral in \eqref{ff_2} in the same frame of reference
as the integral of \eqref{ff_1}. Substituting Eqs.~(\ref{eq:kq},\ref{eq:kppr})
in \eqref{ubig} converts $U^i(k)$ to $\widetilde{U}^i(x)$, and we obtain the
one-dimensional integral
\beq
c^{i,n+1}(2)=\frac{r(p^2)}{32 \pi p^2}\,\Theta(p^2-(m+m_{\pi})^2)
\int_{-1}^1\!dx\,
\frac{\widetilde{U}^i(x)}{\alpha((p^\prime-k)^2)[(p^\prime-k)^2-
\xi^2((p^\prime-k)^2)]},
\eqlab{ff_2x}
\eeq
which is evaluated numerically.

\subsubsection{Construction of a $\gamma \pi N N$ vertex}

Let us consider the last two integrals in \eqref{sys} or rather the pole
contributions associated with them. We cut the nucleon and pion
propagators in the loops of the last two diagrams in Fig.~1. 
The pole contribution of the last integral equals zero for the following reason.
By putting the pion and the nucleon in this integral 
on the mass shell, the following product of 
delta- and theta-functions emerges:
\beq
\delta((p^{\prime}-k)^2-m^2)\,
\Theta(p^{\prime}_0-k_0)\, \delta(k^2-m_{\pi}^2)\, \Theta(k_0),
\eqlab{cut_redund}
\eeq
where in addition $p^{\prime 2}=m^2$. This
combination can only be non-zero if $m_{\pi}^2-2 p^{\prime} k = 0$. But using
\eqref{kppr}, it is clear that the last condition cannot be met
for any $x \in [-1,1]$, and hence the pole contribution to the last
integral in \eqref{sys} vanishes, $\Gamma_{\mu,I}(4)=0$.

Consider next the pole contribution $\Gamma_{\mu,I}(3)$ to the second last 
integral in \eqref{sys} (the lower picture in Fig.~2). 
It contains a ``contact" $\gamma \pi N N$ vertex. We build such a vertex
based on the
dressed half-off-shell $\pi N N$ vertices 
Eqs.~(\ref{eq:pinnver},\ref{eq:pinnver_out}). 
The latter can be rewritten as 
\begin{equation}
\Gamma^5_{\alpha}(m,p)=
\tau_{\alpha}\,\gamma^5 \Big(g_{12}(p^2)+ \frac{\vslash{p}}{m} g_2(p^2)\Big)
\eqlab{4}
\end{equation}
and
\begin{equation}
\Gamma^5_{\alpha}(p^\prime,m)=
\tau_{\alpha}\,\Big(g_{12}(p^{\prime 2})+ \frac{\vslash{p}^\prime}{m} 
g_2(p^{\prime 2})\Big)\,\gamma^5,
\eqlab{4_out}
\end{equation}
where we have suppressed the irrelevant arguments of the
form factors. Also, as before, we retain only 
the real parts of the form
factors, with the notation $g_{12} \equiv Re\,(G_1-G_2)$ and 
$g_2 \equiv Re\,G_2$.  

Let $\psi(x)$ and ${\phi}^{\alpha}(x)$
denote the nucleon spinor field and the pion pseudoscalar field, where
the latter explicitly bears the isospin index.
Then one can write an action integral $S$
corresponding to a vertex which reduces to 
\eqref{4} or \eqref{4_out} if the outgoing or the incoming nucleon is on-shell,
respectively. We write this action as the sum of two functionals, $S=I+J$, where
\beq
I = {\displaystyle -i \int\! d^4x \overline{\psi} \gamma^5 \tau_{\alpha}
\phi^{\alpha} [f(-{\Box}) \psi] -i \int\! d^4x
\overline{[f(-{\Box}) \psi]}
\phi^{\alpha} \tau_{\alpha} \gamma^5 \psi }
\eqlab{act_ps}
\eeq
and
\beq
J = {\displaystyle \int\! d^4x \overline{\psi} \gamma^5 \tau_{\alpha}
\phi^{\alpha} [g(-{\Box}) \vslash{\pd} \psi] - \int\! d^4x
\overline{[g(-{\Box}) \vslash{\pd} \psi]}
\phi^{\alpha} \tau_{\alpha} \gamma^5 \psi },
\eqlab{act_pv}
\eeq
with the D'Alambertian
$\Box \equiv {\partial}^2 = \partial_{\mu} \partial^{\mu}$, and
\beq
f(-{\Box}) = g_{12}(-{\Box})-\frac{g_1(m^2)}{2},\;\;\; 
g(-{\Box}) = \frac{g_2(-{\Box})}{m}.
\eqlab{fg_g12}
\eeq
The functional $S$  is hermitian by construction.
$f(-{\Box})$ is to be understood as a formal series
expansion in powers of $(-{\Box})$, corresponding to an expansion of
$f(p^2)$ in powers of $p^2$. Note that, in general, $S$ contains
higher (i.e.\ not only first) derivatives of the fields, thus it corresponds to
a non-local action \cite{pais}.

Taking this action functional as a starting point, one can construct a
$\gamma \pi N N$ vertex by using the procedure of minimal substitution,
see Appendix A.1 for details. 
The final result for the $\gamma \pi N N$ vertex can be written as
\begin{eqnarray}
M_\alpha^\mu(p^\prime,p,q)&=&-\tau_\alpha\hat{e}\Big\{ {\ds \frac{2p^\mu+q^\mu}
{(p+q)^2-p^2}}\big[ \Gamma^5(m,p+q)-\Gamma^5(m,p) \big] \nonumber \\
&&+\gamma^5
{\ds \frac{g_2((p+q)^2)}{m}}\big[ \gamma^\mu - \vslash{q}\,
{\ds \frac{2p^\mu+q^\mu}
{(p+q)^2-p^2}} \big] \Big\}  \nonumber \\
&&-\hat{e}\tau_\alpha\Big\{ {\ds \frac{2p^{\prime \mu}-q^\mu}
{(p^\prime-q)^2-p^{\prime 2}}}\big[ \Gamma^5(p^\prime-q,m)-
\Gamma^5(p^\prime,m)\big] \nonumber \\
&&+\big[ \gamma^\mu + {\ds \frac{2p^{\prime \mu}-q^\mu}
{(p^\prime-q)^2-p^{\prime 2}}}\,\vslash{q} \big]{\ds \frac{g_2((p^\prime-q)^2)}
{m}}\,\gamma^5 \Big\}, \eqlab{gampinn_cont}
\end{eqnarray}
where 
Eqs.~(\ref{eq:cont_ps},\ref{eq:cont_pv}) and 
Eqs.~(\ref{eq:4},\ref{eq:4_out},\ref{eq:fg_g12}) have been used. 
We emphasize that the photon four-momentum $q_\mu$ has been taken {\it incoming}
throughout the construction of the contact term. The
$\gamma \pi N N$ vertex with the outgoing photon
four-momentum is obtained from \eqref{gampinn_cont} 
by reversing the signs in front of $q_\mu$. 

It should be stressed that \eqref{gampinn_cont} is not a unique form for the
contact term in question. However, any other form that can be obtained in the 
framework of minimal substitution will differ from the given one by a term
transverse to the four-momentum of the photon, and hence only the 
longitudinal part of the contact term is unambiguous. 
This is demonstrated in Appendix A.1 for an
alternative $\gamma \pi N N$ vertex which is derived if one interchanges the
positions of $\vslash{\pd}$ and $g(-{\Box})$ in the functional $J$. Using 
Eqs.~(\ref{eq:cont_ps},\ref{eq:cont_pv_2}) and
Eqs.~(\ref{eq:4},\ref{eq:4_out},\ref{eq:fg_g12}),
this alternative contact term can be written 
\begin{eqnarray}
(M_{alt})\,_\alpha^\mu(p^\prime,p,q)&=&-\tau_\alpha\hat{e}
\Big\{ {\ds \frac{2p^\mu+q^\mu}
{(p+q)^2-p^2}}\big[ \Gamma^5(m,p+q)-\Gamma^5(m,p) \big] \nonumber \\
&&+\gamma^5
{\ds \frac{g_2(p^2)}{m}}\big[ \gamma^\mu - \vslash{q}\,
{\ds \frac{2p^\mu+q^\mu}
{(p+q)^2-p^2}} \big] \Big\} \nonumber  \\
&&-\hat{e}\tau_\alpha\Big\{ {\ds \frac{2p^{\prime \mu}-q^\mu}
{(p^\prime-q)^2-p^{\prime 2}}}\big[ \Gamma^5(p^\prime-q,m)-
\Gamma^5(p^\prime,m)
\big]  \nonumber \\
&&+\big[ \gamma^\mu + {\ds \frac{2p^{\prime \mu}-q^\mu}
{(p^\prime-q)^2-p^{\prime 2}}}\,\vslash{q} \big]{\ds \frac{g_2(p^{\prime 2})}
{m}}\,\gamma^5 \Big\}. \eqlab{gampinn_cont_2}
\end{eqnarray}

Minimal substitution was also employed in Ref. \cite{ohta}, where 
expressions similar to Eqs.~(\ref{eq:gampinn_cont},\ref{eq:gampinn_cont_2})
were obtained. The present
method and that of \cite{ohta} are essentially equivalent, a technical
difference being that we explicitly calculate 
variational derivatives of the
action functionals Eqs.~(\ref{eq:act_ps},\ref{eq:act_pv}) 
and extract the contact vertex from the variations of these functionals. 
As can be seen from the lower diagram in Fig. 2, only contact
terms with all external legs on-shell are required in our model. 

\subsubsection{Calculating $\Gamma_{\mu, I}(3)$}

Having constructed the $\gamma \pi N N$ vertex, the pole 
contribution $\Gamma_{\mu, I}(3)$ to the third integral on the right-hand 
side of \eqref{sys} (the lower diagram in Fig.~2) can be calculated. 
Using the notation introduced above, we have
\begin{eqnarray}
\Gamma_{\mu, I}(3)&=&-\frac{3-\tau_3}{2}\,\frac{1}{8\pi^2}
\int\!d^4k \gamma_5 \Big\{(2p_\mu-2k_\mu-q_\mu)\Big[
\frac{g_{12}((p-k-q)^2)-g_{12}(m^2)}{(p-k-q)^2-m^2)} \nonumber \\
&& +\frac{\vslash{p}-\vslash{k}}{m}\,\frac{g_2((p-k-q)^2)-g_2(m^2)}
{(p-k-q)^2-m^2}\Big]+\frac{\gamma_\mu}{m}g_2((p-k-q)^2)\Big\} \nonumber \\
&&\times(\vslash{p}-\vslash{k}+m)\gamma_5
\Big\{g_{12}(p^2)+\frac{\vslash{p}}{m}g_2(p^2)\Big\} \nonumber \\
&&\times \delta((p-k)^2-m^2)\,
\Theta(p_0-k_0)\, \delta(k^2-m_{\pi}^2)\, \Theta(k_0) \nonumber \\
&& -\frac{3}{2}(1+\tau_3)\frac{1}{8\pi^2}\int\!d^4k\Big\{
(2p^\prime_\mu+q_\mu)\Big[\frac{g_{12}((p^\prime+q)^2)-g_{12}(m^2)}
{(p^\prime+q)^2-m^2} \nonumber \\
&&+\frac{g_2((p^\prime+q)^2)-g_2(m^2)}{(p^\prime+q)^2-m^2}
\frac{\vslash{p}^\prime}{m}\Big] 
+\frac{\gamma_\mu}{m}g_2((p^\prime+q)^2)\Big\}\gamma_5
(\vslash{p}-\vslash{k}+m)\gamma_5 \nonumber \\
&&\times \Big\{g_{12}(p^2)+\frac{\vslash{p}}{m}g_2(p^2)\Big\} 
\delta((p-k)^2-m^2)\,
\Theta(p_0-k_0)\, \delta(k^2-m_{\pi}^2)\, \Theta(k_0). \eqlab{pole_3}
\end{eqnarray}
(Please note that, since in the $\gamma N N$ vertex 
we take the photon four-momentum outgoing, we have used here the contact vertex
\eqref{gampinn_cont} where all $q_\mu$ are replaced with $-q_\mu$.) 
Another expression
for $\Gamma_{\mu, I}(3)$ is obtained if, instead of \eqref{gampinn_cont}, 
one uses the alternative contact term \eqref{gampinn_cont_2}
(to save writing,
we do not give the explicit formula here). The choice of the contact term
has an influence on the nucleon-photon form factors, 
as will be described when discussing results of the calculations.

Applying \eqref{coeff} with $v_\mu=v_\mu(3)=
\Lambda_{+}(p^\prime)\Gamma_{\mu,I}(3)$, one obtains for the contributions 
to the imaginary parts of the form factors:
\beq
{\displaystyle c^i(3)=-\frac{r(p^2)}{64\pi p^2}\,\Theta(p^2-(m+m_{\pi})^2)\,
\Bigg[(3-\tau_3)\int_{-1}^1\!dx\widetilde{W}^i_1(x)
+3(1+\tau_3)\int_{-1}^1\!dx\widetilde{W}^i_2(x)\Bigg], }
\eqlab{ff_3}
\eeq
where $r(p^2)$ is defined as in \eqref{ff_1}.
Here, $\widetilde{W}^i_{1,2}(x)$ are obtained by using \eqref{kq} in the
functions
\begin{eqnarray}
W^i_1(k)&=&\sum_{j=1}^{6} (E^{-1})^i_j \Bigg\langle\,
(\theta^j)^\mu \mbox{\Large ,}\,\Lambda_{+}(p^\prime)
\gamma_5 \Bigg\{(2p_\mu-2k_\mu-q_\mu)\Big[
{\ds \frac{g_{12}((p-k-q)^2)-g_{12}(m^2)}{(p-k-q)^2-m^2)}} \nonumber \\
&&+{\ds\frac{\vslash{p}-\vslash{k}}{m}\,\frac{g_2((p-k-q)^2)-g_2(m^2)}
{(p-k-q)^2-m^2}\Big]+\frac{\gamma_\mu}{m}}g_2((p-k-q)^2)\Bigg\}
(\vslash{p}-\vslash{k}+m)\gamma_5  \nonumber \\
&&\times \Big\{g_{12}(p^2)+{\ds \frac{\vslash{p}}{m}}g_2(p^2)\Big\} 
\Bigg\rangle  \eqlab{wbig1}
\end{eqnarray}
and  
\begin{eqnarray}
W^i_2(k)&=&\sum_{j=1}^{6} (E^{-1})^i_j \Bigg\langle\,
(\theta^j)^\mu \mbox{\Large ,}\,\Lambda_{+}(p^\prime)
\Bigg\{ (2p^\prime_\mu+q_\mu)\Big[{\ds \frac{g_{12}((p^\prime+q)^2)-
g_{12}(m^2)}{(p^\prime+q)^2-m^2}} \nonumber \\
&&+{\ds \frac{g_2((p^\prime+q)^2)-g_2(m^2)}{(p^\prime+q)^2-m^2}\,
\frac{\vslash{p}^\prime}{m}}\Big] 
+{\ds \frac{\gamma_\mu}{m}}g_2((p^\prime+q)^2)\Bigg\}\gamma_5
(\vslash{p}-\vslash{k}+m)\gamma_5  \nonumber \\
&&\times \Big\{g_{12}(p^2)+
{\ds \frac{\vslash{p}}{m}}g_2(p^2)\Big\} \Bigg\rangle, \eqlab{wbig2}
\end{eqnarray}
where the pairing $\langle\;\;\;,\;\;\rangle$ has been defined in
\eqref{dual_act}.

Having thus calculated the pinching pole contribution to the loop integrals
in \eqref{sys}, the imaginary parts of the form factors are found from 
\eqref{identif}, with $c^i=c^i(1)+c^i(2)+c^i(3)$.
Gauge invariance of the full expression for the half-off-shell vertex is proved
in Appendix C.  

\subsubsection{Calculating the real parts of the form factors using dispersion
relations, and normalization of the vertex}

Dispersion relations allow one to construct the real parts of
the form factors from their imaginary parts. In the present model, we implement
the dispersion relations in the iterative procedure in the following way.
First, we note that the unsubtracted dispersion relations of the form
\beq
Re f(x)=\frac{\mathcal{P}}{\pi}\int_{x_0}^{\infty}\!dx^\prime 
\frac{Im f(x^\prime)}{x^\prime - x}
\eqlab{dr_gen}
\eeq
can be written for a function $f(z)$ which vanishes as $z \rightarrow \infty$
and is analytic in the complex 
$z$-plane ($x=Re\,z$) cut along the real axis from 
$x=x_0$ to infinity.
Bincer proved \cite{bincer} that the half-off-shell form factors 
$F_i^{\pm}(p^2)$ are analytic in the complex plane with the cut from 
the pion threshold 
$w^2_{th}=(m+m_{\pi})^2$ to infinity. In our model, the imaginary parts of the
form factors vanish at infinity, as will be seen from the results
of the calculations. 
For the n$^{th}$ iteration, we utilize the following dispersion relations 
for the form factors $F_2^{\pm}(p^2)$:
\beq
Re (F_2^{s,v})^n(p^2)=\kappa_B^{s,v}+\frac{\mathcal{P}}{\pi}
\int_{w_{th}^2}^{\infty}\!dp^{\prime 2}\,\frac{Im (F_2^{s,v})^n(p^{\prime 2})}
{p^{\prime 2}-p^2},
\eqlab{dr_f2}                       
\eeq
where, to keep the expression transparent, we have dropped the superscripts 
$^{\pm}$ of the form factors on both sides of the equation. 
The constants $\kappa_B^{s,v}$ originate from 
the first term on the right-hand side of \eqref{sys}. Note that,
according to \eqref{ver_bare}, $\kappa_B^{s,v}$ are chosen the same for the
form factors $F_2^{+}(p^2)$ and $F_2^{-}(p^2)$. They are fixed
by the requirement that the vertex reproduces the physical anomalous magnetic
moment when both nucleons are on-shell. In other words,
the following normalization condition is imposed for the converged form 
factors:
\beq
F_2^{+,s}(m^2)=-0.06,\;\;\;F_2^{+,v}(m^2)=1.85.
\eqlab{normal_f2}
\eeq
%

Now we turn to the dispersion relations for the form factors $F_1^{\pm}(p^2)$.
The behaviour of the real parts of $F_1^{\pm}(p^2)$ at infinity can be inferred
from the Ward-Takahashi identity \eqref{wti_gen}.  
To this end, it is convenient to write down the inverse nucleon propagator 
\eqref{prop_inv} in the form
\beq
S^{-1}(p)=\vslash{p}-m-\Big[A(p^2)\vslash{p}+B(p^2)m-
(Z_2-1)(\vslash{p}-m)-Z_2\delta m\Big].
\eqlab{prop_inv1}
\eeq
Here, the functions $A(p^2)$ and $B(p^2)$ parametrize the contribution to the
nucleon self-energy from
the pion loop dressing, and $Z_2$ and $\delta m$ are real renormalization
constants fixed so that the propagator has a pole at the nucleon mass with a
unit residue. Now let the left- and
right-hand sides of \eqref{wti_gen} act on the positive-energy spinor 
$\oln{u}(p^\prime)$ from the left. Using \eqref{finon} for the half-off-shell
vertex and \eqref{prop_inv1} for the inverse propagator and taking into account 
that $\oln{u}(p^\prime) S^{-1}(p^\prime)=0$, the following expressions for
the form factors $(F_1^{\pm})^{s,v}(p^2)$ in terms of the self-energy 
functions $A$ and $B$ can be obtained:
\beq
(F_1^{-})^{s,v}(p^2)=\frac{1}{2}[Z_2-A(p^2)],
\eqlab{wti_fm}
\eeq
\beq
(F_1^{+})^{s,v}(p^2)=-\frac{m}{p^2-m^2}[mB(p^2)+Z_2(m-\delta m)]+
\frac{p^2+m^2}{2(p^2-m^2)}[Z_2-A(p^2)].
\eqlab{wti_fp}
\eeq 
As mentioned above, in this model we use the nucleon self-energy calculated in
Ref.\cite{us_pion}. One relevant feature of the functions $A(p^2)$ and $B(p^2)$
constructed there is that they vanish at infinity, and thus  
the functions
$(F_1^\pm)^{s,v}(p^2)-Z_2/2$ vanish for $p^2 \rightarrow \infty$,
as follows from Eqs.~(\ref{eq:wti_fm},\ref{eq:wti_fp}). In addition,
these functions possess the same analyticity properties as the form factors
$(F_1^{\pm})^{s,v}(p^2)$ themselves. Therefore, for the n$^{th}$ iteration, 
the form factors $(F_1^{\pm})^{s,v}(p^2)$ obey the dispersion relations 
\beq
Re (F_1^{s,v})^n(p^2)=\frac{Z_2}{2}+\frac{\mathcal{P}}{\pi}
\int_{w_{th}^2}^{\infty}\!dp^{\prime 2}\,
\frac{Im (F_1^{s,v})^n(p^{\prime 2})}{p^{\prime 2}-p^2},
\eqlab{dr_f1}
\eeq
omitting the superscripts $^\pm$ of the form factors.
%

In terms of the parametrization of the propagator \eqref{prop_inv}, the
relations in Eqs.~(\ref{eq:wti_fm},\ref{eq:wti_fp}) read
\beq
(F_1^{-})^{s,v}(p^2)=\frac{\alpha(p^2)}{2}
\eqlab{wti_fm_alxi}
\eeq
and
\beq
(F_1^{+})^{s,v}(p^2)=\frac{\alpha(p^2)m}{p^2-m^2}\Big[
\frac{p^2+m^2}{2m}-\xi(p^2) \Big],
\eqlab{wti_fp_alxi}
\eeq
where $\lim_{p^2\rightarrow m^2}(F_1^{+})^{s,v}(p^2)$ is finite 
because $\lim_{p^2\rightarrow m^2}\xi(p^2)=m$ due to the correct location of
the pole of the renormalized propagator.

\section{Results for the form factors}

Using the iteration procedure described in the previous section, a solution
was obtained for the half-off-shell nucleon-photon form factors. We considered
the procedure converged at iteration $n$ if all the results of this
iterations were identical up to five significant digits with those of iterations
$n+1,\ldots,n+10$. With this convergence criterion, we needed 20 
iterations to obtain the solution. It is noteworthy that the dominant
contribution to the form factors is given by the first integral on the
right-hand side of \eqref{sys}. Since this integral is an inhomogeneity of the
equation, already the first iteration contains the bulk of the magnitude of the
form factors. This, however, does not mean that the other integrals on the
right-hand side of \eqref{sys} are of minor importance. In particular, they are
crucial for the vertex to satisfy the Ward-Takahashi identity.

As stated above, all results presented in this section are obtained using
the half-off-shell $\pi N N$ vertex and the nucleon self-energy calculated in
Ref.\cite{us_pion}. In general, the half-off-shell $\pi N N$
vertex, \eqref{pinnver}, contains 
both pseudovector and pseudoscalar couplings. The solution for the
nucleon-pion form factors depends on the choice for the bare vertex
in \cite{us_pion}. However,
the half-width of the constructed pseudovector form factor is bound from above,
and this feature
persists independent of the choice of the bare form factor (called the 
"cut-off function" in \cite{us_pion}). For the present
calculation of the half-off-shell form factors in the $\gamma N N$ vertex, we
chose that solution for the nucleon self-energy and the $\pi N N$ vertex 
in which the bare pseudovector form factor is given by Eq.~(23) of
Ref.\cite{us_pion}, with the maximal half-width 
$\Lambda^2 = 1.28 \mbox {GeV}^2$. We also did the calculation using the other
choice of the bare $\pi N N$ vertex, given by Eq.~(24) of Ref.\cite{us_pion},
with the half-width $\Lambda^2 = 1.33 \mbox {GeV}^2$ (not shown).
We found that the results for the
nucleon-photon form factors do not depend significantly on the choice of the 
$\pi N N$ vertex.

In Fig.~5 the imaginary and real parts of the form factors $F_2^+(p^2)$ (the
solid line) and $F_2^-(p^2)$ (the dotted line) are shown for the case of the
proton-photon vertex. One can see that the slope of $Im\,F_2^-(p^2)$ at the pion
threshold, $p^2=(m+m_{\pi})^2$, is much steeper as compared to that of
$Im\,F_2^+(p^2)$. As a consequence of this, we obtain a pronounced cusp-like
behaviour of $Re\,F_2^-(p^2)$ at the threshold.

The form factors in Fig.~5 are calculated using the $\gamma \pi N N$ contact
term of \eqref{gampinn_cont} when evaluating the third diagram in Fig.~2, as
described in Section III.B.5. An alternative form of the 
$\gamma \pi N N$ vertex is given by \eqref{gampinn_cont_2}. As is shown in
Appendix A.1, the difference $\Delta^\mu$
between these two contact terms, \eqref{cont_pv_diff}, is transverse to the
photon four-momentum. To illustrate the influence of the different choices of
the contact terms on the $\gamma N N$ vertex, in Fig.~6 we show the
form factors $F_2^{\pm}(p^2)$ calculated using the alternative contact term.
From a comparison of Figs.~5 and 6, it follows that the different choices of the
contact term affect mainly the form factor $F_2^-(p^2)$. The form factor
$F_2^+(p^2)$ is normalized at $p^2=m^2$ to the physical anomalous magnetic 
moment of the nucleon and is only slightly sensitive to the choice of the
contact term. 

The results for the form factors $F_2^{\pm}(p^2)$ in the neutron-photon vertex
are shown in Figs.~7 and 8 for the two choices of the $\gamma \pi N N$ vertex.
The conclusions drawn above for the proton apply qualitatively to this case as
well.

As describeded above, we normalize the vertex so that 
Eqs.~(\ref{eq:normal_f2}) are
fulfilled by the converged form factors $(F_2^{+})^{s,v}$. This is achieved by
adjusting the bare renormalization constants $\kappa_B^{s,v}$ defined in 
\eqref{ver_bare}. Specifically, $\kappa_B^s=0.03$ and $\kappa_B^v=1.51$ for the
case of the calculation in which the contact term \eqref{gampinn_cont} 
is employed,
and $\kappa_B^s=0$ and $\kappa_B^v=1.6$ if the contact term 
\eqref{gampinn_cont_2} is used instead.

Since the $\gamma N N$ vertex obeys the Ward-Takahashi identity, the form
factors $F_1^{\pm}(p^2)$ are uniquely determined by the functions parametrizing
the nucleon propagator. We checked numerically that 
Eqs.~(\ref{eq:wti_fm_alxi},\ref{eq:wti_fp_alxi}) are fulfilled by the converged
vertex, whereas a vertex obtained at any iteration before the 
full convergence has been achieved does not satisfy these constraints.
One of the consequences of the Ward-Takahashi identity is that
$(F_1^{\pm})^s=(F_1^{\pm})^v$ and therefore
$F_1^{\pm}=0$ for the case of the neutron-photon vertex. 
The from factors
$F_1^{\pm}(p^2)$ in the proton-photon vertex are depicted in Fig.~9. Since the 
two forms of the contact $\gamma \pi N N$ term differ only by a 
part transverse to the photon four-momentum, the $\gamma N N$ vertex
calculated with both of them will obey
the Ward-Takahashi identity. Therefore, contrary to $F_2^{\pm}(p^2)$,
the form factors $F_1^{\pm}(p^2)$ do not depend on the choice of the 
$\gamma \pi N N$ vertex.

\section{Consistency of the model with a K-matrix formalism}

In this section we will describe the consistency of the present
half-off-shell form factors and nucleon self-energy 
with a coupled channel K-matrix 
approach 
\cite{gouds,korsch,korschtim}. 
In particular, we consider a simplified version
of the K-matrix formalism with only the nucleon, pion and photon degrees of 
freedom, since these are the particles that have been included in our model.
Also, only the one-pion threshold discontinuities are taken into account in both
the K-matrix approach in question and the present model.
An important point to be addressed here is the following. In the iterative
procedure applied in the model, only the real parts of the form factors and the
nucleon self-energy from the iteration n are retained to calculate the imaginary
parts for the iteration n+1. We will show that by doing so, we avoid double
counting of the pole contributions to the one-particle reducible loop diagrams
generated in the K-matrix approach. 

Suppose we want to consider simultaneously pion-nucleon scattering, 
pion photoproduction and Compton scattering. Then the scattering matrix
has two indices corresponding to the channel in the initial and
final state, ${\mathcal{T}}_{c' c}$, where the indices can be
$\pi$ or $\gamma$ for the channels $\pi N$ or $\gamma N$, respectively.
The Bathe-Salpeter equation for the scattering matrix can be written as
\beq
{\mathcal{T}}_{c' c}\,=\,V_{c' c}\, +\sum_{c''}V_{c' c''}\,
{\mathcal G}_{c''}\,{\mathcal T}_{c'' c}\,,
\eqlab{k1}
\eeq
where $V_{c' c}$ is the sum of all irreducible diagrams describing the process
$c \rightarrow c'$ and ${\mathcal{G}}_{c''}$ is the free two-body 
propagator pertinent to the channel $c''$. ${\mathcal{G}}_{c''}$ contains
the on-shell contribution $i\delta_{c''}$ which is imaginary, according to
Cutkosky rules, and the principal value (off-shell) part 
${\mathcal {G}}^P_{c''}$ which is real,
\beq
{\mathcal{G}}_{c''}\,=\,{\mathcal{G}}^P_{c''}\,+\,i\delta_{c''}.
\eqlab{k2}
\eeq
The K-matrix can be defined by the equation
\beq
K_{c' c}\,=\,V_{c' c}\,+\,\sum_{c''}V_{c' c''}\,{\mathcal{G}}^P_{c''}\,
K_{c'' c}\,.
\eqlab{k3}
\eeq
According to this formula, the loop diagrams contributing to the K-matrix
contain only the principal value part of the two-particle propagator. The 
remaining pole contribution enters explicitly in the equation for the 
T-matrix expressed in the K-matrix,
\beq
{\mathcal{T}}_{c' c}\,=\,K_{c' c}\,+\,\sum_{c''}K_{c' c''}\,i\delta_{c''}\,
{\mathcal{T}}_{c'' c}\, ,
\eqlab{k4}
\eeq
which can be obtained from the three previous equations.
A formal solution of this equation reads (suppressing the channel indices)
\beq
{\mathcal T}= \frac{1}{1-K i\delta}\,K ,
\eqlab{k5}
\eeq
from which it follows that if $K$ is hermitian, the S-matrix,
$S=1+2i\mathcal{T}$, will be unitary.

In the remainder of this section we show that it is possible to
construct a K-matrix as a sum of skeleton diagrams in which the half-off-shell
vertices and the nucleon propagator from the present model are used. This
K-matrix is thus 
a solution of \eqref{k3}, 
where the kernel $V_{c c'}$ is a sum of tree diagrams. Consequently, the
T-matrix found from \eqref{k5} with this K-matrix will be a solution of the
system of Eqs.~(\ref{eq:k3},\ref{eq:k4}). 

We construct the K-matrix in the following way. Let the entry 
$K_{\gamma \gamma}$, pertinent to Compton scattering, be equal to the sum of
the skeleton diagrams depicted in Fig.~10. The pion photoproduction entry 
$K_{\gamma \pi}$ is chosen as in Fig.~3, and the pion scattering entry 
$K_{\pi \pi}$ is the sum of the s- and u-type diagrams with the intermediate
nucleon exchange. In all these diagrams, the dressed nucleon propagator and the
nucleon-pion and nucleon-photon half-off-shell vertices, as well as the contact
terms, calculated in our model are substituted. 

To interpret this K-matrix in 
terms of \eqref{k3}, let us choose $V_{c' c}$ as follows. $V_{\gamma \gamma}$ 
is the sum of the s- and u-type diagrams for Compton scattering, 
where the free nucleon propagator and the bare nucleon-photon vertex are used.
$V_{\gamma \pi}$ is the sum of the nucleon s- and u-exchange diagrams and the
pion exchange diagram in the t-channel, plus the four-point vertex 
deriving from the minimal
substitution in the pseudovector pion-nucleon coupling (the free propagators
and bare vertices are used here).
Finally, $V_{\pi \pi}$ is the s-channel plus the 
u-channel nucleon exchange diagrams, again with the bare nucleon-pion vertices
and the free nucleon propagator. Now one can consider an iterative solution of
\eqref{k3} as a formal expansion of $K_{c' c}$ in powers of $V_{c' c}$. 
Let us concentrate, for definiteness, on the Compton scattering matrix element
up to second order in $V_{c' c}$, denoted as
$K^{(2)}_{\gamma \gamma}$. It is
natural to
neglect the terms suppressed by two powers of the electromagnetic coupling
constant. This leaves us with
\beq
K^{(2)}_{\gamma \gamma}=V_{\gamma \gamma}+V_{\gamma \pi}\,
{\mathcal{G}}^P_{\pi}\,V_{\pi \gamma}.
\eqlab{k6}
\eeq
The set of diagrams corresponding to the right-hand side of this equation is
depicted in Fig.~11. The notation $ss$, $su$ etc. for the loop diagrams refer 
to their structure
in terms of the s-, u- , t-channel and contact tree diagrams 
of which $V_{\gamma \pi}$ consists. 
The index ${\it Re}$ at the loops indicates that only the principal
value integrals are taken into account, in accordance with \eqref{k6}. 
In other words,
the self-energy functions and form factors parametrizing these loops are real
functions. One can see that the one-particle reducible diagrams
in Fig.~11 (diagrams {\it ss, su, st, sc, us, uc, ts, cs} and {\it cu}) 
are part of the dressing of
the nucleon propagator and half-off-shell nucleon-photon vertices, in
exact correspondence with our model. Therefore, the sum of these diagrams and 
$V_{\gamma \gamma}$ contributes to
the s- and u-channel parts of $K_{\gamma \gamma}$, represented by the two upper
diagrams in Fig.~10. The other, one-particle irreducible, diagrams in 
Fig.~11 are part of a dressed contact term. This contact term and the 
contact term constructed by the minimal substitution, shown by 
the lower diagram in Fig.~10, play the same role in that they both ensure 
the gauge invariance of $K_{\gamma \gamma}$. However, those parts of a 
$\gamma \gamma N N$ vertex which are gauge invariant by themselves cannot be
unambiguously determined by the minimal substitution procedure. This ambiguity 
is quite analogous to that encountered in constructing the 
$\gamma \pi N N$ vertex, where the term orthogonal to the photon 
four-momentum cannot be uniquely determined.

The above description of $K_{\gamma \gamma}$ in terms of
\eqref{k3} has an illustrative character since only terms up to
second order in $V_{c c'}$ have been discussed. 
Considering higher orders in the same fashion would 
involve full non-perturbative solutions for the nucleon propagator
and the vertices. 
The fact that only principal value parts of the loop integrals (or,
equivalently, only the real parts of the form factors and self-energy 
functions) are taken into account in the iterative procedure for the vertices 
and the nucleon propagator is consistent with \eqref{k3} for the K-matrix. 
In particular, inclusion
of the imaginary parts would lead to double counting of the pole parts of the
one-particle reducible loop
diagrams contributing to the T-matrix. These pole parts are generated 
by \eqref{k4} and therefore should not be included explicitly in the K-matrix
through the imaginary parts of the form factors and self-energy functions.

\section{Application in Compton scattering}

\subsection{$\gamma \gamma N N$ vertex}

Motivated by an application to Compton scattering, described in a following
section, we first construct a four-point $\gamma \gamma N N$ vertex 
(contact term) needed to provide current conservation in the process.
Similar to the construction of the $\gamma \pi N N$ contact term,
the method of minimal substitution is applied. All the details of the
derivation can be found in Appendices A.2 and B.

Assume that the initial photon has four-momentum $k$ and polarization index 
$\nu$, and 
the initial nucleon has four-momentum $p$. The final photon has four-momentum 
$-q$ and polarization index $\mu$,
and the final nucleon has four-momentum $p^\prime$. Thus, we have the following
four-momentum conservation condition: $p^\prime=p+k+q$.
Since both incoming and outgoing nucleons are on the mass shell in Compton 
scattering, we need
only the matrix element of the contact $\gamma \gamma N N$ vertex
\eqref{ct_gamgam} between the
positive-energy spinors of the incoming and outgoing nucleons:
\begin{eqnarray}
\oln{u}(p^\prime)\,M^{ct}_{\mu \nu}(q,k)\,u(p)&=&
\oln{u}(p^\prime)\,i\hat{e}^2\,
\Bigg\{ \Big[\frac{\alpha((p+q)^2)m+\beta((p+q)^2)}{2[(p+q)^2-m^2]}-
\frac{\alpha((p+k)^2)m+\beta((p+k)^2)}{2[(p+k)^2-m^2]}\Big] \nonumber \\
&& \times\Big[\frac{(p_\mu+p^\prime_\mu-k_\mu)(p_\nu+p^\prime_\nu+q_\nu)}
{(p+q)^2-m^2} - \frac{(p_\mu+p^\prime_\mu+k_\mu)(p_\nu+p^\prime_\nu-q_\nu)}
{(p+k)^2-m^2}\Big] \nonumber \\
&&+g_{\mu \nu}\,\Big[\frac{\alpha((p+k)^2)m+\beta((p+k)^2)}{(p+k)^2-m^2}+
\frac{\alpha((p+q)^2)m+\beta((p+q)^2)}{(p+q)^2-m^2}\Big] \nonumber \\
&&+\frac{\alpha((p+k)^2)-\alpha(m^2)}{2[(p+k)^2-m^2]}\,
\Big[(p_\mu+p^\prime_\mu+k_\mu)\gamma_\nu +
(p_\nu+p^\prime_\nu-q_\nu)\gamma_\mu\Big] \nonumber \\
&&+\frac{\alpha((p+q)^2)-\alpha(m^2)}{2[(p+q)^2-m^2]}\,
\Big[(p_\nu+p^\prime_\nu+q_\nu)\gamma_\mu +
(p_\mu+p^\prime_\mu-k_\mu)\gamma_\nu\Big]  \nonumber \\
&&+\frac{H(p+k)^2-H(m^2)}{(p+k)^2-m^2}\,\Big[ [\vslash{q},\gamma_\mu]
(p_\nu+p^\prime_\nu-q_\nu)+
(p_\mu+p^\prime_\mu+k_\mu)[\vslash{k},\gamma_\nu] \Big] \nonumber \\
&&+\frac{H(p+q)^2-H(m^2)}{(p+q)^2-m^2}\,\Big[ [\vslash{k},\gamma_\nu]
(p_\mu+p^\prime_\mu-k_\mu)+
(p_\nu+p^\prime_\nu+q_\nu)[\vslash{q},\gamma_\mu] \Big] \nonumber \\
&&+F((p+k)^2)\Big[[\vslash{q},\gamma_\mu] \gamma_\nu+
\gamma_\mu [\vslash{k},\gamma_\nu]\Big] \nonumber \\
&&+F((p+q)^2)\Big[[\vslash{k},\gamma_\nu] \gamma_\mu+
\gamma_\nu [\vslash{q},\gamma_\mu]\Big]  \Bigg\}\,u(p),  
\eqlab{ct_me}
\end{eqnarray} 
where the notation introduced in 
Eqs.~(\ref{eq:prop_inv_ms},\ref{eq:f_f2},\ref{eq:g_f2}) has been used, and 
$H(p^2) \equiv F(p^2)m+G(p^2)$.

If Compton scattering is
described by a scattering amplitude $M_{\mu \nu}(q,k)$, then gauge invariance
requires \cite{weinb}
\beq
q^\mu \,\oln{u}(p^\prime)M_{\mu \nu}(q,k)u(p) = 
k^\nu\,\oln{u}(p^\prime) M_{\mu \nu}(q,k)u(p) =0.
\eqlab{gauge_inv_compt}
\eeq
We write the full amplitude in terms of three contributions, 
as depicted in Fig.~6:
\beq
M_{\mu \nu}(q,k) =M_{\mu \nu}^s(q,k)+M_{\mu \nu}^u(q,k)+M_{\mu \nu}^{ct}(q,k).
\eqlab{compt_split}
\eeq
The pole contributions in this formula (``class A" diagrams, in terms of
Ref. \cite{gelgold}) are given by the s-channel skeleton diagram
\beq
\oln{u}(p^\prime)M_{\mu \nu}^s(q,k)u(p)=
\oln{u}(p^\prime)(-i e^2)\Gamma_\mu(m,p^\prime-q)S(p^\prime-q)
\Gamma_\nu(p+k,m)u(p),
\eqlab{compt_s}
\eeq
and the u-channel skeleton diagram
\beq
\oln{u}(p^\prime)M_{\mu \nu}^u(q,k)u(p)=
\oln{u}(p^\prime)(-i e^2)\Gamma_\nu(m,p^\prime-k)S(p+q)\Gamma_\mu(p+q,m)u(p),
\eqlab{compt_u}
\eeq
which are constructed in terms of the irreducible half-off-shell $\gamma N N$
vertex and the dressed renormalized nucleon propagator.
$M_{\mu \nu}^{ct}(q,k)$ denotes the contact
term (comprising ``class B" diagrams) whose matrix element is given by 
\eqref{ct_me}. 
Contracting the sum of \eqref{compt_s} and \eqref{compt_u} with $q^\mu$ and
$k^\nu$ gives, respectively,
\beq
q^\mu \,\oln{u}(p^\prime)\Big[M^s_{\mu \nu}(q,k)+M^u_{\mu \nu}(q,k)\Big]u(p)=
-i\hat{e}^2\,\oln{u}(p^\prime)\Big[\Gamma_\nu(m,p^\prime-k)-
\Gamma_\nu(p+k,m)\Big]u(p)
\eqlab{contr_q}
\eeq
and 
\beq
k^\nu \,\oln{u}(p^\prime)\Big[M^s_{\mu \nu}(q,k)+M^u_{\mu \nu}(q,k)\Big]u(p)=
-i\hat{e}^2\,\oln{u}(p^\prime)\Big[\Gamma_\mu(m,p^\prime-q)-
\Gamma_\mu(p+q,m)\Big]u(p),
\eqlab{contr_k}
\eeq
where the Ward-Takahashi identity \eqref{wti_gen} has been used.
Using Eqs.~(\ref{eq:ver_fg_out},\ref{eq:ver_fg_in}) for the half-off-shell
vertices and \eqref{ct_me} for the matrix element of the contact term, 
it is straightforward to show that the right-hand sides of
\eqref{contr_q} and \eqref{contr_k} are equal to 
$-q^\mu\,\oln{u}(p^\prime)M^{ct}_{\mu \nu}(q,k)u(p)$ and
$-k^\nu\,\oln{u}(p^\prime)M^{ct}_{\mu \nu}(q,k)u(p)$, respectively, 
thus proving Eqs.~(\ref{eq:gauge_inv_compt}).

\subsection{Compton cross section}

In this section, using the K-matrix constructed as described above, 
we calculate the $\gamma N$ scattering cross section as a function of the
energy of the incident photon. Since only nucleon, pion and photon 
are included in the present calculation, we do not compare the 
calculated cross section with experimental data. For a definitive comparison
with experiment, other important degrees of freedom, such as the
$\Delta$-resonance, would have to be included. This extension of the model 
is in progress.   

The result for the forward scattering is
shown by the solid line in Fig.~12. For comparison, the dotted line shows 
the cross section calculated using the K-matrix built with the bare vertices 
and the free nucleon propagator,
$K_{c' c}=V_{c' c}$, thereby neglecting the principal value parts of the loop 
integrals  contributing to the T-matrix (see Eqs.~(\ref{eq:k3},\ref{eq:k4})).
We checked numerically that both these calculations are gauge invariant.
At low photon energies, the two cross sections converge to the same limit,
the Thomson cross section. This is a consequence of the low-energy theorem 
\cite{gelgold,low}. At higher energies, however, there is a difference 
between the two calculations: about 5 \% at $E_{\gamma}=145$ MeV, which is just
below the pion production threshold, and up to 10 \% above the threshold. 
It turns out that the bulk of this 
difference is due to the inclusion of the form factors $F_2^{\pm}$.  
Also, this effect of 
including the principal parts of the loop integrals in the T-matrix 
(by using the dressed vertices and the dressed nucleon propagator 
in the K-matrix) depends on the kinematics of the process. 
In particular, the difference between
the two cross sections at $\theta_{\gamma}=180^o$ below the pion threshold
has the opposite sign 
compared to the case of the forward scattering, and  
it is about two to ten times 
smaller in absolute value, depending on the photon energy.

The cusp-like structure of the cross section at the pion production threshold
is a well-known consequence of unitarity of the scattering matrix and 
has been observed experimentally (see, e.g., \cite{berg} and references 
therein). Opposed to that, the dashed line in Fig.~12 shows the cross 
section calculated based on the T-matrix equal to the sum of tree-diagrams 
$V_{\gamma \gamma}$. The corresponding S-matrix will not be unitary,
and the cross section will not, therefore, exhibit the unitary cusp.

\subsubsection{The low-energy behaviour of the cross section}

Throughout this section, we choose $p$ ($p^\prime$) and $k$ ($k^\prime$) to be
the four-momenta of the initial 
(final) proton and photon, respectively, so that the four-momentum conservation 
in this process is $p+k=p^\prime+k^\prime$. 
It is convenient to work in the laboratory frame, where
the initial proton is at rest, i.e.\ $\{ p_\mu \}=(m,\overrightarrow{0})$.

According to the low-energy theorem for Compton scattering \cite{gelgold,low}, 
the leading term in the expansion of the cross section in powers of the small
photon energy $\omega$ is the Thomson cross section, and it is the only
term that can be determined model independently, based on Lorentz and gauge
invariance and crossing symmetry. The other terms are of even powers
in the energy, and their structure depends on details of the model utilized
for the description of the process. Thus, one has
\beq
\frac{d\sigma}{d\Omega}=\left(\frac{d\sigma}{d\Omega}\right)_{\!\!Th}\!\!\!(1+
c_2 \omega^2+\ldots),
\eqlab{let}
\eeq
where the Thomson cross section 
\beq
\left(\frac{d\sigma}{d\Omega}\right)_{\!\!Th}\!\!\!=\frac{\alpha^2}{m^2},
\eqlab{cs_thoms}
\eeq 
with $\alpha \approx 1/137$ being the electromagnetic coupling constant.

In this section we calculate the coefficient $c_2$ of the quadratic 
term in the 
low-energy expansion \eqref{let}. We consider the
case where the Compton scattering amplitude is described by the sum of the
s- and u-channel diagrams and the contact $\gamma \gamma N N$ term, 
as shown in Fig.~11. It is a good
approximation to the unitary amplitude below the pion threshold, because the
difference between this amplitude and the fully unitarized one is 
suppressed by
powers of the electromagnetic coupling constant. 
To simplify the following analytical calculations, we consider
only the form factors $F_1^{\pm}(p^2)$ in the
half-off-shell $\gamma N N$ vertices. 
Also, the dressed nucleon propagator
is used in the s- and u-channel diagrams. The Ward-Takahashi identity implies
that $F_1^{\pm}(p^2)$ can be expressed in terms of the functions 
$\alpha(p^2)$ (not to
be confused with the fine structure constant $\alpha$) and $\xi(p^2)$
parametrizing the renormalized dressed nucleon propagator, 
see Eqs.~(\ref{eq:wti_fm_alxi},\ref{eq:wti_fp_alxi}).
The term quadratic in the photon energy will therefore 
be expressed in $\alpha(p^2)$ and $\xi(p^2)$. If one used bare 
vertices and the free nucleon propagator, no contact diagram would arise 
and the amplitude would be described by the s- and u-channel diagrams only. 
In that case, one would obtain just the Thomson cross section 
in the low-energy limit.

Since the calculated deviation of the cross section from the Thomson limit 
is largest for the forward scattering, we will consider this case in more
detail. We fix the photon scattering
angle $\theta_{\gamma}=0^o$ and hence $k^\prime=k$ and $p^\prime=p$. We also
choose the gauge in which the polarizations of the initial and final
photons have the form $\{ \epsilon_\mu \}=(0,\overrightarrow{\epsilon})$ and
$\{ \epsilon^\prime_\mu \}=(0,\overrightarrow{\epsilon}^\prime)$, and 
the photon four-momenta $\{ k_\mu \}=\{ k^\prime_\mu \}=
(\omega,0,0,\omega)$, so  
$k\cdot \epsilon=k^\prime \cdot \epsilon^\prime =0$. 
It follows from this choice of the gauge and the frame of reference that 
the scalar products of all four-momenta in the problem
with both $\epsilon$ and $\epsilon^\prime$ vanish. The Mandelstam variables in
the laboratory frame are $s=(p+k)^2=m^2+2m\omega$ and 
$u=(p-k^\prime)^2=m^2-2m\omega$.

The expression for the matrix element of the amplitude reads
\beq
{\mathcal{M}}(k,k)={\mathcal{M}}_s(k,k)+
{\mathcal{M}}_u(k,k)+{\mathcal{M}}_{ct}(k,k),
\eqlab{cs_amp_gen}
\eeq
where
\begin{eqnarray}
{\mathcal{M}}_s(k,k)&=&-ie^2\,\oln{u}(p^\prime)\,
\vslash{\epsilon}^\prime\, 
\left\{ F_1^+(s)\,\frac{\vslash{p}+\vslash{k}+m}{2m}+F_1^-(s)\,
\frac{-\vslash{p}-\vslash{k}+m}{2m} \right\}\,\frac{\vslash{p}+\vslash{k}+\xi(s)}
{\alpha(s)[s-\xi^2(s)]} \nonumber \\
&& \times\left\{ \frac{\vslash{p}+\vslash{k}+m}{2m}\,F_1^+(s)
+\frac{-\vslash{p}-\vslash{k}+m}{2m}\,F_1^-(s) \right\}\,\vslash{\epsilon}\,u(p),
\eqlab{cs_ms}
\end{eqnarray}
\begin{eqnarray}
{\mathcal{M}}_u(k,k)&=&-ie^2\,\oln{u}(p^\prime)\,\vslash{\epsilon}\,
 \left\{F_1^+(u)\,\frac{\vslash{p}-\vslash{k}+m}{2m}+F_1^-(u)\,
\frac{-\vslash{p}+\vslash{k}+m}{2m} \right\}\,
\frac{\vslash{p}-\vslash{k}+\xi(u)}
{\alpha(u)[u-\xi^2(u)]} \nonumber \\
&& \times\left\{ \frac{\vslash{p}-\vslash{k}+m}{2m}\,
F_1^+(u)+\frac{-\vslash{p}+\vslash{k}+m}{2m}\,F_1^-(u) \right\}\,
\vslash{\epsilon}^\prime \,u(p)
\eqlab{cs_mu}
\end{eqnarray}    
and
\begin{eqnarray}
{\mathcal{M}}_{ct}(k,k)&=&ie^2\,\oln{u}(p^\prime)\,\left\{ 
(\epsilon \cdot
\epsilon^\prime)\,\Big[\frac{m\alpha(s)-\alpha(s)\xi(s)}{s-m^2}+
\frac{m\alpha(u)-\alpha(u)\xi(u)}{u-m^2}\Big] \right. \nonumber \\
&&\left. -\frac{\widetilde{F}_2^+(s)-
\widetilde{F}_2^-(s)}{2m^2}\,\vslash{\epsilon}^\prime\,\vslash{k}\,
\vslash{\epsilon}\,
+\frac{\widetilde{F}_2^+(u)-\widetilde{F}_2^-(u)}{2m^2}\,
\vslash{\epsilon}\,\vslash{k}\,
\vslash{\epsilon}^\prime\, \right\}\,u(p). 
\eqlab{cs_mct}
\end{eqnarray}
In the last formula we used \eqref{ct_me} (where, according to the chosen
kinematics, $q=k$ and $q^\prime = -k^\prime$) with
$\beta(p^2)=-\alpha(p^2)\xi(p^2)$ and \eqref{f_f2} in which 
$F_2^{\pm}(p^2)=0$. Thus, the contact term in \eqref{cs_mct} is consistent with
the half-off-shell $\gamma N N$ vertices in which only the form factors 
$F_1^{\pm}(p^2)$ are taken into account. 
Using Eqs.~(\ref{eq:wti_fm_alxi},\ref{eq:wti_fp_alxi}) and 
Eqs.~(\ref{eq:f2pl_tilde},\ref{eq:f2mi_tilde}) and introducing the notation
$x(p^2) \equiv \xi(p^2)-m$, we obtain 
\begin{eqnarray}
{\mathcal{M}}(k,k)&=&\frac{ie^2}{2m\omega}\,\oln{u}(p^\prime)\,\Bigg\{
\vslash{\epsilon}^\prime\,\vslash{k}\,\vslash{\epsilon}\,
\Big[ \frac{\alpha(s)x(s)}{\omega}-\alpha(m^2) \Big] -
\vslash{\epsilon}\,\vslash{k}\,\vslash{\epsilon}^\prime\,
\Big[ \frac{\alpha(u)x(u)}{\omega}+\alpha(m^2) \Big] \nonumber \\
&&+[\vslash{\epsilon}^\prime,\vslash{\epsilon}]\,\frac{\alpha(s)x(s)+
\alpha(u)x(u)}{2} \Bigg\}\,u(p). \eqlab{cs_m_2}
\end{eqnarray}
Note that \eqref{cs_m_2} is explicitly crossing symmetric, i.e.\ symmetric
under the replacements $k \longleftrightarrow -k$ (and
consequently, $\omega \longleftrightarrow -\omega$ and 
$s \longleftrightarrow u$) and 
$\vslash{\epsilon} \longleftrightarrow \vslash{\epsilon}^\prime$.

The differential cross section, averaged (summed) over projections of the
spins of the initial (final) proton and the polarizations of the initial 
(final) photon is obtained in a standard way, by squaring the absolute value of 
\eqref{cs_m_2} and evaluating traces of products of
the $\gamma$-matrices. The result can be written in the form
\beq
\frac{d\sigma}{d\Omega}=\left(\frac{d\sigma}{d\Omega}\right)_{\!\!Th}\!\!\!
\Big[ \alpha(m^2)+{\mathcal{A}}(\omega) \Big]^2,
\eqlab{comp_cs_2}
\eeq
where 
\beq
{\mathcal{A}}(\omega)=\frac{\alpha(u)x(u)-\alpha(s)x(s)}{2\omega}.
\eqlab{a}
\eeq

To analyze the low-energy expansion of the cross section \eqref{comp_cs_2},
we first note that from the fact that the propagator has a simple pole at 
$\vslash{p}=m$ with a unit residue, it follows that 
\beq
x(m^2)=0,\;\;\;\;\;\;\;\;\;\;\;
2m \left. \frac{dx}{d (p^2)}\right|_{p^2=m^2}\!\!\!\!\!
=1-\frac{1}{\alpha(m^2)}, 
\eqlab{alx_m}
\eeq 
in terms of \eqref{prop_inv} with $\xi(p^2)=m+x(p^2)$. Using 
Eqs.~(\ref{eq:alx_m}), we obtain from \eqref{a} that
${\mathcal{A}}(0)=1-\alpha(m^2)$. 
Further, since ${\mathcal{A}}(\omega)$ is an even function of $\omega$, we have
$\left.(d^k {\mathcal{A}}/d\omega^k)\right|_{\omega=0}=0$, for any odd $k$. 
The last 
property of the function ${\mathcal{A}}(\omega)$ ensures that the Compton 
cross section \eqref{comp_cs_2} is crossing symmetric.
Thus, the low-energy expansion of the
function ${\mathcal{A}}(\omega)$ contains only even powers of $\omega$, 
\beq
{\mathcal{A}}(\omega)=1-\alpha(m^2)+\frac{1}{2}
\left.\frac{d^2{\mathcal{A}}}{d\omega^2}\right|_{\omega=0}\!\!\!\!\!\!\omega^2
+O(\omega^4).
\eqlab{a_exp}
\eeq
Substituting \eqref{a_exp} in \eqref{comp_cs_2} gives the low-energy
expansion of the cross section in the form of \eqref{let}, where 
$c_2=\left.(d^2 A/d\omega^2)\right|_{\omega=0}$.
Using \eqref{a}, the coefficient $c_2$ can be found based
on the nucleon self-energy calculated in \cite{us_pion}, with the result
$c_2 \approx 1.21\cdot 10^{-6} \mbox{Mev}^{-2}$. Otherwise, 
one can infer
$c_2$ by comparing the calculation of the Compton cross section
(below the pion threshold) with the bare vertices and the free nucleon
propagator, on the one hand, and the dressed vertices and propagator, on the
other. The value obtained in such a way is 
$c_2 \approx 1.18\cdot 10^{-6} \mbox{Mev}^{-2}$. 
These two numbers are in agreement with each other.

Now let us return to the general case, depicted in Fig.~12 (the photon energy 
$\omega \equiv E_{\gamma}$), in which both 
$F_1^{\pm}$ and $F_2^{\pm}$ are kept in the vertex.
It is known that the coefficient at the $\omega^2$-term in the low-energy 
expansion 
can be related to the electric and magnetic polarizabilities of the proton 
$\oln{\alpha}$ and $\oln{\beta}$ \cite{polar}: in terms of \eqref{let},
the model-dependent part of the coefficient $c_2$ can be written as
\beq
-\frac{m}{2\alpha}\Big[(\oln{\alpha}+\oln{\beta})(1+cos \theta)^2+
(\oln{\alpha}-\oln{\beta})(1-cos \theta)^2\Big].
\eqlab{c2_pol}
\eeq
Thus, this coefficient is 
given solely by the sum $\oln{\alpha}+\oln{\beta}$
of the polarizabilities in the case of the forward
scattering, and by their difference $\oln{\alpha}-\oln{\beta}$ in the case of
the backward scattering. 
Comparing the cross sections shown by the 
solid and dashed curves in Fig.~12 (below the pion threshold), 
we can extract the 
effect on $\left(\oln{\alpha}+\oln{\beta}\right)$ due to the inclusion of
the principal value parts of the nucleon-pion loop integrals in the T-matrix,
\beq
\Delta\left(\oln{\alpha}+\oln{\beta}\right)=-4.76\cdot 10^{-5} fm^3.
\eqlab{apb_dif}
\eeq
The effect on the
difference of the polarizabilities can be inferred from the similar calculations
at $\theta_{\gamma}=180^o$,
\beq
\Delta\left(\oln{\alpha}-\oln{\beta}\right)=2.97\cdot 10^{-5} fm^3.
\eqlab{amb_dif}
\eeq 
As pointed out above, the form factors $F_2^{\pm}$ give the dominant 
contribution (about 90 \%) to these effects.

\section{Conclusions}

The electromagnetic interaction of off-shell nucleons is an important ingredient
in the description of many hadronic processes at low and intermediate energies.
Compared to the on-shell $\gamma N N$ vertex, the structure of the 
off-shell nucleon-photon vertex is more complicated.  
In this paper we consider the situation where only one of the nucleons
in the vertex is off the mass shell. We have developed a non-perturbative 
model to calculate the form factors in such a vertex. The key element of the
model is an integral equation which describes dressing of the vertex with 
an infinite number of pion loops. 
In the solution procedure we take advantage of unitarity and analyticity
considerations in that dispersion relations \cite{bincer} are utilized to find 
the real parts of the form factors from their imaginary parts. The latter, 
in turn, 
are obtained by applying cutting rules \cite{cut}, with only the 
one-pion-nucleon discontinuities of the loop integrals taken into account. 
The dependence of the form factors on the four-momentum squared of the 
off-shell nucleon 
deviates from a monopole- (or dipole-) like shape adopted often in 
phenomenological applications. In particular, a characteristic feature of our 
results is a cusp-like structure of the form
factors in the vicinity of the one-pion threshold, especially conspicuous for
the magnetic form factors corresponding to negative-energy states of the 
off-shell nucleon.

One of the important requirements for the electromagnetic vertex is obeying the
Ward-Takahashi identity, which relates the vertex with the nucleon propagator
\cite{takah}.
Our model is consistent with this condition, thus yielding
a vertex satisfying the Ward-Takahashi identity.
In this respect, a crucial ingredient of the equation is a four-point 
$\gamma \pi N N$ term. In a theory with
nucleon-pion form factors, the presence of
such a term is necessitated by the requirement that the photon field be 
included consistently with the principle of local gauge invariance. 
We construct a $\gamma \pi N N$ vertex using the prescription of
minimal substitution, in terms of an interaction Lagrangian with higher
derivatives.
Terms in the contact vertex which are transverse to the
photon four-momentum cannot be uniquely determined if the vertex is built using
minimal substitution. As an example of such an ambiguity, we have
constructed two contact terms with different transverse components. 
We used these two contact terms in the calculation of the 
half-off-shell nucleon-photon vertex and found 
that the negative-energy magnetic form factors are influenced noticeably by the 
choice of the contact term, while the effect on the positive-energy 
form factors is rather small.

It should be emphasized that off-shell vertices (as any general 
Green's functions, for that matter) depend not only on the model used to 
calculate them, but also on the representation of fields in the Lagrangian.
In contrast, the measurable physical observables are obtained from 
the scattering matrix and are therefore oblivious to the representation of the
Lagrangian (see, e.g., \cite{kamef}). Even though information on the
half-off-shell vertices cannot be unambiguously extracted from experiment,
they are important for the calculation of observables.

Our approach to the calculation of the observables in scattering processes can
be viewed as a two-stage scheme. First, by constructing the dressed
half-off-shell vertices and nucleon propagator, we account for the real 
parts of the loop diagrams contributing to the T-matrix. From the thus
constructed
effective interaction, the kernel K in the K-matrix approach is built
using only skeleton diagrams. 
We argued that
the present model can be consistently
applied in a coupled-channel K-matrix approach to Compton scattering, 
pion photoproduction and pion scattering  (where, apart from the dressed 
nucleon-photon 
vertex, also the dressed nucleon propagator and
the dressed half-off-shell nucleon-pion vertex, obtained in a previous paper, 
are needed). 
We focused our attention on a 
comparison of the observables obtained without and with the use of the dressed
vertices and propagator in the K-matrix. In particular, the cross section of
real Compton scattering was calculated, keeping only the nucleon, 
pion and photon degrees of freedom in the K-matrix model. To ensure the gauge
invariance of the full Compton amplitude, a contact
$\gamma \gamma N N$ term is added. 
The latter corresponds to the 
particular representation of the class A diagrams 
(see \cite{gelgold} and, for a recent discussion, 
\cite{scherkorkoch,scher}). 
We have constructed such a term
based on the dressed nucleon propagator and the half-off-shell $\gamma N N$
vertex using the minimal substitution prescription.
We have shown that the full gauge invariant and crossing symmetric 
Compton amplitude thus constructed gives the cross section 
satisfying the low-energy theorem \cite{gelgold,low}. 
The effect 
of the using the dressed vertices and nucleon propagator can be up to about 
10 \% for the cross section, depending on the kinematics. 
As is known, the next-to-leading order term in a low-energy expansion of the 
cross section can be related to the electric 
and magnetic polarizabilities of the proton. By comparing the
cross sections calculated with and without the use of the form factors and
the nucleon self-energy, we obtained the change of the
polarizabilities due to the inclusion of the real parts of the nucleon-pion loop
integrals. In a simplified
case where the vertices do not contain the magnetic form factors, and for the
forward scattering, we expressed the
next-to-leading order term in the
functions parametrizing the dressed nucleon propagator.
    
As mentioned above, additional degrees of freedom can be included in
the framework of the present model. 
The model is being currently extended along these lines.

\acknowledgements

This work is part of the research program of the ``Stichting voor
Fundamenteel Onderzoek der Materie'' (FOM) with financial support
from the ``Nederlandse Organisatie voor Wetenschappelijk
Onderzoek'' (NWO). We  would like to thank
Alex Korchin and Rob Timmermans for discussions.

\appendix
\section{The technique of minimal substitution in dressed vertices}
\subsection{The minimal substitution performed in coordinate space.
Example of a $\gamma \pi N N$ vertex}

Consider first only the functional \eqref{act_ps}.  
The corresponding vertex has the structure
\beq
\tau_{\alpha}\,\gamma^5 [f(p^2)+f(p^{\prime 2})]
\eqlab{pion_ps_in}
\eeq

Suppose now we introduce an external photon field $A_{\mu}(x)$.
A standard way to make a theory invariant under the local gauge transformations
\beq
\begin{array}{l}
\psi(x)\,\longrightarrow\,\exp(-i \hat{e} \alpha(x))\,\psi(x),\\
A_{\mu}(x) \longrightarrow\, A_{\mu}(x) + \partial_{\mu} \alpha(x),
\end{array}
\eqlab{gauge}
\eeq
is the minimal substitution procedure which consists in replacing all
derivatives in the Lagrangian with their covariant counterparts,
\beq
{\partial}^{\prime}_{\mu}=\partial_{\mu}+i \hat{e} A_{\mu},
\eqlab{der_cov}
\eeq
where $\hat{e} = e \hat{e}_N \equiv e(1+\tau_3)/2$. 
We are going to apply this prescription to
the integrand of the functional \eqref{act_ps}. The D'Alambertian gets replaced
with
\beq
{\Box}+2i\hat{e}A_{\mu}\partial^{\mu}+
i\hat{e}(\partial^{\mu}A_{\mu})-\hat{e}^2 A_{\mu} A^{\mu}.
\eqlab{dal_cov}
\eeq
Correspondingly, due to the presence of the electromagnetic field,
the functional $I$ undergoes the transformation
\beq
I \longrightarrow I^{\prime}=I+\delta I,
\eqlab{itild}
\eeq
with the variation
\beq
\delta I = \int\!d^4x \left( \frac{\delta I^{\prime}}{\delta A_{\mu}}
A_{\mu} + O(A^2) \right),
\eqlab{ivar}
\eeq
where it is tacitly understood that the variational derivative
$\delta I^{\prime}/\delta A_{\mu}$ is taken at the ``point" $A_{\mu}=0$.
In \eqref{ivar}, $O(A^2)$ denotes all terms involving two or more photon
fields. In addition to the original $\pi N N$ vertex, the transformed
action $I^{\prime}$ contains the vertex corresponding to the first term
in the integrand of \eqref{ivar}. It comprises two nucleon fields, one pion
field and one photon field, i.e.\ it is the four-point vertex we are
after.
Thus, the problem of deriving a contact $\gamma \pi N N$ vertex has been 
re-expressed in terms of finding the variational derivative of 
$I^{\prime}$ with respect to $A_{\mu}$.
Incidentally, we note that the
last term can be neglected in the covariant D'Alambertian \eqref{dal_cov},
because the part of the variational derivative coming from this term will vanish
for $A_{\mu}=0$. So we will use only the relevant part of the covariant
D'Alambertian, which reads
\beq
{\Box}^{\prime} = {\Box} + 2i\hat{e}A_{\mu}\partial^{\mu}+
i\hat{e}(\partial^{\mu}A_{\mu}).
\eqlab{dal_tild}
\eeq
The transformed action has the form of \eqref{act_ps} with ${\Box}$
replaced by ${\Box}^{\prime}$. It can be written as
\beq
I^{\prime} = \int\!d^4 x F(A_{\mu},\partial_{\nu} A_{\mu},
\partial_{\nu}\partial_{\lambda} A_{\mu},\ldots),
\eqlab{itil}
\eeq
where we have explicitly indicated only the dependence of the integrand $F$ on 
the photon
field and its derivatives. If $F$ depend on $A_\mu$ and its derivatives of 
the order up to and including $N$, then, by the Euler-Lagrange formula, 
the variational derivative of $I^\prime$ equals
\beq
\frac{\delta I^{\prime}}{\delta A_{\mu}}=
\frac{\partial F}{\partial A_{\mu}} +
\sum_{n=1}^{N} (-)^n \pd_{\alpha_1} \ldots \pd_{\alpha_n}\,\frac{\partial F}
{\partial(\partial_{\alpha_1} \ldots \partial_{\alpha_n} A_{\mu})}.
\eqlab{varder}
\eeq

To give examples of the procedure outlined, we are going to build contact
terms for the functionals $I_1$ and $I_2$ which are defined according to
\eqref{act_ps} with $f(-{\Box})={-\Box}$ and
$f(-{\Box})={\Box}^2$, respectively.
Then we will generalize these results for the case of
$f(-{\Box})={-\Box}^k$, for any natural $k$,
which will allow us to formulate the contact vertex for the general form of the
function $f(-{\Box})$.

First, let $f(-{\Box})=-\Box=-{\partial}^2$. Then the corresponding
$\pi N N$ vertex \eqref{pion_ps_in} with the incoming off-shell nucleon
has the form
\beq
\tau_{\alpha} \gamma^5 [p^2+p^{\prime 2}]
\eqlab{piver_1}
\eeq
In this case, the action in the presence of the
electromagnetic field reads 
\beq
I^{\prime}_1 = \int\!d^4 x F_1(A_{\mu},\partial_{\nu} A_{\mu}),
\eqlab{Iprime_1}
\eeq
where
\begin{eqnarray}
F_1&=&i \overline{\psi} \gamma^5 \tau_{\alpha} \phi^{\alpha}
[{\partial}^2+2i\hat{e}A_{\mu}\partial^{\mu}+i\hat{e}(\partial^{\mu}A_{\mu})]
\psi \nonumber \\
&&+i \overline{\psi}[\overleftarrow{{\partial}^2}-2i\hat{e}
\overleftarrow{{\partial}^{\mu}}A_{\mu}-
i\hat{e}(\partial^{\mu}A_{\mu})] \tau_{\alpha} \phi^{\alpha} \gamma^5 \psi.
\eqlab{F_1} \end{eqnarray}
By \eqref{varder},
\begin{eqnarray}
{\displaystyle \frac{\delta I^{\prime}_1}{\delta A_{\mu}}} A_{\mu} &=& i \Big\{
2i\tau_{\alpha}\hat{e}\oln{\psi}\gamma^5\phi^{\alpha}(\partial^{\mu}\psi)-
2i\hat{e}\tau_{\alpha}(\overline{\psi}\overleftarrow{{\partial}^{\mu}})
\phi^{\alpha}\gamma^5\psi \nonumber \\
&&-i\tau_{\alpha}\hat{e}
(\partial^{\mu}[\overline{\psi}\gamma^5\phi^{\alpha}\psi])+
i\hat{e}\tau_{\alpha}([\overline{\psi}\phi^{\alpha}\gamma^5\psi]
\overleftarrow{{\partial}^{\mu}}) \Big\} A_{\mu}. \eqlab{d1}
\end{eqnarray}
Substituting this in \eqref{ivar} and integrating the last two terms by parts,
we obtain for the variation of $I_1$,
\begin{eqnarray}
\delta I_1 &= &i {\ds \int}\!d^4x \Big\{ 2i\tau_{\alpha}\hat{e}
\overline{\psi}\gamma^5
\phi^{\alpha}(\partial^{\mu}\psi)A_{\mu}-2i\hat{e}\tau_{\alpha}(\overline{\psi}
\overleftarrow{{\partial}^{\mu}})\phi^{\alpha}\gamma^5\psi A_{\mu} \nonumber \\
&&+i\tau_{\alpha}\hat{e}\overline{\psi}\gamma^5\phi^{\alpha}\psi
(\partial^{\mu}A_{\mu})-i\hat{e}\tau_{\alpha}\overline{\psi}\phi^{\alpha}
\gamma^5\psi(\partial^{\mu}A_{\mu}) \Big\}.
\end{eqnarray}
The four-point vertex corresponding to this functional is
\beq
-\gamma^5\{\tau_{\alpha}\hat{e}(2p^{\mu}+q^{\mu})+\hat{e}\tau_{\alpha}
(2p^{\prime \mu}-q^{\mu})\},
\eqlab{contver_1}
\eeq
where $q^{\mu}$ is the four-momentum of the incoming photon. 

Now let $f(-{\partial}^2)={\Box}^2={\partial}^4$. Then the initial
$\pi N N$ vertex is
\beq
\tau_{\alpha}\gamma^5 [p^4+p^{\prime 4}]
\eqlab{piver_2}
\eeq
and we have to take the variational derivative of the action
\beq
I^{\prime}_2 = \int\!d^4 x F_2(A_{\mu},\partial_{\nu} A_{\mu},
\partial_{\nu}\partial_{\lambda}\partial_{\rho} A_{\mu}),
\eqlab{Iprime_2}
\eeq
where
\begin{eqnarray}
F_2&=&-i \overline{\psi} \gamma^5 \tau_{\alpha} \phi^{\alpha}
[{\partial}^4+4i\hat{e}A_{\mu}\partial^{\mu}\partial^2+2i\hat{e}
(\pd^\mu A_\mu)\pd^2+2i\hat{e}(\pd^2 A_\mu)\pd^\mu  \nonumber \\
&&+i\hat{e}(\pd^2 \pd^\mu A_\mu)+2i\hat{e}(\pd^\lambda \pd^\mu
A_\mu)\pd_\lambda+
4i\hat{e}(\pd^\lambda A_\mu)\pd^\mu \pd_\lambda]\psi  \nonumber \\
&&-i \overline{\psi}[\overleftarrow{{\partial}^4}-4i\hat{e}
\overleftarrow{\pd^2} \overleftarrow{{\partial}^{\mu}}A_{\mu}-
2i\hat{e}\overleftarrow{\pd^2}(\pd^\mu A_\mu)-2i\hat{e}\overleftarrow{\pd^\mu}
(\pd^2 A_\mu) \nonumber \\
&&-i\hat{e}(\pd^2 \pd^\mu
A_\mu)-2i\hat{e}\overleftarrow{\pd_\lambda}(\pd^\lambda \pd^\mu A_\mu)-
4i\hat{e}\overleftarrow{\pd_\lambda} \overleftarrow{\pd^\mu}(\pd^\lambda
A_\mu)]\tau_{\alpha} \phi^{\alpha}\gamma^5 \psi. \eqlab{F_2}
\end{eqnarray}
Again, applying \eqref{varder}, we find
\begin{eqnarray}
\frac{\delta I^{\prime}_2}{\delta A_{\mu}} A_{\mu} &=& -i \Big\{
4i\tau_{\alpha}\hat{e}\oln{\psi}\gamma^5\phi^{\alpha}(\partial^{\mu}\pd^2 \psi)-
4i\hat{e}\tau_{\alpha}(\overline{\psi}
\overleftarrow{\pd^2}\overleftarrow{{\partial}^{\mu}})
\phi^{\alpha}\gamma^5\psi
-2i\tau_{\alpha}\hat{e}(\pd^\mu \overline{\psi}\gamma^5 \phi^\alpha(\pd^2
\psi)) \nonumber \\
&&-4i\tau_\alpha \hat{e}(\pd_\lambda \oln{\psi}\gamma^5 \phi^\alpha
(\pd^\lambda \pd^\mu \psi))+2i\hat{e}\tau_\alpha(\pd^\mu
(\oln{\psi}\olft{\pd^2})\phi^\alpha \gamma^5 \psi)+4i\hat{e}\tau_\alpha
(\pd_\lambda(\oln{\psi}\olft{\pd^\lambda}\olft{\pd^\mu})\phi^\alpha
\gamma^5 \psi) \nonumber \\
&&+2i\tau_\alpha \hat{e}(\pd^2 \oln{\psi}\gamma^5 \phi^\alpha(\pd^\mu \psi))+
2i\tau_\alpha \hat{e}(\pd^\lambda \pd^\mu \oln{\psi}\gamma^5 \phi^\alpha
(\pd_\lambda \psi))-2i\hat{e}\tau_\alpha(\pd^2(\oln{\psi}\olft{\pd^\mu})
\phi^\alpha \gamma^5 \psi) \nonumber \\
&&-2i\hat{e}\tau_\alpha(\pd^\lambda
\pd^\mu(\oln{\psi}\olft{\pd_\lambda})\phi^\alpha \gamma^5 \psi) -
i\tau_\alpha\hat{e}(\pd^2\pd^\mu\oln{\psi}\gamma^5\phi^\alpha\psi)+
i\hat{e}\tau_\alpha(\pd^2\pd^\mu\oln{\psi}\gamma^5\phi^\alpha\psi)
\Big \} A_\mu.  \eqlab{d2}
\end{eqnarray}
Upon substituting this derivative in \eqref{ivar} and integrating the second to
the sixth terms by parts once, the seventh to the tenth ones twice, and the
last two terms thrice, we obtain the following four-point vertex:
\beq
-\gamma^5\{\tau_{\alpha}\hat{e}(2p^{\mu}+q^{\mu})[(p+q)^2+p^2]+
\hat{e}\tau_{\alpha}
(2p^{\prime \mu}-q^{\mu})[(p^{\prime}-q)^2+p^{\prime 2}]\}.
\eqlab{contver_2}
\eeq

Next, let the $\pi N N$ form factor in \eqref{pion_ps_in} be a monomial
$f(p^2)=p^{2k}$, $k$ being a natural number. It corresponds to
$f(-\Box)=(-{\Box})^k$ in \eqref{act_ps}. Following the same steps as for
the two above considered cases, we obtain for the four-point vertex:
\beq
-\gamma^5\{\tau_{\alpha}\hat{e}(2p^{\mu}+q^{\mu})
\sum_{l=0}^{k-1}p^{2l}(p+q)^{2(k-l-1)}\,
+\hat{e}\tau_{\alpha}(2p^{\prime \mu}-q^{\mu})\sum_{l=0}^{k-1}
p^{\prime 2l}(p^{\prime}-q)^{2(k-l-1)}\}.
\eqlab{contver_k}
\eeq
A generic function $f(p^2)$ in \eqref{pion_ps_in} can be formally
expanded in powers of $p^2$,
\beq
f(p^2)=\sum_k a_k p^{2k}.
\eqlab{f_expans}
\eeq
Using \eqref{contver_k} and the identity
\beq
\sum_{l=0}^{k-1} x^{2l} y^{2(k-l-1)} = \frac{y^{2k}-x^{2k}}{y^2-x^2},
\eqlab{telescope}
\eeq
the corresponding $\gamma \pi N N$ vertex can be written
\beq
-\gamma^5 \Big\{ \tau_\alpha \hat{e}(2p^{\mu}+q^{\mu})
\frac{f((p+q)^2)-f(p^2)}{(p+q)^2-p^2}+\hat{e}\tau_\alpha
(2p^{\prime \mu}-q^{\mu})\frac{f((p^{\prime}-q)^2)-f(p^{\prime 2})}
{(p^{\prime}-q)^2-p^{\prime 2}} \Big\}.
\eqlab{cont_ps}
\eeq

As a following application, consider the functional in \eqref{act_pv}.
The corresponding $\pi N N$ vertex reads
\beq
\tau_\alpha [\gamma^5 g(p^2) \vslash{p}+\vslash{p}^{\prime} g(p^2) \gamma^5],
\eqlab{pv_off_in}
\eeq
Following the procedure described above, we can write 
down the expression for the
contact term resulting from the minimal substitution in the functional 
$J$ of \eqref{act_pv}:
\begin{eqnarray}
&&-\tau_\alpha \hat{e}\gamma^5\Big\{ (2p^{\mu}+q^{\mu})\vslash{p}
{\displaystyle \frac{g((p+q)^2)-g(p^2)}{(p+q)^2-p^2}}+
\gamma^\mu g((p+q)^2) \Big\} \nonumber \\
&&-\hat{e}\tau_\alpha \Big\{ (2p^{\prime \mu}-q^{\mu})
{\displaystyle \frac{g((p^{\prime}-q)^2)-g(p^{\prime 2})}
{(p^{\prime}-q)^2-p^{\prime 2}}}
\vslash{p}^{\prime}+\gamma^\mu g((p^{\prime}-q)^2)
\Big\}\gamma^5.\eqlab{cont_pv}
\end{eqnarray}

One remark is in order here. Along with $J$ in \eqref{act_pv},
an equivalent form of the action functional could be written, corresponding
to the vertex \eqref{pv_off_in},
\beq
K = {\displaystyle \int\! d^4x \overline{\psi} \gamma^5 \tau_{\alpha}
\phi^{\alpha} [\vslash{\pd}g(-{\Box}) \psi] - \int\! d^4x
\overline{[\vslash{\pd} g(-{\Box}) \psi]}
\phi^{\alpha} \tau_{\alpha} \gamma^5 \psi }.
\eqlab{act_pv2}
\eeq
Since the operators $\Box$ and $\vslash{\pd}$ commute, $K=J$.
Switching on the electromagnetic field amounts to the replacements
$\vslash{\pd} \longrightarrow \vslash{\pd}^{\prime}$ and
$\Box \longrightarrow \Box^{\prime}$, 
see Eqs.~(\ref{eq:der_cov},\ref{eq:dal_tild}), and
correspondingly $J \longrightarrow J^{\prime}$ and
$K \longrightarrow K^{\prime}$. Now we note that
$K^{\prime}\ne J^{\prime}$, owing to the fact that $\Box^{\prime}$ and 
$\vslash{\pd}^{\prime}$ do not commute,
\beq
[\Box^{\prime},\,\vslash{\pd}^{\prime}]=i\hat{e}(\pd^2\vslash{A})+2i\hat{e}
(\pd_\mu \vslash{A})\pd^\mu-2i\hat{e}(\vslash{\pd}A_\mu)\pd^\mu-
i\hat{e}(\vslash{\pd}\pd^\mu A_\mu)+O(A^2).
\eqlab{commut}
\eeq
One can build the contact term corresponding to the initial
functional $K$ in \eqref{act_pv2}. It has the form
\begin{eqnarray}
&&-\tau_\alpha \hat{e}\gamma^5\Big\{ (2p^{\mu}+q^{\mu})(\vslash{p}+\vslash{q})
{\displaystyle \frac{g((p+q)^2)-g(p^2)}{(p+q)^2-p^2}}+
\gamma^\mu g(p^2) \Big\} \nonumber \\
&&-\hat{e}\tau_\alpha \Big\{ (2p^{\prime \mu}-q^{\mu})
{\displaystyle \frac{g((p^{\prime}-q)^2)-g(p^{\prime 2})}
{(p^{\prime}-q)^2-p^{\prime 2}}}
(\vslash{p}^{\prime}-\vslash{q})+\gamma^\mu g(p^{\prime 2}) \Big\}\gamma^5.
\eqlab{cont_pv_2} 
\end{eqnarray}

The difference between the vertices \eqref{cont_pv} and \eqref{cont_pv_2} equals
\begin{eqnarray}
\Delta^\mu&=&-\tau_\alpha \hat{e}\gamma^5\Big\{ (2p^{\mu}+q^{\mu})\vslash{q}
{\ds \frac{g((p+q)^2)-g(p^2)}{(p+q)^2-p^2}}+
\gamma^\mu [g(p^2)-g((p+q)^2)] \Big\} \nonumber \\
&&-\hat{e}\tau_\alpha \Big\{ (2p^{\prime \mu}-q^{\mu})
{\ds \frac{g((p^{\prime}-q)^2)-g(p^{\prime 2})}
{(p^{\prime}-q)^2-p^{\prime 2}}}
(-\vslash{q})+\gamma^\mu [g(p^{\prime 2})-g((p^{\prime}-q)^2)] \Big\}\gamma^5,
\eqlab{cont_pv_diff}
\end{eqnarray}
and it is trivial to show that it is orthogonal to the photon four-momentum,
 $q\cdot \Delta =0$. This presents one example
of the known ambiguity in constructing such contact vertices:
the terms orthogonal to the photon four-momentum cannot
be uniquely determined if one applies the minimal
substitution prescription.

\subsection{The minimal substitution performed in momentum space}

The technique applied here is essentially 
equivalent to that used above to build the 
$\gamma \pi N N$ vertex, except all manipulations now will be done directly in
momentum space, without resorting explicitly to variational derivatives of 
an action functional with higher derivatives.
A similar method of the minimal 
substitution in nucleon-pion form factors was employed by Ohta \cite{ohta}.

The minimal substitution in momentum space amounts to the following replacement
of the nucleon four-momentum: 
$P_\mu \longrightarrow \widetilde{P}_\mu=P_\mu-\hat{e}A_\mu$.
$P_\mu$ has to be considered as an operator
acting on the right.
If in a given term $P_\mu$ is the rightmost operator, the result of its action
is the (incoming) nucleon four-momentum $p_\mu$ which has 
c-number components.
Therefore, the minimal substitution in $\vslash{p}$ results in the
familiar electromagnetic vertex $-\hat{e}\gamma_\mu$, which we formally
denote as the mapping
\beq
\vslash{p}\longmapsto -\hat{e}\gamma_\mu.
\eqlab{ms_psl} 
\eeq

The electromagnetic field $A_\mu$
carries the four-momentum $q_\mu$ (throughout this appendix, all photon 
four-momenta are assumed to flow inwards all the vertices).
When acting
directly on the electromagnetic field, $P_\mu$ gives: $(P \cdot A)=A \cdot q$.
If we have, for example, $P^2$ on the left from the
scalar product of the electromagnetic field $A_\mu$ with a 
c-number vector $v_\mu$, the result will be
\beq
P^2 A\cdot v=A\cdot v\, (p+q)^2. 
\eqlab{p2_scpr}
\eeq 
More generally, considering any function $f(p^2)$ in the sense of the
formal expansion \eqref{f_expans}, one obtains
\beq
f(P^2) A\cdot v=A\cdot v\,f((p+q)^2).
\eqlab{f_scpr}
\eeq
Under the minimal
substitution, the nucleon four-momentum squared changes as
\beq
P^2 \longrightarrow \widetilde{P}^2=p^2-2\hat{e}A\cdot p-
\hat{e}A\cdot q +O(A^2)=p^2-\hat{e}\,A^\mu(2p_\mu+q_\mu)+O(A^2).
\eqlab{ms_p2}
\eeq
Collecting the coefficients of the terms linear in $A^\mu$ results in 
the mapping
\beq
p^2 \longmapsto -\hat{e}(2p_\mu+q_\mu).
\eqlab{p2_map}
\eeq

To generalize this for an arbitrary function
$f(p^2)$, we first consider the following combination:
\begin{eqnarray}
P^2\, \widetilde{P}^2&=&p^4-2\hat{e}q^2\,A\cdot p-2\hat{e}A\cdot p\, p^2-4\hat{e}
q\cdot p\, A\cdot p-\hat{e}q^2\, A\cdot q-\hat{e}A\cdot q\, p^2 \nonumber \\
&&-2\hat{e}A\cdot q \,q\cdot p+O(A^2) 
=p^4-\hat{e}\,A^\mu(2p_\mu+q_\mu)(p+q)^2+O(A^2), \eqlab{ms_p4}
\end{eqnarray}
where \eqref{ms_p2} has been used. The next step is to find the result of the
minimal substitution in the monomials $p^{2n},\,n=1,2,3\ldots\;$. Using 
Eqs.~(\ref{eq:p2_scpr},\ref{eq:ms_p2},\ref{eq:ms_p4}), we have
\begin{eqnarray}
P^{2n}\longrightarrow \widetilde{P}^{2n}&=&p^{2n}-[2\hat{e}A\cdot p+
\hat{e}A\cdot q]
\,\stackrel{n-1}{\overbrace{P^2\cdots P^2}}-
P^2\,[2\hat{e}A\cdot p+\hat{e}A\cdot q]\,
\stackrel{n-2}{\overbrace{P^2\cdots P^2}} \nonumber \\
&&-\ldots-\stackrel{n-1}{\overbrace{P^2\cdots P^2}}\,[2\hat{e}A\cdot p+
\hat{e}A\cdot q]+O(A^2) \nonumber \\
&=&p^{2n}-\hat{e}\,A^\mu(2p_\mu+q_\mu)\,{\displaystyle \sum_{m=0}^{n-1}}
(p+q)^{2m}\,p^{2(n-1-m)}+O(A^2), \eqlab{ms_p2n}
\end{eqnarray}
leading to the first part of \eqref{contver_k} in the case of the 
$\gamma \pi N N$ vertex.
Thus, 
\beq
p^{2n}\longmapsto -\hat{e}\,(2p_\mu+q_\mu)\,\frac{(p+q)^{2n}-p^{2n}}
{(p+q)^2-p^2},
\eqlab{p2n_map}
\eeq
where \eqref{telescope} has been used. Having established this and in
view of the expansion \eqref{f_expans}, it is
trivial to write the result of the minimal substitution for an arbitrary
function of $p^2$,
\beq
f(p^2)\longmapsto -\hat{e}\,(2p_\mu+q_\mu)\,\frac{f((p+q)^2)-f(p^2)}
{(p+q)^2-p^2},
\eqlab{f_map}
\eeq
leading to the first term in \eqref{cont_ps}. 
 
To find how the product $f(p^2)\vslash{p}$ changes under the minimal
substitution, we make use of
Eqs.~(\ref{eq:ms_psl},\ref{eq:f_map}) and \eqref{f_scpr}, as well as
\eqref{f_expans},
\begin{eqnarray}
f(P^2)\pslash{P}&\longrightarrow &f(\widetilde{P}^2)\widetilde{\pslash{P}}
\nonumber \\
&=&f(p^2)\vslash{p}-\hat{e}A^\mu(2p_\mu+q_\mu)\,
{\displaystyle \frac{f(p+q)^2)-f(p^2)}{(p+q)^2-p^2}}\,
\vslash{p}-\hat{e}A^\mu \gamma_\mu\,f((p+q)^2)+O(A^2), \eqlab{ms_fpsl}
\end{eqnarray}
and hence
\beq
f(p^2)\vslash{p}\longmapsto -\hat{e}\,(2p_\mu+q_\mu)\,\frac{f(p+q)^2)-f(p^2)}
{(p+q)^2-p^2}\,\vslash{p}-\hat{e}\,\gamma_\mu\,f((p+q)^2).
\eqlab{fpsl_map}
\eeq

Now one can do the minimal substitution of one photon
with the four-momentum $q_\mu$. Suppose that, using the same framework,
we want to add the second photon that carries the four-momentum $k_\nu$ flowing
inwards. Then, in addition to the formulas given above, we need to have the
result of the minimal substitution in a generic function $g(p \cdot q)$.
In the manner absolutely
analogous to deriving \eqref{f_map}, one can obtain  
\beq
g(p\cdot q)\longmapsto -\hat{e}\,q_\mu \frac{g((p+k)\cdot q)-g(p\cdot q)}
{k\cdot q}.
\eqlab{gpdq_map}
\eeq
An example of this formula, relevant to the construction of a
$\gamma \gamma N N$ vertex, is given by  
\beq
\frac{1}{(p+q)^2-p^2}\longmapsto -\hat{e}\,\frac{2q_\mu}{[(p+q+k)^2-(p+k)^2]
[(p+q)^2-p^2]}.
\eqlab{den_map}
\eeq 
Using Eqs.~(\ref{eq:f_map},\ref{eq:gpdq_map}) and \eqref{f_scpr}, we find that 
the minimal substitution in the product of a function $f$ of $p^2$ and a 
function $g$ of $p\cdot q$ gives
\begin{eqnarray}
f(p^2)g(p\cdot q)&\longmapsto& -\hat{e}\Big\{ (2p_\mu+k_\mu)
{\ds \frac{f((p+k)^2)-f(p^2)}{(p+k)^2-p^2}}\,g(p\cdot q) \nonumber \\
&& +q_\mu f((p+k)^2)\,
{\ds \frac{g((p+k)\cdot q)-g(p\cdot q)}{k\cdot q}} \Big\}. \eqlab{fg_map}
\end{eqnarray}
Next, it is trivial to show that
\beq
p_\mu\longmapsto -\hat{e}g_{\mu \nu}.
\eqlab{pmu_map}
\eeq

Suppose now that, instead of the operator $P_\mu$ considered above, we 
are interested in the operator 
$\olft{P}^{\,\prime}_\mu$ corresponding to the four-momentum of the outgoing 
nucleon. 
$\olft{P}^{\,\prime}_\mu$ acts to the left, yielding from the leftmost position
$p^\prime_\mu$, the c-number four-momentum of the outgoing nucleon. 
Under the minimal substitution, $\olft{P}^{\,\prime}_\mu$ changes as
$\olft{P}^{\,\prime}_\mu \longrightarrow \widetilde{\olft{P}}^{\,\prime}_\mu=
\olft{P}^{\,\prime}_\mu-\hat{e}A_\mu$.
Repeating the steps similar to those
that led to \eqref{f_map}, one obtains the result of the minimal substitution
in a function of the four-momentum squared of the outgoing nucleon:
\beq
f(p^{\prime 2})\longmapsto -\hat{e}\,(2p^\prime_\mu-q_\mu)\,
\frac{f((p^\prime-q)^2)-f(p^{\prime 2})}
{(p^\prime-q)^2-p^{\prime 2}},
\eqlab{f_prime_map}
\eeq
leading to the second term in \eqref{cont_ps}.
Also, the formulas analogous to 
Eqs.~(\ref{eq:gpdq_map},\ref{eq:fg_map},\ref{eq:pmu_map}) read
\beq
g(p^\prime\cdot q)\longmapsto \hat{e}\,q_\mu \frac{g((p^\prime-k)\cdot q)-
g(p^\prime\cdot q)}{k\cdot q},
\eqlab{gpdq_map_pr}
\eeq 
\begin{eqnarray}
g(p^\prime\cdot q)f(p^{\prime 2})&\longmapsto& -\hat{e}\Big\{ 
(2p^\prime_\mu-k_\mu)
{\ds \frac{f((p^\prime-k)^2)-f(p^{\prime 2})}{(p^\prime-k)^2-p^{\prime 2}}}\,
g(p^\prime\cdot q) \nonumber \\
&& -q_\mu f((p^\prime-k)^2)\,
{\ds \frac{g((p^\prime-k)\cdot q)-g(p^\prime\cdot q)}{k\cdot q}} \Big\}
\eqlab{fg_map_pr} 
\end{eqnarray}
and
\beq
p^\prime_\mu\longmapsto -\hat{e}g_{\mu \nu}.
\eqlab{pmu_map_pr}
\eeq

\section{The $\gamma \gamma N N$ vertex}
\subsection{The minimal substitution in the inverse nucleon propagator}

First we describe the minimal substitution of a photon with  
four-momentum $q_\mu$ in the function
\beq
i S^{-1}(p)=i[\alpha(p^2)\vslash{p}+\beta(p^2)]
\eqlab{prop_inv_ms}
\eeq
which is, up to a factor $i$, the inverse nucleon 
propagator \eqref{prop_inv}, with $\beta(p^2)= -\alpha(p^2)\xi(p^2)$. 
We can use Eqs.~(\ref{eq:f_map},\ref{eq:fpsl_map}) to write the 
nucleon-photon vertex obtained by the minimal substitution in 
\eqref{prop_inv_ms},
\begin{eqnarray}
-ie\Gamma^{min\,(1)}_\mu(p+q,p)&=&-i\hat{e}\left\{ (2p_\mu+q_\mu)\,{\ds \frac
{\alpha((p+q)^2)\vslash{p}+\beta((p+q)^2)-\alpha(p^2)\vslash{p}-\beta(p^2)}
{(p+q)^2-p^2}} \right. \nonumber \\
&&\left. + \gamma_\mu\,\alpha((p+q)^2) \right\}. \eqlab{gam_min}
\end{eqnarray}
This vertex satisfies the Ward-Takahashi identity \eqref{wti_gen}, which is
obvious if one rewrites \eqref{gam_min} in the form 
\beq
\Gamma^{min\,(1)}_\mu(p^\prime,p)=\hat{e}_N\left\{ \frac{2p_\mu+q_\mu}
{(p+q)^2-p^2}\,
[S^{-1}(p^\prime)-S^{-1}(p)]
+\alpha((p+q)^2)[\gamma_\mu-\vslash{q}\,\frac{2p_\mu+q_\mu}{(p+q)^2-p^2}]
\right\}.
\eqlab{gam_min_wti}
\eeq
(We stress that here the photon four-momentum $q=p^\prime-p$ is incoming, 
and therefore the sign of the the right-hand side of the Ward-Takahashi 
identity is reverse compared to \eqref{wti_gen}.)
In principle, both nucleons can be off-shell in this vertex.

Instead of \eqref{prop_inv_ms} for the inverse nucleon propagator, 
one could start from the form with the terms $\vslash{p}$ and $\alpha(p^2)$
interchanged, and proceed along the above lines to build a nucleon-photon 
vertex. The result then can be written 
\begin{eqnarray}
-ie\Gamma^{min\,(2)}_\mu(p^\prime,p^\prime-q)&=&-i\hat{e}\left\{ 
(2p^\prime_\mu-q_\mu)\,{\ds \frac
{\vslash{p}^\prime \alpha(p^{\prime 2})+\beta(p^{\prime 2})-
\vslash{p}^\prime \alpha((p^\prime-q)^2)-\beta((p^\prime-q)^2)}
{p^{\prime 2}-(p^\prime-q)^2}} \right. \nonumber \\
&&\left. +\gamma_\mu\,\alpha((p^\prime-q)^2) \right\}. \eqlab{gam_min_2}
\end{eqnarray}
This vertex differs form \eqref{gam_min} only by a part transverse to the photon
four-momentum, which is clear upon rewriting \eqref{gam_min_2} in the form
\beq
\Gamma^{min\,(2)}_\mu(p^\prime,p)=\hat{e}_N\left\{ \frac{2p_\mu+q_\mu}
{(p+q)^2-p^2}\,
[S^{-1}(p^\prime)-S^{-1}(p)]
+\alpha(p^2)[\gamma_\mu-\vslash{q}\,\frac{2p_\mu+q_\mu}{(p+q)^2-p^2}]
\right\},
\eqlab{gam_min_wti_2}
\eeq
to be compared with \eqref{gam_min_wti}. 

At this point we note that (a half of) the sum of the inverse propagator 
\eqref{prop_inv_ms} and the one with interchanged $\vslash{p}$ and 
$\alpha(p^2)$ corresponds to a hermitian action functional in coordinate
space. Starting from such a form of the propagator, one obtains the vertex 
\beq
-ie\Gamma_\mu^{min}(p^\prime,p)=\frac{-ie}{2}
\Big[\Gamma_\mu^{min\,(1)}(p^\prime,p)+
\Gamma_\mu^{min\,(2)}(p^\prime,p)\Big].
\eqlab{gam_min_herm}
\eeq

To obtain the half-off-shell vertex $\Gamma^{min}_\mu(m,p)$ with the outgoing 
on-shell nucleon, we apply \eqref{gam_min_herm} to the positive-energy spinor 
$\oln{u}(p^\prime)$ to the left, where $p^\prime=p+q$. First, using 
$\oln{u}(p^\prime)\vslash{p}^\prime=
\oln{u}(p^\prime)m$ as well as the anticommutator relation for the
$\gamma$-matrices, it can be shown that
\beq
\oln{u}(p^\prime)\,(p_\mu+p^\prime_\mu)=\oln{u}(p^\prime)\,
(\gamma_\mu \vslash{p}+
\gamma_\mu m-i\sigma_{\mu \lambda}q^\lambda),
\eqlab{p+ppr_1}
\eeq
by virtue of which we have
\begin{eqnarray}
\oln{u}(p^\prime)\Gamma^{min}_\mu(m,p)&=&\oln{u}(p^\prime)\,\hat{e}_N\Bigg\{
\gamma_\mu\, {\displaystyle \frac{\alpha(p^2)p^2+\beta(p^2)m}{p^2-m^2}}\,
+\gamma_\mu{\displaystyle \frac{\vslash{p}}{m}}\, 
{\displaystyle \frac{\alpha(p^2)m^2+\beta(p^2)m}{p^2-m^2}} \nonumber \\
&&+{\displaystyle i\frac{\sigma_{\mu \lambda}q^\lambda}{2m}}\,m
{\displaystyle \frac{\alpha(p^2)m+2\beta(p^2)-\beta(m^2)}
{m^2-p^2}}\,+
{\displaystyle i\frac{\sigma_{\mu
\lambda}q^\lambda}{2m}}{\displaystyle \frac{\vslash{p}}{m}}\,m^2
{\displaystyle \frac{\alpha(p^2)-\alpha(m^2)}{m^2-p^2}} \Bigg\}.
\eqlab{gmin_finon}
\end{eqnarray}
This can be cast in the form of \eqref{finon} (note that since the
photon here is incoming, we have a plus sign in front of the magnetic term), 
\beq
\overline{u}(p^{\prime})\,\Gamma^{min}_{\mu}(m,p) =
\overline{u}(p^{\prime}) \sum_{l={\pm}}
\left\{ \gamma_{\mu} F_1^l(p^2) +
i \frac{\sigma_{\mu \nu}q^{\nu}}{2m} \widetilde{F}_2^l(p^2) \right\}\, 
\Lambda_l(p),
\eqlab{finon_min}
\eeq
where the form factors $F_1^{\pm}(p^2)=
(F_1^{\pm})^s(p^2)+\tau_3(F_1^{\pm})^v(p^2)$ are 
found from the Ward-Takahashi identities 
Eqs.~(\ref{eq:wti_fm_alxi},\ref{eq:wti_fp_alxi}) with $\beta(p^2)=
-\alpha(p^2)\xi(p^2)$,
and 
\beq 
(\widetilde{F}_2^{+})^{s,v}(p^2)=(F_1^{-})^{s,v}(p^2)-(F_1^{+})^{s,v}(p^2),
\eqlab{f2pl_tilde}
\eeq
\beq
(\widetilde{F}_2^{-})^{s,v}(p^2)=\frac{2m^2}{m^2-p^2}(F_1^{-})^{s,v}(m^2)-
\frac{m^2+p^2}{m^2-p^2}(F_1^{-})^{s,v}(p^2)-(F_1^{+})^{s,v}(p^2).
\eqlab{f2mi_tilde}
\eeq

If the initial nucleon is on-shell, we have to apply \eqref{gam_min}
to the spinor $u(p)$ to the right. Taking advantage of the identity
\beq
(p_\mu+p^\prime_\mu)u(p)=(\vslash{p}^\prime \gamma_\mu+m\gamma_\mu-
i\sigma_{\mu \lambda}q^\lambda)u(p),
\eqlab{p+ppr_2}
\eeq
the result for this half-off-shell vertex can be written
\beq
\Gamma^{min}_{\mu}(p^{\prime},m)\,u(p) =
\sum_{k=\pm} \Lambda_k(p^{\prime})\,
\left\{ \gamma_{\mu} F_1^{k}(p^{\prime 2}) +
i \frac{\sigma_{\mu \nu}q^{\nu}}{2m} 
\widetilde{F}_2^{k}(p^{\prime 2}) \right\}\,u(p).
\eqlab{initon_min}
\eeq

The functions $\widetilde{F}_2^\pm(p^2)$ are expressed in terms of 
$F_1^\pm(p^2)$ 
and, therefore, determined by the Ward-Takahashi identity. 
In general, however, the magnetic form factors $F_2^\pm(p^2)$ are not restricted
by the Ward-Takahashi identity. 
Let us introduce two auxiliary functions $F(p^2)$ and
$G(p^2)$. Then we can write down the most general form of the 
half-off-shell vertex with the outgoing nucleon on-shell as
\beq
\oln{u}(p^\prime)\Gamma_\mu(m,p)=\oln{u}(p^\prime)\Big(
\Gamma^{min}_\mu(m,p)+[\vslash{q},\gamma_\mu]\{F(p^2)\vslash{p}+
G(p^2)\}\Big),
\eqlab{ver_fg_out}
\eeq
and of the half-off-shell vertex with the incoming nucleon on-shell as
\beq
\Gamma_{\mu}(p^{\prime},m)\,u(p)=\Big(\Gamma^{min}_{\mu}(p^{\prime},m)
+\{\vslash{p}^\prime F(p^{\prime2})+
G(p^{\prime 2})\}[\vslash{q},\gamma_\mu]\Big)\,u(p).
\eqlab{ver_fg_in}
\eeq
Now we equate \eqref{ver_fg_out} and \eqref{finon} 
(with the plus sign in front of the term proportional to 
$\sigma_{\mu \nu}q^\nu$) to find the functions
$F(p^2)$ and $G(p^2)$ in terms of the magnetic
form factors $(F_2^{\pm})^{s,v}(p^2)$,
\beq
(F)^{s,v}(p^2)=\frac{(F_2^{+})^{s,v}(p^2)-(F_2^{-})^{s,v}(p^2)-
(\widetilde{F}_2^{+})^{s,v}(p^2)+(\widetilde{F}_2^{-})^{s,v}(p^2)}{8m^2},
\eqlab{f_f2}
\eeq
\beq
(G)^{s,v}(p^2)=\frac{(F_2^{+})^{s,v}(p^2)+(F_2^{-})^{s,v}(p^2)-
(\widetilde{F}_2^{+})^{s,v}(p^2)-(\widetilde{F}_2^{-})^{s,v}(p^2)}{8m},
\eqlab{g_f2}
\eeq
where $(\widetilde{F}_2^{\pm})^{s,v}(p^2)$ are given by
Eqs.~(\ref{eq:f2pl_tilde},\ref{eq:f2mi_tilde}). 

\subsection{The minimal substitution in the $\gamma N N$ vertex}

Let the second photon field carry the four-momentum $k_\nu$ flowing inwards.
Using results established in Appendix A.2, the two-nucleon--two-photon contact 
term based on \eqref{gam_min_herm} can be written 
(the four-momentum conservation in this formula is $p^\prime=p+q+k$)
\beq
M^{ct,min}_{\mu \nu}(q,k)=\frac{1}{2}\Big[M^{ct,min\,(1)}_{\mu \nu}(q,k)+
M^{ct,min\,(2)}_{\mu \nu}(q,k) \Big],
\eqlab{ph_ph_ct_min_herm}
\eeq
where
\begin{eqnarray}
M^{ct, min\,(1)}_{\mu \nu}(q,k)&=&i\hat{e}^2\Bigg\{ {\ds \frac{\alpha(p^{\prime
2})
\vslash{p}+\beta(p^{\prime 2})-\alpha((p+q)^2)\vslash{p}-\beta((p+q)^2)}
{[p^{\prime 2}-(p+q)^2][(p+q)^2-p^2]}}\,(2p_\nu+2q_\nu+k_\nu)(2p_\mu+q_\mu)
\nonumber \\
&&{\ds -\frac{\alpha(p+k)^2)\vslash{p}+\beta((p+k)^2)-\alpha(p^2)
\vslash{p}-\beta(p^2)}{[(p+k)^2-p^2][(p+q)^2-p^2]}}\,(2p_\nu+k_\nu)
(2p_\mu+q_\mu) \nonumber \\
&&-{\ds \frac{\alpha(p^{\prime 2})\vslash{p}+\beta(p^{\prime 2})-
\alpha((p+k)^2)\vslash{p} -\beta((p+k)^2)}{[p^{\prime2}-(p+k)^2]
[(p+q)^2-p^2]}}\,2q_\nu(2p_\mu+q_\mu) \nonumber \\
&&{\ds +\frac{\alpha(p^{\prime 2})\vslash{p}+\beta(p^{\prime 2})-
\alpha((p+k)^2)\vslash{p} -\beta((p+k)^2)}{p^{\prime 2}-(p+k)^2}}\,
2g_{\mu \nu}   \nonumber \\
&&+{\ds \frac{\alpha(p^{\prime 2})-\alpha((p+k)^2)}{p^{\prime 2}-(p+k)^2}}\,
(2p_\mu+2k_\mu+q_\mu)\gamma_\nu  \nonumber \\ 
&&+{\ds \frac{\alpha(p^{\prime 2})-\alpha((p+q)^2)}
{p^{\prime 2}-(p+q)^2}}\,(2p_\nu+2q_\nu+k_\nu)\gamma_\mu \Bigg\} 
\eqlab{ph_ph_ct_min}
\end{eqnarray}
and
\begin{eqnarray}
M^{ct, min\,(2)}_{\mu \nu}(q,k)&=&i\hat{e}^2\Bigg\{
(2p^\prime_\mu-q_\mu)(2p^\prime_\nu-k_\nu)
{\ds \frac{\vslash{p}^\prime\alpha((p^\prime-k)^2)+\beta((p^\prime-k)^2)-
\vslash{p}^\prime\alpha(p^{\prime 2})-\beta(p^{\prime 2})}
{[p^{\prime 2}-(p^\prime-q)^2][(p^\prime-k)^2-p^{\prime 2}]}} \nonumber \\ 
&&-(2p^\prime_\mu-q_\mu)(2p^\prime_\nu-2q_\nu-k_\nu) 
{\ds \frac{\vslash{p}^\prime \alpha(p^2)+\beta(p^2)-
\vslash{p}^\prime \alpha((p^\prime-q)^2)-\beta((p^\prime-q)^2)}
{[p^{\prime 2}-(p^\prime-q)^2][p^2-(p^\prime-q)^2]}} \nonumber \\
&&-2q_\nu(2p^\prime_\mu-q_\mu){\ds \frac{\vslash{p}^\prime 
\alpha((p^\prime-k)^2))+\beta((p^\prime-k)^2))-
\vslash{p}^\prime \alpha(p^2) -\beta(p^2)}
{[(p^\prime-k)^2-p^2][p^{\prime 2}-(p^\prime-q)^2]}} \nonumber \\
&&+2g_{\mu \nu}{\ds \frac{\vslash{p}^\prime \alpha((p^\prime-k)^2)+
\beta((p^\prime-k)^2)-
\vslash{p}^\prime \alpha(p^2) -\beta(p^2)}
{(p^\prime-k)^2-p^2} }   \nonumber \\
&&+\gamma_\nu (2p^\prime_\mu-2k_\mu-q_\mu)
{\ds \frac{\alpha((p^\prime-k)^2)-\alpha(p^2)}
{(p^\prime-k)^2-p^2} }   \nonumber \\ 
&&+\gamma_\mu(2p^\prime_\nu-2q_\nu-k_\nu)
{\ds \frac{\alpha((p^\prime-q)^2)-\alpha(p^2)}
{(p^\prime-q)^2-p^2} } \Bigg\}. \eqlab{ph_ph_ct_min_2}
\end{eqnarray}

Hitherto, we have done the minimal substitution of the second photon only in the
vertex \eqref{gam_min_herm}. However, as was stressed above, this vertex does
not reproduce the most general magnetic terms for the half-off-shell cases.
Therefore, to complete the construction of the contact term, we have to add
to \eqref{ph_ph_ct_min_herm} the results of the minimal substitution of 
the second photon in those terms in
Eqs.~(\ref{eq:ver_fg_out},\ref{eq:ver_fg_in}) which contain
the auxiliary functions 
$F(p^2)$ and $G(p^2)$ given by Eqs.~(\ref{eq:f_f2},\ref{eq:g_f2}). 
This additional term, which we denote $M^{ct, add}_{\mu \nu}(q,k)$,
is built using the techniques described in Appendix A.2. It reads
\begin{eqnarray}
M^{ct, add}_{\mu \nu}(q,k)&=&i\hat{e}^2\Bigg\{
[\vslash{q},\gamma_\mu]\,\Big[ \frac{F((p+k)^2)\vslash{p}+
G((p+k)^2)-F(m^2)\vslash{p}-G(m^2)}{(p+k)^2-p^2}
\nonumber \\
&&\times (p_\nu+p^\prime_\nu-q_\nu)+F((p+k)^2)\gamma_\nu \Big] 
\nonumber \\ 
&&+\Big[ \frac{\vslash{p}^\prime F((p^\prime-k)^2)+
G((p^\prime-k)^2)-\vslash{p}^\prime F(m^2)-
G(m^2)}{(p^\prime-k)^2-p^{\prime 2}} \nonumber \\
&&\times (p_\nu+p^\prime_\nu+q_\nu)+
F((p^\prime-k)^2)\gamma_\nu\Big]\,
[\vslash{q},\gamma_\mu] \Bigg\}. \eqlab{m_add}
\end{eqnarray}
  
One requirement for the contact term is that it has to
be crossing symmetric, i.e.\ invariant under the simultaneous transformations
$\mu \longleftrightarrow \nu$ and $q \longleftrightarrow k$ \cite{gelgold}.
The matrix element of the ``minimal" contact term
\eqref{ph_ph_ct_min_herm} between the positive-energy spinors of the incoming
and outgoing nucleons is crossing symmetric, as can be verified explicitly.
This is a consequence of the fact that $M^{ct,min}_{\mu \nu}(q,k)$ is built
starting from an explicitly hermitian inverse nucleon propagator. 
Contrary to that, $M^{ct, add}_{\mu \nu}(q,k)$ 
has to be explicitly crossing symmetrized, i.e.\ the term 
$M^{ct, add}_{\nu \mu}(k,q)$ has to be added. 

Putting all the pieces together, we write the matrix element of the resulting 
$\gamma \gamma N N$ vertex between the positive-energy spinors of the
incoming and outgoing nucleons,
\beq 
\oln{u}(p^\prime)\,M^{ct}_{\mu \nu}(q,k)\,u(p)=\oln{u}(p^\prime)\,\left[
M^{ct,min}_{\mu \nu}(q,k)+M^{ct, add}_{\mu \nu}(q,k)+
M^{ct, add}_{\nu \mu}(k,q)\right]\,u(p),
\eqlab{ct_gamgam}
\eeq
with its explicit form given in \eqref{ct_me}.

\section{Consistency of the method with the Ward-Takahashi identity
for the $\gamma N N$ vertex}

The Ward-Takahashi identity is an important constraint on the nucleon-photon
vertex \cite{takah},
\beq
q\cdot \Gamma(p^\prime,p)=\hat{e}_N \big[ S^{-1}(p)-S^{-1}(p^\prime) \big],
\eqlab{wti_gen}
\eeq
with the photon four-momentum $q^\mu=p^\mu-p^{\prime \mu}$.
In this appendix we show that if we assume the validity of the Ward-Takahashi
identity for the $\gamma N N$ vertex on the right-hand
side of \eqref{sys}, then the present method of solution is consistent with
this assumption, i.e.\ the left-hand side of \eqref{sys}
will also obey \eqref{wti_gen}.

To this end, we first construct a pion photoproduction amplitude
and prove its gauge invariance.
Consider the following amplitude for the pion photoproduction process
(here both nucleons and the pion are on-shell, and the amplitude is written up
to a factor of $e$, the elementary charge),
\beq
T_\alpha^\mu = \sum_{i=1}^{4} T_{i,\alpha}^\mu,
\eqlab{pigam_prod}
\eeq
with
\beq
\oln{u}(p^\prime)T_{1,\alpha}^\mu u(p)=\oln{u}(p^\prime)\, 
\tau_\alpha \Gamma^5(m,p^\prime-k) 
S(p^\prime-k) \Gamma^\mu(p^\prime-k,m) \,u(p),
\eqlab{t1}
\eeq
\beq
\oln{u}(p^\prime)T_{2,\alpha}^\mu u(p)=\oln{u}(p^\prime)\, 
\Gamma^\mu(m,p^\prime+q) S(p^\prime+q)
\tau_\alpha \Gamma^5(p^\prime+q,m) \,u(p),
\eqlab{t2}
\eeq
\beq
\oln{u}(p^\prime)T_{3,\alpha}^\mu u(p)=\oln{u}(p^\prime)\,\tau_\beta\,\gamma^5\,
g\,D(k-q) V_{\beta \alpha}^\mu(k-q,k) \,u(p),
\eqlab{t3}
\eeq
\begin{eqnarray}
\oln{u}(p^\prime)T_{4,\alpha}^\mu u(p)&=&
\oln{u}(p^\prime) \left( -\tau_\alpha\hat{e}_N\Big\{ {\ds \frac{2p^\mu-q^\mu}
{(p-q)^2-p^2}}\big[ \Gamma^5(m,p-q)-\Gamma^5(m,m) \big] \right. \nonumber \\
&&+\gamma^5
{\ds \frac{g_2((p-q)^2)}{m}}\big[ \gamma^\mu + \vslash{q}\,
{\ds \frac{2p^\mu-q^\mu}{(p-q)^2-p^2}} \big] \Big\}  \nonumber \\
&& -\hat{e}_N\tau_\alpha\Big\{ {\ds \frac{2p^{\prime \mu}+q^\mu}
{(p^\prime+q)^2-p^{\prime 2}}}\big[ \Gamma^5(p^\prime+q,m)-\Gamma^5(m,m)
\big] \nonumber \\
&&\left. +\big[ \gamma^\mu - {\ds \frac{2p^{\prime \mu}+q^\mu}
{(p^\prime+q)^2-p^{\prime 2}}}\,\vslash{q} \big]
{\ds \frac{g_2((p^\prime+q)^2)}{m}}\,\gamma^5 \Big\} \right) u(p) ,\eqlab{t4} 
\end{eqnarray}
where the definitions given in Eqs.~(\ref{eq:pinnver}-\ref{eq:wtipion}) have 
been used. 
Eqs.~(\ref{eq:t1}-\ref{eq:t4})) correspond, 
respectively, to the u-, s-, t-channel and contact diagrams for the process
$N\pi \longrightarrow N\gamma$. The four-momentum of the pion $k$ is taken
incoming, the four-momentum of the photon $q$ is taken outgoing,
$p$ and $p^\prime$ are the momenta of the incoming and outgoing nucleons,
respectively; so four-momentum conservation reads
$p+k=p^\prime+q$.  The $\pi N N$ vertex in \eqref{t3} is taken as
$\tau_\beta \gamma^5 g$, where $g$ is the pion-nucleon coupling constant.
The reason for this is that the nucleon-pion vertex with the two
on-shell nucleons and an off-shell pion (carrying four-momentum $k$)
reduces to the form (see \eqref{pinnver})
$\tau_\alpha \gamma^5 G_1(m^2,m^2,k^2)$, where the form factor is
normalized so that $G_1(m^2,m^2,m_\pi^2)=g$. However, as mentioned above,
we have obtained a very weak dependence of the nucleon-pion vertex on the
pion four-momentum squared. Thus, we take $G_1(m^2,m^2,k^2)=g$
for all $k^2$, which
explains the form of the $\pi N N$ vertex in \eqref{t3}. 
The diagrammatic form of
Eqs.~(\ref{eq:t1}-\ref{eq:t4}) is shown in Fig.~3.

To prove gauge invariance of the pion photoproduction amplitude 
\eqref{pigam_prod}, i.e.\ to prove that \cite{weinb}
\beq
q_\mu \oln{u}(p^\prime)T^\mu_\alpha u(p) = 0,
\eqlab{gauge_piprod}
\eeq
we use \eqref{wti_gen} and the fact that if a nucleon is
on-shell with four-momentum $p$, then $\oln{u}(p)S^{-1}(p)=0$ and
$S^{-1}(p)u(p)=0$ since the propagator has a pole at the nucleon mass.
First, for $T_{1,\alpha}$ we have
\beq
q_\mu \oln{u}(p^\prime)T_{1,\alpha}^\mu u(p) = -\tau_\alpha \hat{e}_N
\oln{u}(p^\prime)\Gamma^5(m,p^\prime-k)u(p).
\eqlab{gauge_t1}
\eeq
Next,
\beq
q_\mu \oln{u}(p^\prime)T_{2,\alpha}^\mu u(p) = \hat{e}_N \tau_\alpha
\oln{u}(p^\prime)\Gamma^5(p^\prime+q,m)u(p),
\eqlab{gauge_t2}
\eeq
and
\beq
q_\mu \oln{u}(p^\prime)T_{3,\alpha}^\mu u(p) = 
(\hat{e}_\pi)_{\alpha \beta} \tau_\beta \oln{u}(p^\prime)\gamma^5 g u(p).
\eqlab{gauge_t3}
\eeq
Finally, for the contact term,
\begin{eqnarray}
q_\mu \oln{u}(p^\prime)T_{4,\alpha}^\mu u(p) &=& \tau_\alpha \hat{e}_N
\oln{u}(p^\prime)\big[ \Gamma^5(m,p^\prime-k) - \gamma^5 g \big]u(p)\nonumber\\
&&-\hat{e}_N \tau_\alpha
\oln{u}(p^\prime)\big[ \Gamma^5(p^\prime+q,m) - \gamma^5 g \big]u(p),
\eqlab{gauge_t4}
\end{eqnarray}
where we have taken into account the normalization condition for the
nucleon-pion vertex with both nucleons on-shell, 
$\oln{u}(p^\prime)\Gamma^5(m,m)u(p)=
\oln{u}(p^\prime)\Gamma^5(m,m)u(p)=\oln{u}(p^\prime)\gamma^5 g u(p)$.
At this point it may be noted that the terms in the second and the fourth lines
in the contact term \eqref{t4} are orthogonal to the photon momentum $q_\mu$, 
and therefore, although we keep these terms in $T_{4,\alpha}^\mu$, 
their presence 
is not necessary for the amplitude $T_\alpha^\mu$ to be gauge invariant.
Adding Eqs.~(\ref{eq:gauge_t1}-\ref{eq:gauge_t4}) and using
$[\hat{e}_N,\tau_\alpha]=-(\hat{e}_\pi)_{\alpha \beta} \tau_\beta$, 
we obtain the desired result,
\beq
q_\mu \oln{u}(p^\prime)T_\alpha^\mu u(p)= \oln{u}(p^\prime)\Big(
(\hat{e}_\pi)_{\alpha \beta} \tau_\beta \gamma^5 g +
\gamma^5 g [\hat{e}_N,\tau_\alpha] \Big) u(p) = 0.
\eqlab{gauge_t}
\eeq

We are going to use the gauge invariance of the 
pion photoproduction amplitude to show that our solution of \eqref{sys} obeys
the Ward-Takahashi identity for the nucleon-photon vertex.
This can be done in a transparent way with the help of diagrammatic
expressions.
As explained, the pole contribution to the vertex is given by the sum of
diagrams depicted in Fig.~2. (We assume here that the convergence of the
procedure has been reached, and hence the superscript $^n$ can be erased
in the $\gamma N N$ vertex in the loop of the second diagram.)
This sum can be rewritten by adding and
subtracting an additional diagram containing the pole contribution to the
self-energy in the incoming nucleon leg, as shown in Fig.~4 (top). 
Index $I$ on the
left-hand side ({\it l.h.s.}) of this equation indicates that
only the pole contribution to the vertex is considered.
We remind the reader that, as part
of our solution procedure, only real parts of the form factors and the
functions parametrizing the nucleon self-energy enter in the diagrams on the
right-hand side ({\it r.h.s.}). In other words, the imaginary part contribution 
to the vertex is solely due to the explicitly shown cut loops in Fig.~4 (top).
Now let us consider the scalar product of the {\it r.h.s.} with the photon
four-momentum $q_\mu$. To find the
result, it is convenient to rewrite the equation of Fig.~4 (top) as shown in 
Fig.~4 (bottom).
Here, a common sub-diagram, which is a half-off-shell nucleon-pion vertex, 
has been extracted from some of the diagrams on the {\it r.h.s.} of 
Fig.~4 (top), 
and the asterisk indicates that an integration is tacitly understood over 
the phase space of the cut (on-shell) nucleon and pion lines.
Such separation of a sub-diagram is
consistent with the interpretation of Cutkosky rules as a unitarity condition
\cite{cut}. Note also that we have erased the dash on the outgoing
nucleon line, to stress that now the Dirac spinor $\oln{u}(p^\prime)$ is
explicitly identified with this line. The sum of diagrams in the
brackets is just the scattering amplitude
$T_\alpha^\mu$ for pion photoproduction.
As has been shown, this amplitude is
gauge invariant, i.e.\ $q_\mu T_\alpha^\mu=0$. 
Therefore, the scalar product
of $q_\mu$ with the {\it r.h.s.} equals the scalar product of $q_\mu$ with 
only the
first diagram on the right-hand side of Fig.~4 (bottom),
\beq
q_\mu (r.h.s.)^\mu=-q_\mu \,\oln{u}(p^\prime)\Gamma^\mu(m,p) S(p) \Sigma_I(p),
\eqlab{qdotrhs}
\eeq
where $\Sigma_I(p)$ stands for the pole contribution to the nucleon
self-energy. Since we have assumed that the Ward-Takahashi identity holds for
the vertices on the {\it r.h.s.}, we can contract $q_\mu$ with 
$\Gamma^\mu(m,p)$ in \eqref{qdotrhs} and obtain
\beq
q_\mu (r.h.s.)^\mu=-\hat{e}_N \oln{u}(p^\prime) \Sigma_I(p).
\eqlab{qdotrhs_res}
\eeq
This is exactly what one obtains if one contracts $q_\mu$ with the pole
contribution to the general half-off-shell $\gamma N N$ vertex.
Indeed, it follows from \eqref{wti_gen} that
\beq
q_\mu \oln{u}(p^\prime) \Gamma_I^\mu(m,p) = \hat{e}_N \oln{u}(p^\prime)
S_I^{-1}(p),
\eqlab{wti_pole}
\eeq
where subscript $I$ denotes the pole contribution of the relevant entity.
Using the Schwinger-Dyson equation, $S^{-1}(p)=S_0^{-1}(p)-\Sigma(p)$, with
$S_0(p)=(\vslash{p}-m)^{-1}$ being the free nucleon propagator,
and taking into account that the pole contribution to $S_0^{-1}(p)$ is zero,
we see that the right-hand sides of \eqref{wti_pole} and \eqref{qdotrhs_res} are
equal. Thus, we have shown that the procedure used in this model
is consistent with the Ward-Takahashi identity. 


\newpage
\begin{figure}
\epsfxsize 12 cm
\centerline{\epsffile[0 240 594 559]{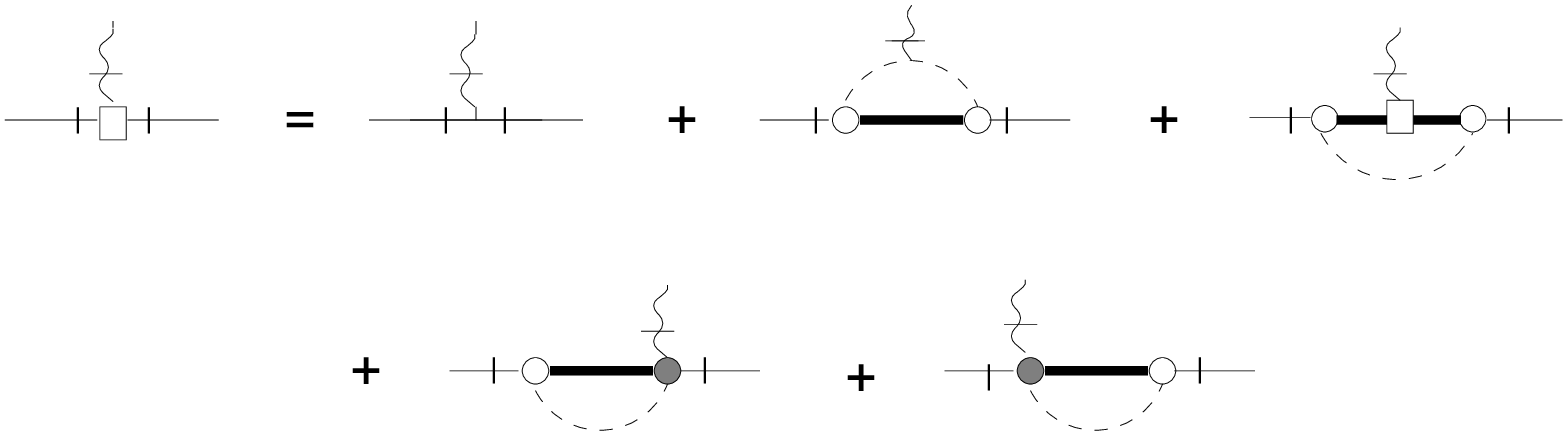}}
\caption[f1]{
The graphical representation of \eqref{sys}. The solid, dashed and wavy lines
are nucleons, pions and photons, respectively. The thick line is the dressed
nucleon propagator. The empty circle (square) is the dressed $\pi N N$
($\gamma N N$) vertex, and the shaded circle stands for the ``contact"
$\gamma \pi N N$ vertex. The dashed external lines are stripped away. 
\label{fig1}}
\end{figure}

\begin{figure}
\epsfxsize 12 cm
\centerline{\epsffile[0 140 593 660]{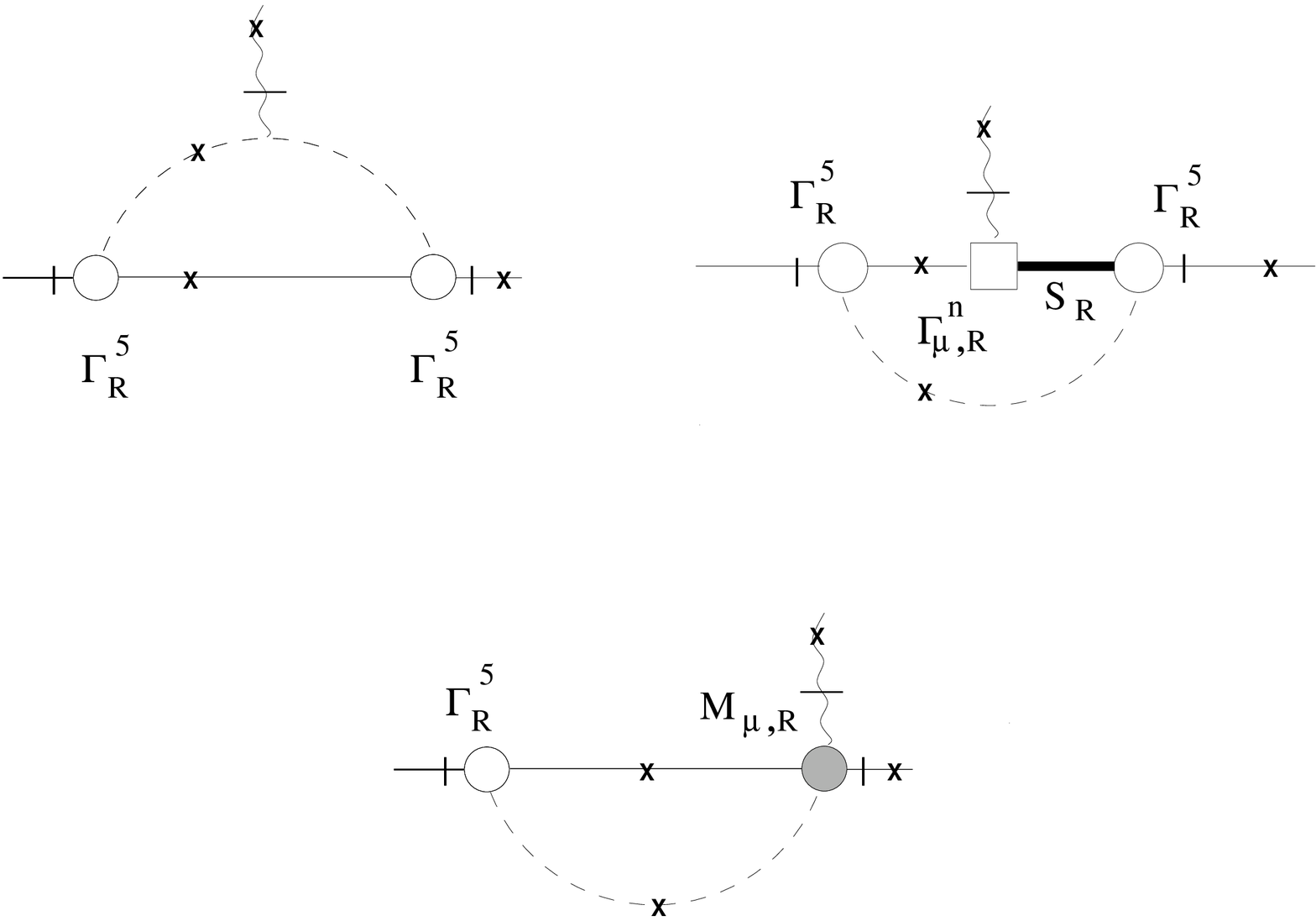}}
\caption[f2]{
The pole contributions of iteration $n+1$ to
the $\gamma N N$vertex. The left upper, the right upper and the lower diagrams
correspond to Eqs~(\ref{eq:pole_1},\ref{eq:pole_2},\ref{eq:pole_3}),
respectively. In addition to the notation explained in Fig.~1, the crossed lines
are on-shell particles. The vertices and the propagator bearing the subscript 
R contain only the real parts of the scalar functions that parametrize them.  
\figlab{f2}}
\end{figure}

\begin{figure}
\epsfxsize 12 cm
\centerline{\epsffile[0 210 594 560]{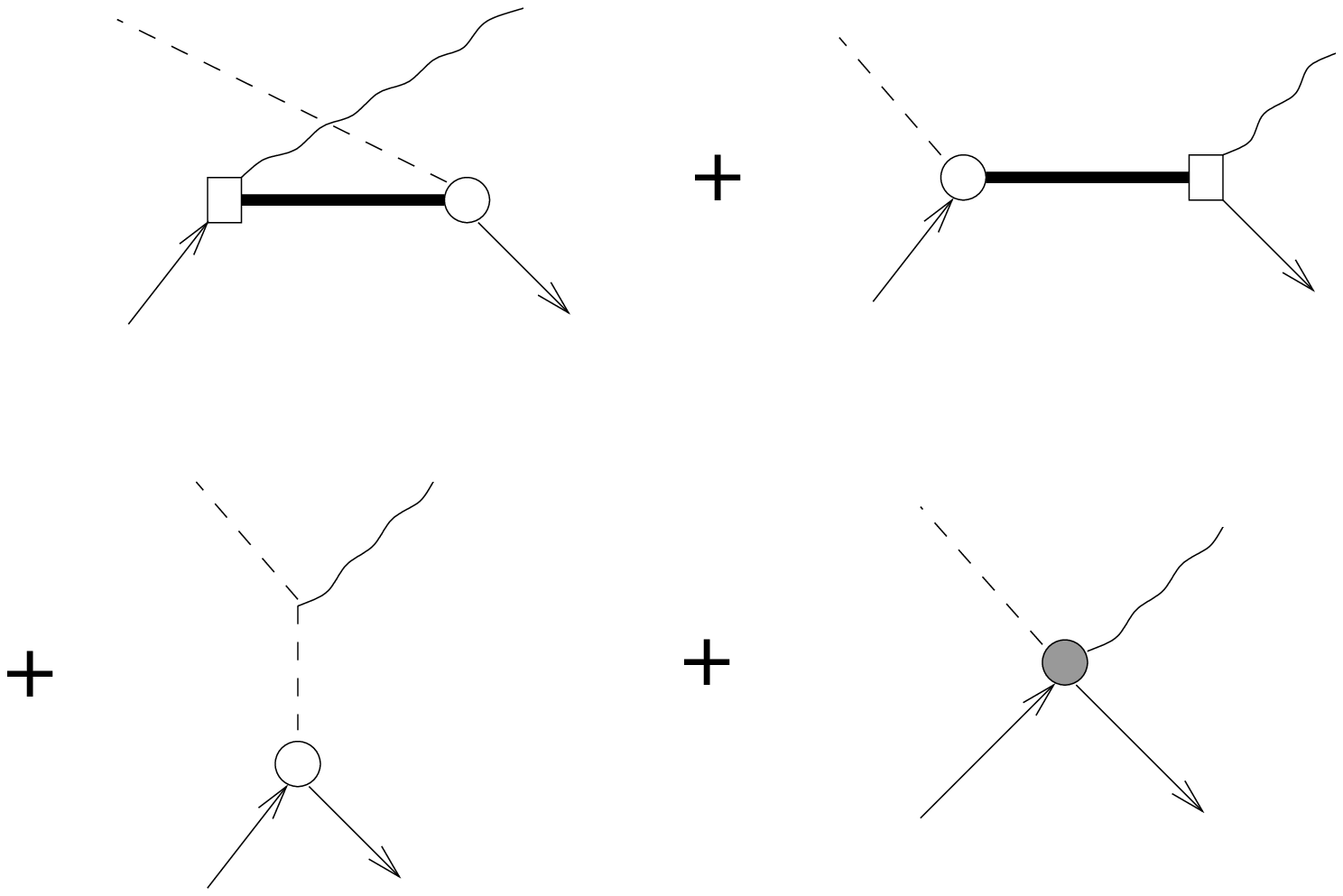}}
\caption[f3]{
The skeleton diagrams for the pion photoproduction amplitude
as given by Eqs.~(\ref{eq:pigam_prod}-\ref{eq:t4}). The notation is as in
Figs.~1 and 2, except all the external lines are on-shell and not 
stripped away.  
\label{fig3}}
\end{figure}

\newpage
\begin{figure}
\epsfxsize 12 cm
\centerline{\epsffile[0 210 594 595]{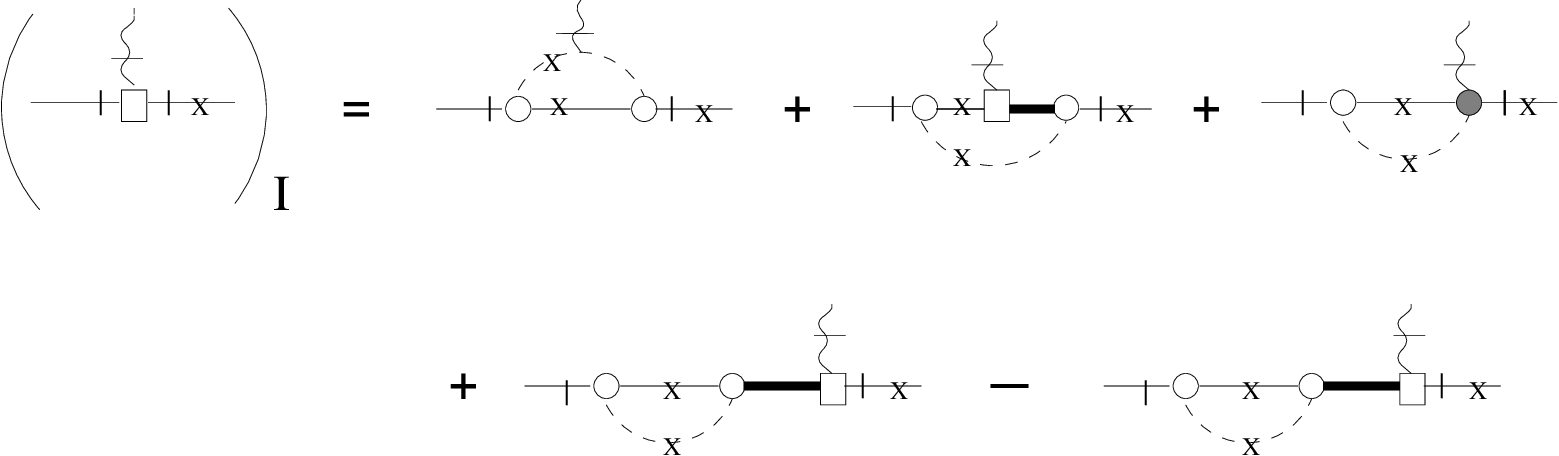}}

\epsfxsize 12 cm
\centerline{\epsffile[0 205 594 560]{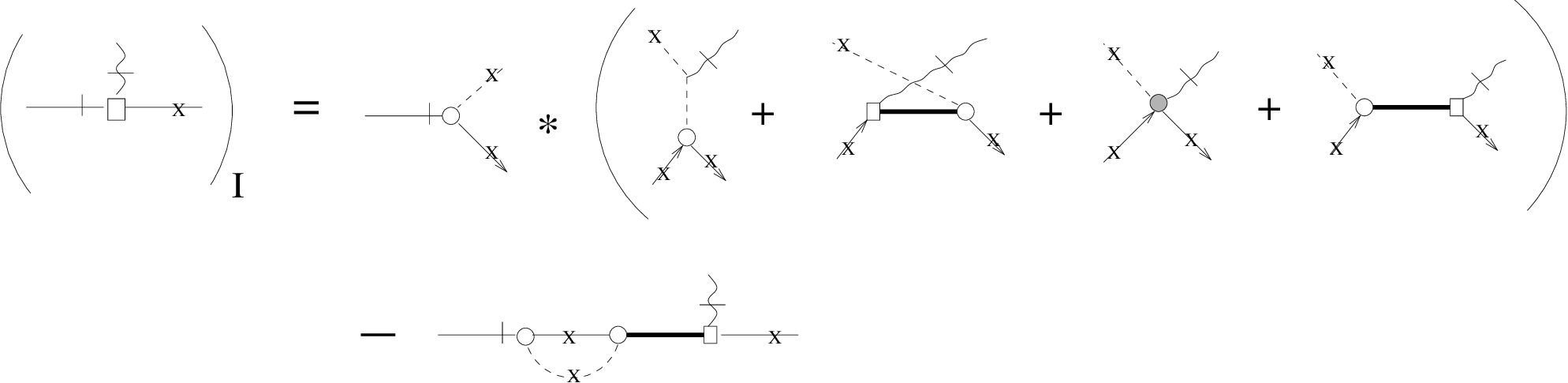}}
\caption[fig4]
{Top:
The diagrammatic equation used in the first part of the proof of the 
consistency of the method with the Ward-Takahashi identity, 
as explained in Section III.B.6. The notation is as in 
Figs.~1 and 2. The pole contribution to the 
$\gamma N N$ vertex (indicated by the subscript I) is shown on the left-hand
side. The right-hand side equals the sum of the diagrams in Fig.~2, where the
convergence is assumed to have been reached.
Bottom:
The second part of the proof of the 
consistency of the method with the Ward-Takahashi identity, 
as explained in Section III.B.6.
It is equivalent to applying both sides of the top equation 
to the positive-energy spinor of the outgoing nucleon. 
The asterisk denotes an integration over the phase space of the
cut nucleon and pion lines.
\label{fig4}}
\end{figure}

\begin{figure}
\epsfxsize 14. cm
\centerline{\epsffile[0 20 594 650]{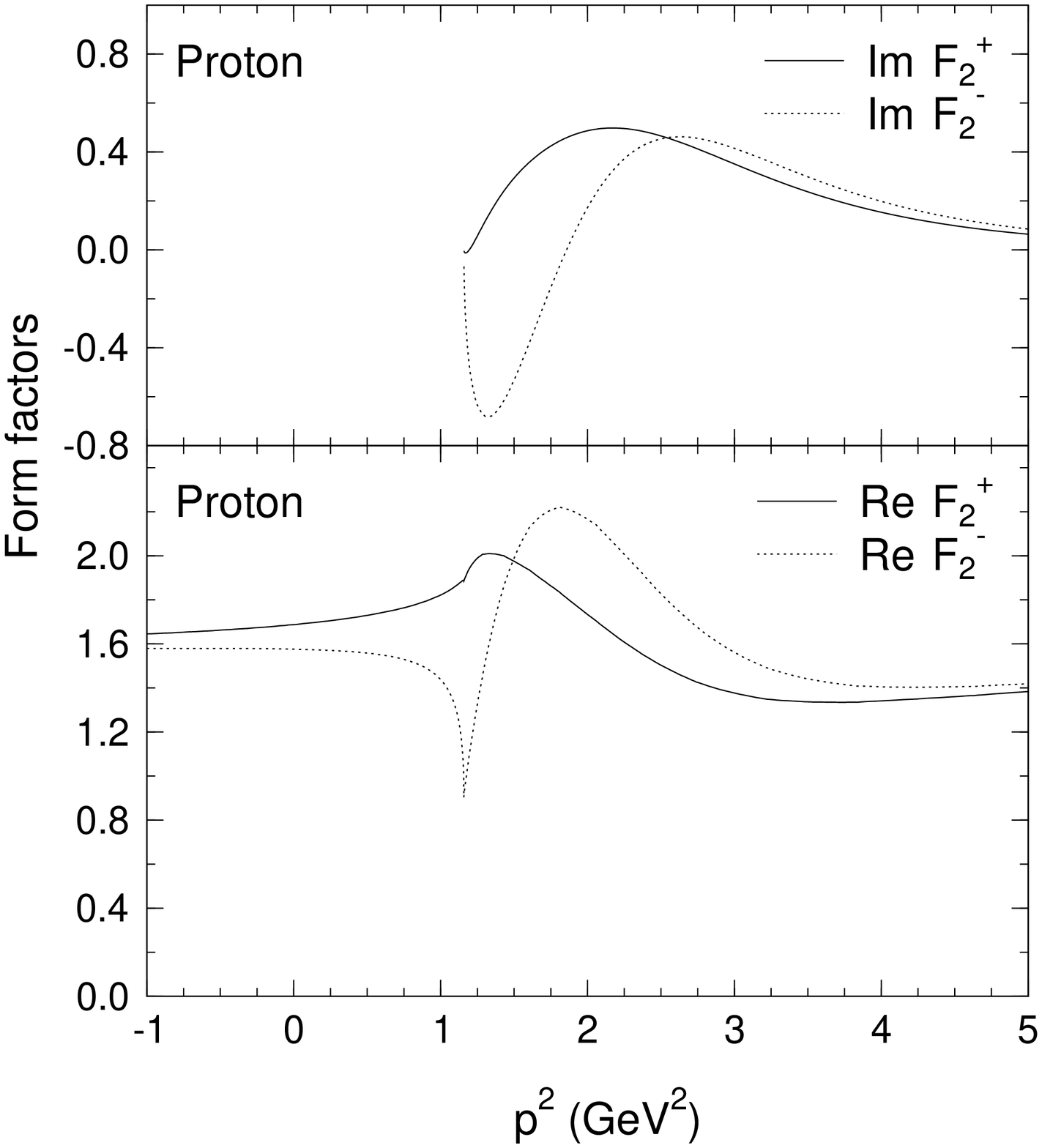}}
\caption[f5]{
The imaginary (the upper panel) and real (the lower panel) parts of the form 
factors $F_2^{+}$ (the solid curves)
and $F_2^{-}$ (the dotted curves) as functions of the
momentum squared of the off-shell proton, defined in \eqref{finon}. 
In this calculation, the contact $\gamma \pi N N$ term of \eqref{gampinn_cont} 
was used. 
\label{fig5}}
\end{figure}

\begin{figure}
\epsfxsize 14. cm
\centerline{\epsffile[0 20 594 650]{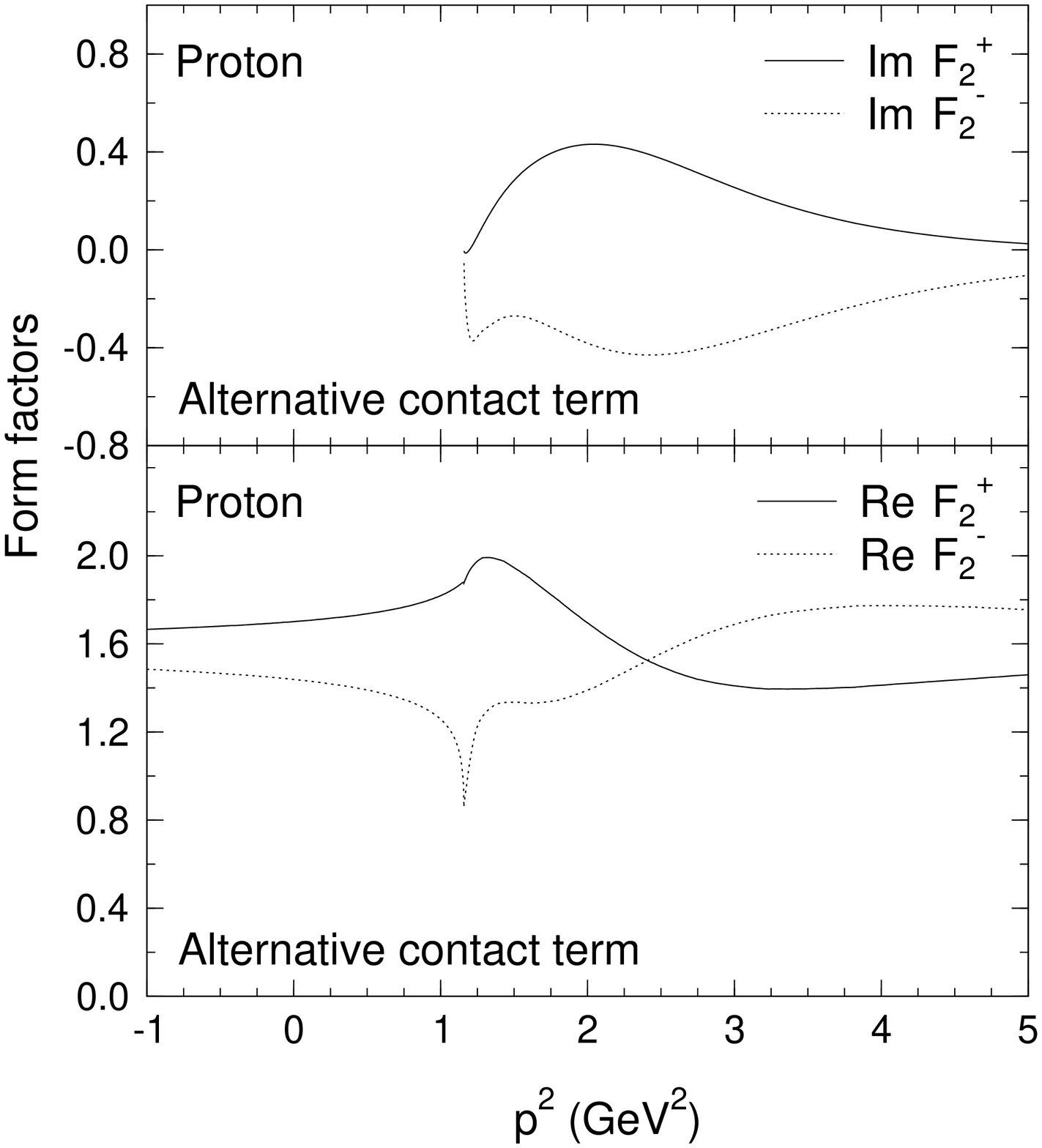}}
\caption[f6]{
The same as in Fig.~5, except the alternative contact $\gamma \pi N N$ term of
\eqref{gampinn_cont_2} was used in this calculation. 
\label{fig6}}
\end{figure}

\begin{figure}
\epsfxsize 14. cm
\centerline{\epsffile[0 20 594 650]{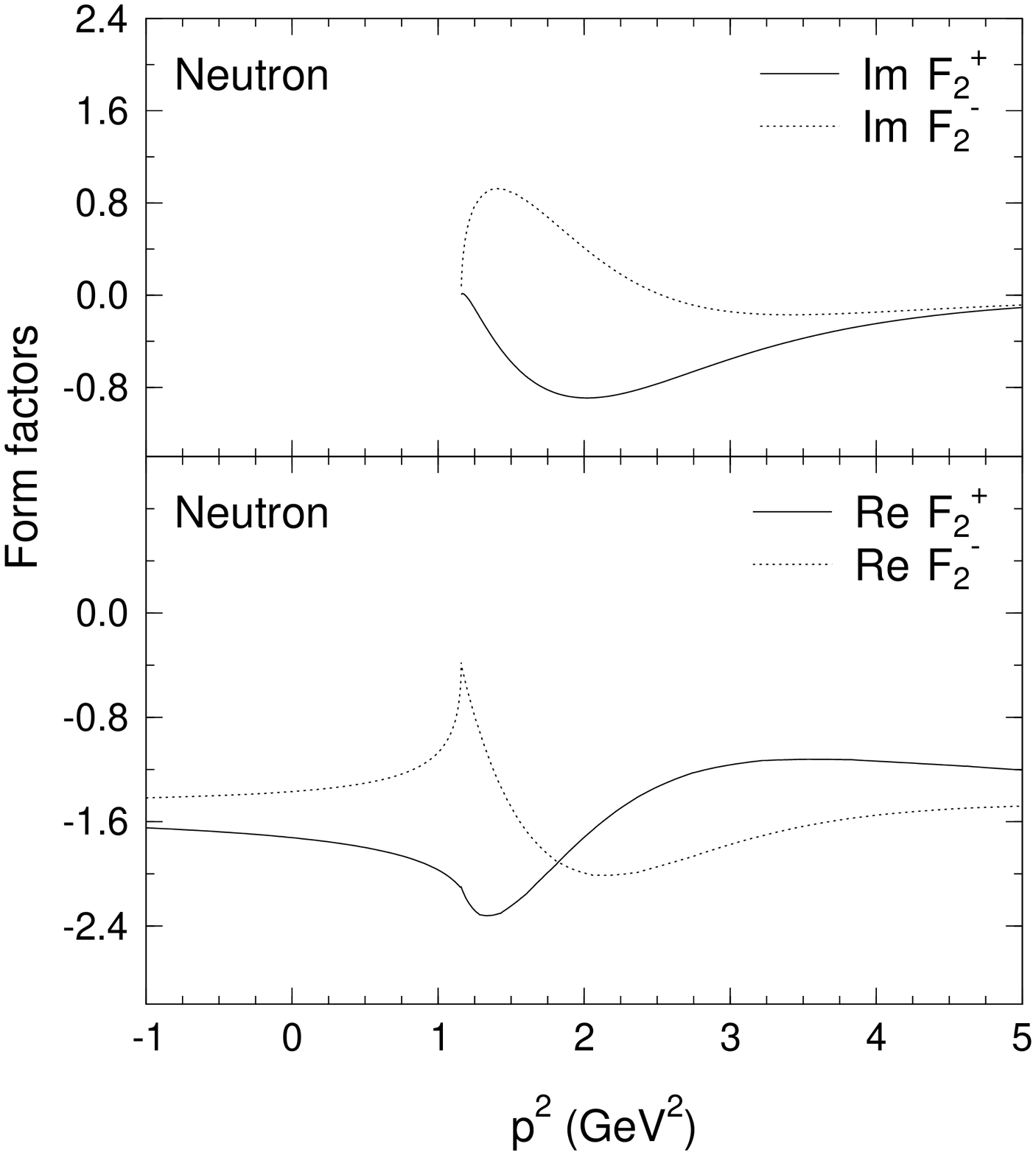}}
\caption[f7]{
The same as in Fig.~5, but for the neutron-photon vertex.
\label{fig7}}
\end{figure}

\begin{figure}
\epsfxsize 14. cm
\centerline{\epsffile[0 20 594 650]{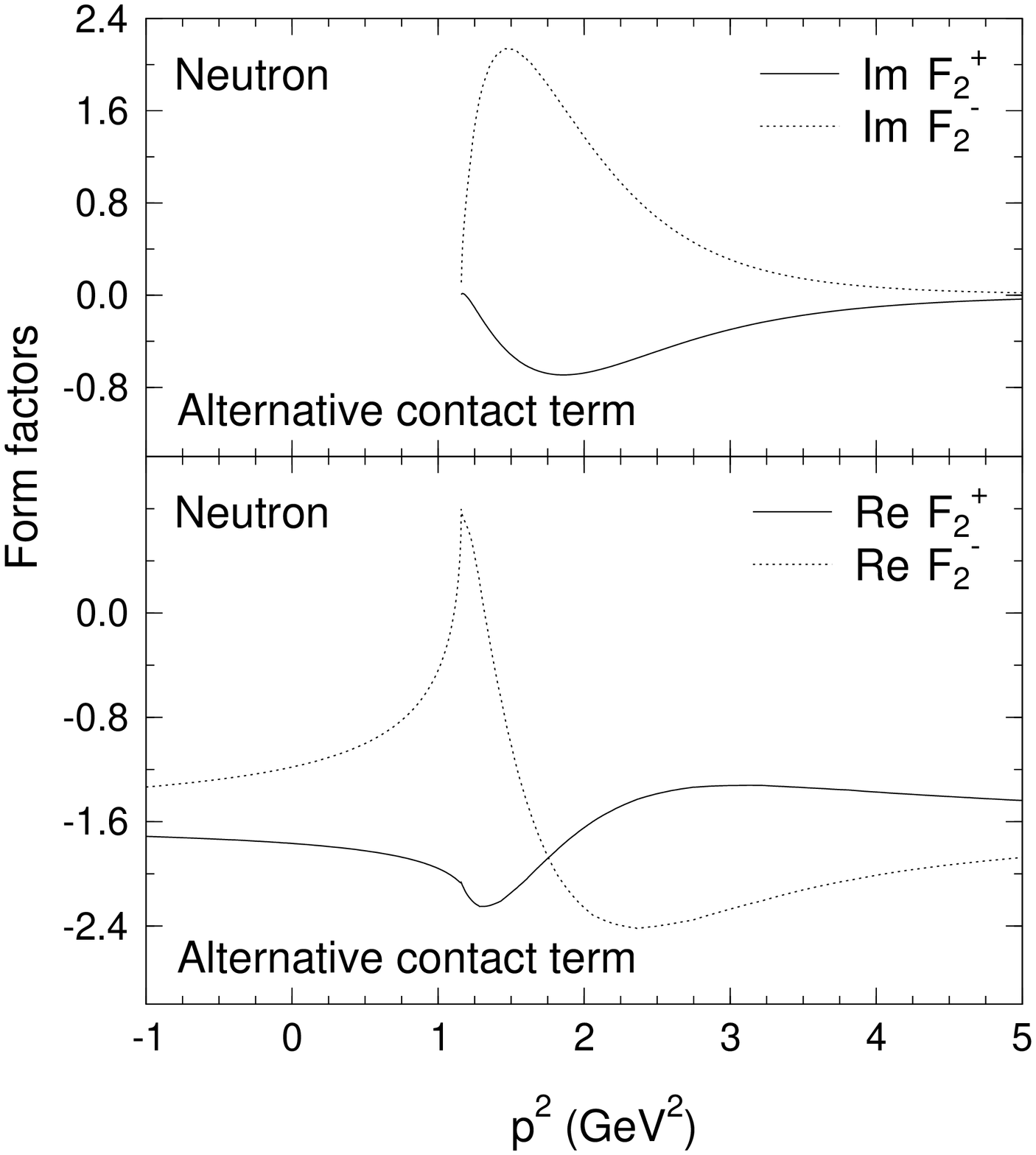}}
\caption[f8]{
The same as in Fig.~6, but for the neutron-photon vertex 
\label{fig8}}
\end{figure}

\begin{figure}
\epsfxsize 14. cm
\centerline{\epsffile[0 20 594 650]{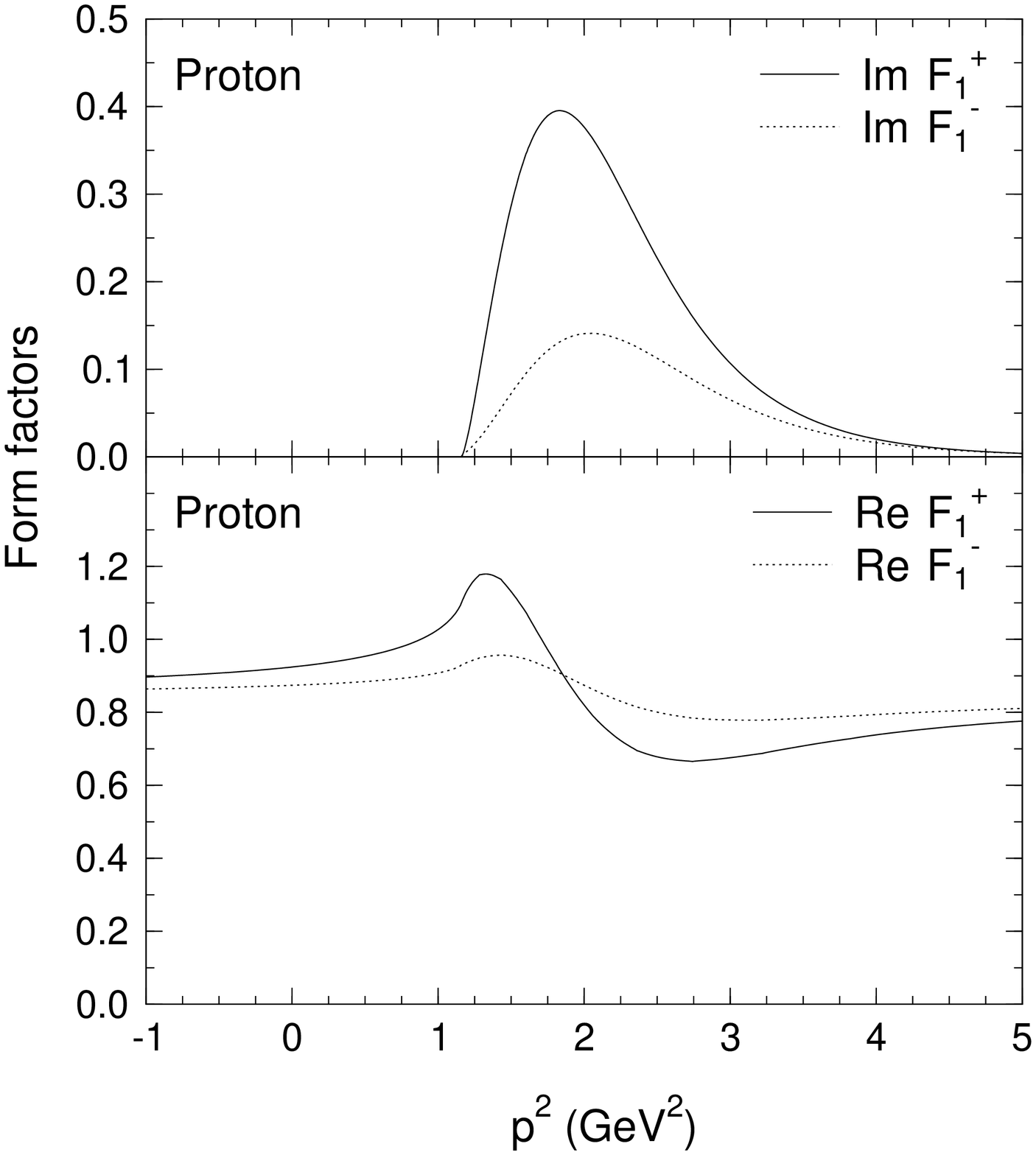}}
\caption[f9]{
The same as in Figs.~5 and 6, but for the form factors $F_1^{\pm}$. 
\label{fig9}}
\end{figure}

\begin{figure}
\epsfxsize 12 cm
\centerline{\epsffile[0 210 594 595]{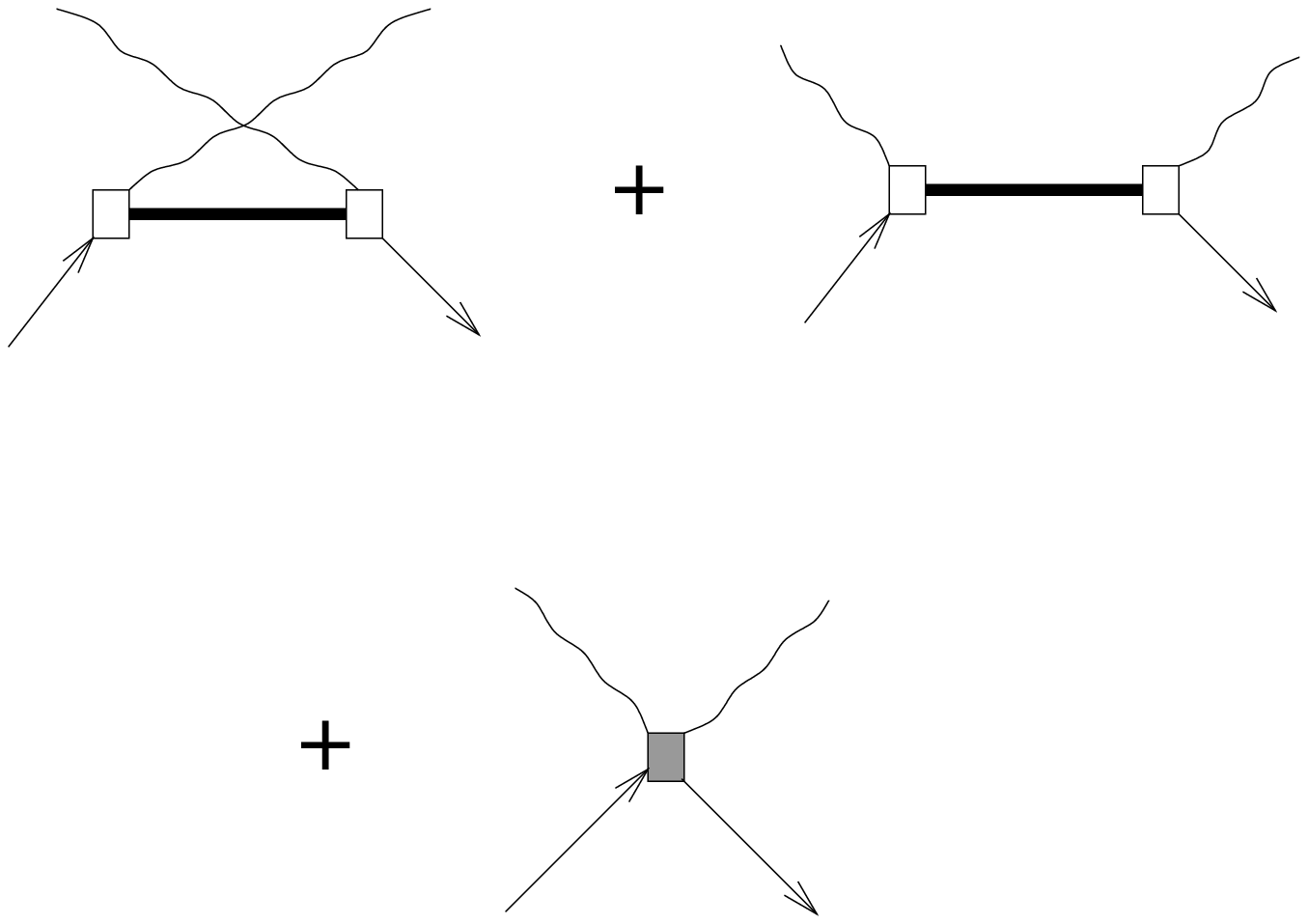}}
\caption[f10]{
The skeleton diagrams for the Compton scattering amplitude as given by
Eqs.~(\ref{eq:compt_split}-\ref{eq:compt_u}) and (\ref{eq:ct_me}). The notation
is as in Figs.~1 and 2, except all the external lines are on-shell and not
stripped away. The shaded square is the contact $\gamma \gamma N N$ term.
\label{fig10}}
\end{figure}

\begin{figure}
\epsfxsize 16 cm
\centerline{\epsffile[0 125 594 685]{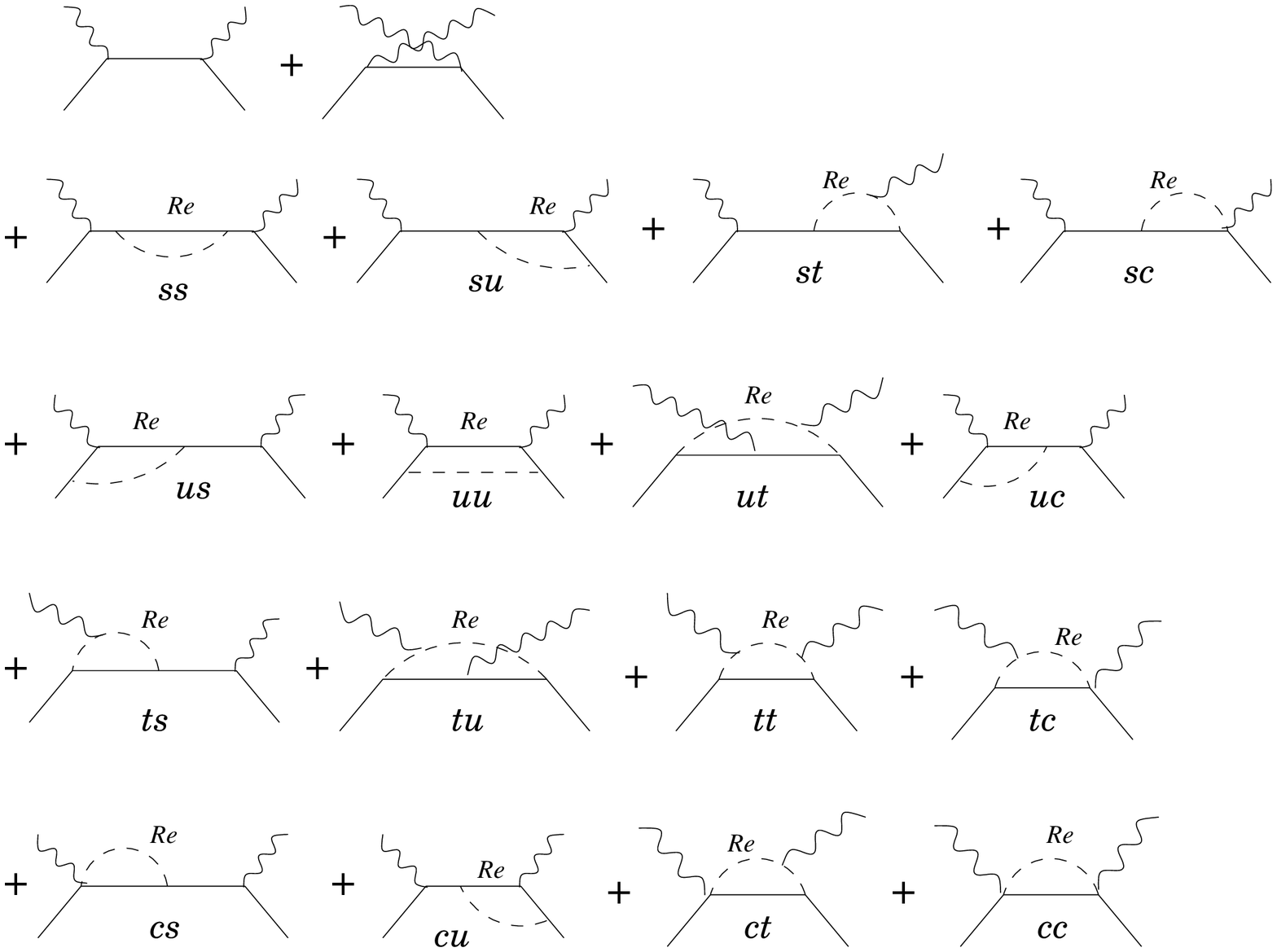}}
\caption[f11]{
The Feynman diagrams forming the K-matrix for Compton scattering up to second 
order in the ``potential" $V_{c' c}$, according to \eqref{k6}. The notation is
as in Fig.~1, and all the external lines are on-shell. The index {\it Re}
shows that
only the principal value part of the loop integrals are included. The
structure of the loop diagrams in terms of the tree diagrams for pion
photoproduction is denoted by {\it ss, su, st} etc., as explained in the text.
\label{fig11}}
\end{figure}

\begin{figure}
\epsfxsize 15.5 cm
\centerline{\epsffile[0 20 594 700]{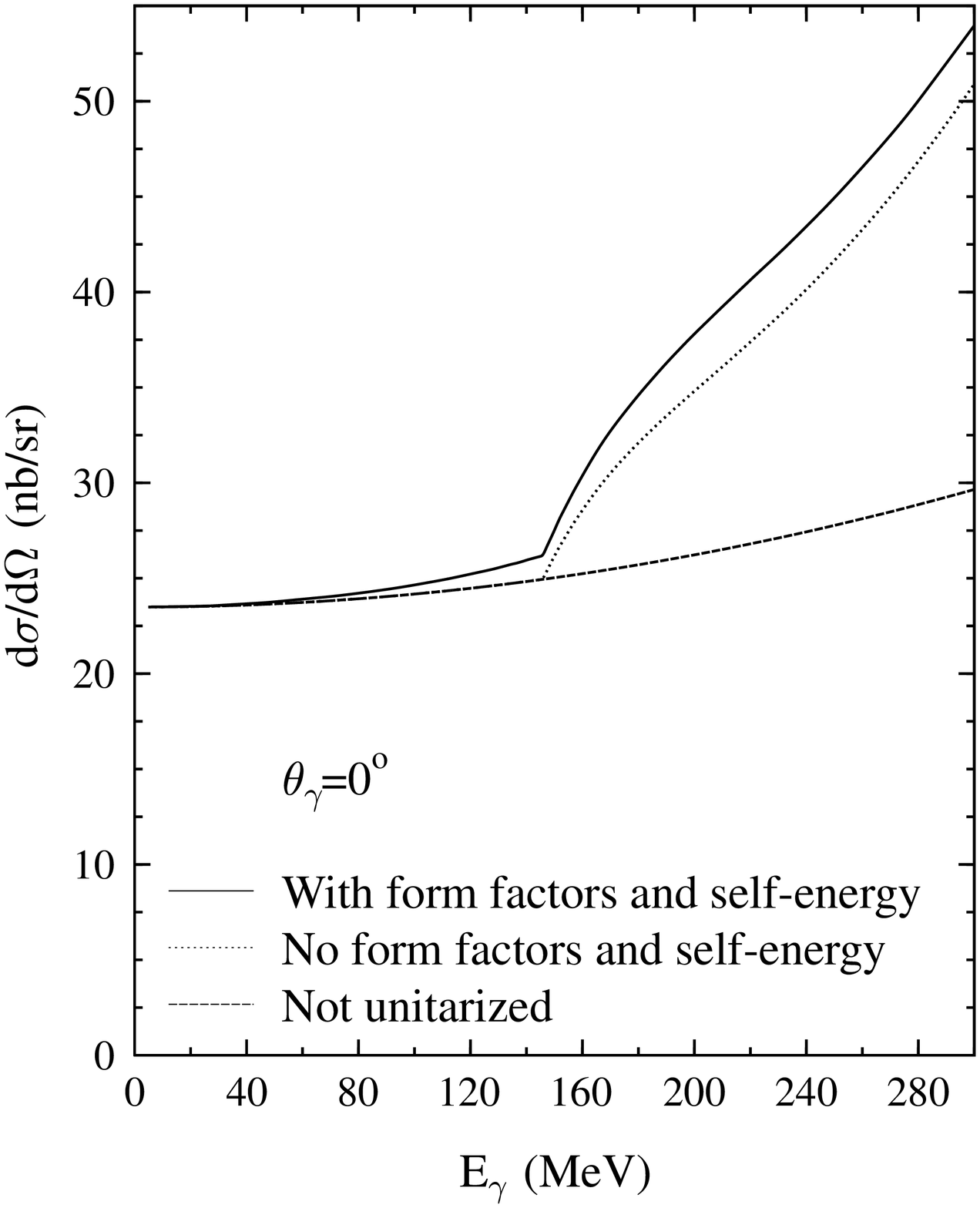}}
\caption[f12]{
The cross section for the Compton scattering in the forward direction as a
function of the photon energy,
obtained in the K-matrix formalism as described Section VI.
The solid line represents the calculation where the dressed (irreducible) 
nucleon-photon and nucleon-pion vertices, as well as the dressed nucleon 
propagator, are included in the K-matrix. The dotted line is the calculation
with the same K-matrix, but where the  bare vertices and the free nucleon 
propagator are used. The dashed line is the cross section calculated using only
the tree-level diagrams.  
\label{fig12}}
\end{figure}

\end{document}